\documentclass[a4paper,fleqn,usenatbib,useAMS]{mnras}

\usepackage[T1]{fontenc}
\usepackage{ae,aecompl}

\usepackage{longtable}
\usepackage{rotating}

\usepackage{graphicx}	
\usepackage{amsmath}	
\usepackage{amssymb}	
\usepackage{multirow,bigdelim,array,pdflscape,booktabs,psfig}
\usepackage{multicol}
\usepackage{newtxtext}
\usepackage[slantedGreek]{newtxmath}

\newcolumntype{C}[1]{>{\centering\arraybackslash}m{#1}}

\usepackage{float}
\usepackage{array}
\usepackage{makecell}
\usepackage{color,natbib}

\usepackage[normalem]{ulem}

\newcommand{\kms}{$\rm{\,km \,s}^{-1}$}

\title[LeMMINGs. III.]{LeMMINGs. III. The {\it e-}MERLIN Legacy Survey of the Palomar sample. Exploring the origin of nuclear radio emission in active and inactive galaxies through the [\ion{O}{iii}] -- radio connection.}

\author[R. D. Baldi et al.]{R.D.~Baldi$^{1,2}$\thanks{E-mail:ranieri.baldi@inaf.it},
D.R.A. Williams$^{3}$,
R.J. Beswick$^{3}$,
I. McHardy$^{2}$,
B.T. Dullo$^{4}$,
J.H. Knapen$^{5,6}$,
\newauthor L. Zanisi$^{2}$, 
M.K. Argo$^{3,7}$,
S. Aalto$^{8}$,  
A. Alberdi$^{9}$,
W.A. Baan$^{10}$,
G.J. Bendo$^{3,11}$,
\newauthor 
D.M. Fenech$^{12}$,
D.A. Green$^{13}$,
H.-R. Kl\"{o}ckner$^{14}$,
E. K\"{o}rding$^{15}$, 
T.J. Maccarone$^{16}$,
\newauthor 
J.M. Marcaide$^{17}$,
I. Mutie$^{18,3}$,
F. Panessa$^{19}$, 
M.A. P\'erez-Torres$^{9}$,
C. Romero-Ca\~nizales$^{20}$,
\newauthor
D.J. Saikia$^{21}$,
P. Saikia$^{22}$, 
F. Shankar$^{2}$, 
R.E. Spencer$^{3}$, 
I.R. Stevens$^{23}$, 
P. Uttley$^{24}$,
\newauthor
E. Brinks$^{25}$, 
S. Corbel$^{26,27}$, 
I. Mart\'i-Vidal$^{28}$, 
C.G. Mundell$^{29}$, 
M. Pahari$^{30}$,
M.J. Ward$^{31}$
}

\pubyear{2020}

\begin{document}
\label{firstpage}
\pagerange{\pageref{firstpage}--\pageref{lastpage}}
\maketitle

\begin{abstract}

What determines the nuclear radio emission in local galaxies? To address this question,  we combine optical [\ion{O}{iii}] line emission, robust black hole (BH) mass estimates, and high-resolution {\it e-}MERLIN 1.5-GHz data, from the LeMMINGs survey, of a statistically-complete sample of 280  nearby, optically active (LINER and Seyfert) and inactive (H{\sc ii} and Absorption line galaxies [ALG]) galaxies. Using [\ion{O}{iii}] luminosity ($L_{\rm [\ion{O}{iii}]}$) as a proxy for the accretion power, local galaxies follow distinct sequences in the optical--radio planes of BH activity, which suggest different origins of the nuclear radio emission for the optical classes. The 1.5-GHz radio luminosity of their parsec-scale cores ($L_{\rm core}$) is found to scale with BH mass ($M_{\rm BH}$) and  [\ion{O}{iii}] luminosity. Below $M_{\rm BH} \sim$10$^{6.5}$ M$_{\odot}$, stellar processes from non-jetted H{\sc ii} galaxies dominate with $L_{\rm core} \propto M_{\rm BH}^{0.61\pm0.33}$ and  $L_{\rm core} \propto L_{\rm [\ion{O}{iii}]}^{0.79\pm0.30}$. Above $M_{\rm BH} \sim$10$^{6.5}$ M$_{\odot}$,
accretion-driven processes dominate with $L_{\rm core} \propto M_{\rm BH}^{1.5-1.65}$ and $L_{\rm core} \propto L_{\rm [\ion{O}{iii}]}^{0.99-1.31}$  for active galaxies: radio-quiet/loud LINERs, Seyferts and jetted H{\sc ii} galaxies always display (although low) signatures of radio-emitting BH activity,  with $L_{\rm 1.5\, GHz}\gtrsim$10$^{19.8}$ W Hz$^{-1}$ and $M_{\rm BH}\gtrsim10^{7}$ M$_{\odot}$, on a broad range of Eddington-scaled accretion rates ($\dot{m}$). Radio-quiet and radio-loud LINERs are powered by low-$\dot{m}$ discs launching sub-relativistic and relativistic jets, respectively. Low-power slow jets and disc/corona winds from moderately high to high-$\dot{m}$ discs account for the compact and edge-brightened jets of Seyferts, respectively. Jetted H{\sc ii} galaxies may host weakly active BHs. Fuel-starved BHs and recurrent activity account for ALG properties. In conclusion, specific accretion-ejection states of active BHs determine the radio production and the optical classification of local active galaxies.

\end{abstract}

\begin{keywords}
  galaxies: active -- galaxies: jet -- galaxies: nuclei -- galaxies:
  star formation -- radio continuum: galaxies
\end{keywords}




\section{Introduction}

Supermassive black holes (SMBHs) are expected to reside at the centre
of all massive galaxies (e.g. \citealt{aller02,marconi04,shankar04}). The accretion on to SMBHs in Active
Galactic Nuclei (AGN) provides a major power source in the Universe and
is believed to regulate the evolution of galaxies, by injecting energy
and momentum (`feedback', \citealt{fabian12}). Such feedback sets the reciprocal relation
between the observed properties of SMBHs (e.g. central velocity dispersion $\sigma$ as a proxy of BH mass) and their host bulges 
(e.g. \citealt{magorrian98,gebhardt00,haring04}). Much of our understanding of AGN
phenomena (feeding and feedback) comes from the study of bright, powerful active SMBHs
that have accreted primarily during typical active periods of the order of a few 10$^{8}$ yr in a radiatively efficient mode, as Seyfert nuclei and QSOs \citep{marconi04,shankar04,cao07}.  During such a phase, a geometrically thin, optically thick,
standard accretion disc (SAD, \citealt{shakura73}), characterised by a high radiative efficiency mode
(mass-to-energy conversion efficiency of $\epsilon \sim$0.1, e.g. 
\citealt{shankar20}), is
needed both to reproduce the currently observed SMBH space density and
large-scale environment and to solve the Soltan argument \citep{soltan82}.

 However, the vast majority of SMBHs in the local Universe are in a quiescent state or, unless activity has been seen in some band, of extremely-low nuclear outputs
 (e.g. \citealt{huchra92,filho06,ho08}): the latter are regarded to host a low-luminosity AGN
(LLAGN, traditionally defined as having H$\alpha$ luminosities $\leq$ 10$^{40}$ erg s$^{-1}$, \citealt{ho97a}).  Unfortunately, the weakness of the radiative signals from
low-activity SMBHs make their study difficult. Problems include: i) confusion with brighter, non-AGN components, ii) obscuration and, iii) selection effects.

The scarcity of deep multi-band studies for large samples of LLAGN means that understanding  SMBH activity in the low luminosity regime is currently limited. Theories indicate that LLAGN are characterized by a
low accretion rate ($\dot{m}$) and/or low radiative efficiency
\citep{ho99,panessa07}. In models of radiatively inefficient
accretion flows (RIAFs; \citealt{narayan94,narayan95,yuan12a,yuan12b,white20}), the kinetic
energy is either advected with the gas into
the SMBH, or channelled into an outflow. The magnetic field combined with `puffed-up' geometrically thick structures could provide a plausible
mechanism for collimating axial outflows, accounting for a large fraction of observed jets in LLAGN \citep{maoz07,mezcua14}. 

Probably the most common
manifestation of LLAGN appears in the form of low-ionisation nuclear
emission-line regions (LINERs; \citealt{heckman80}), which are
detected in the nuclei of a large fraction of nearby galaxies
\citep{ho97b,kauffmann03c}. The current picture of LINERs is a `mixed bag' of objects, with some photoionized by stars, some by AGN,
and some perhaps excited by shocks \citep{allen08,sarzi10,capetti11b,singh13}. By assuming an active BH origin of
the LINER population (or a part of it), the transition from the
Seyfert to LINER regime corresponds to a decrease in the
Eddington ratio (defined as the ratio
  between the bolometric AGN luminosity and the Eddington luminosity, $L_{\rm Bol}$/$L_{\rm Edd}$)  \citep{ho08,kewley06}, accompanied by a
hardening of the ionizing spectrum. At the extremely low end of the Eddington luminosity, active SMBHs with low $\dot{m}$ and low radiative efficiency can even be hidden at the centre of (apparently) inactive galaxies, where star formation (SF) outshines the faint active nucleus. 

\setcounter{table}{0}
\begin{table*}
	\centering
		\caption{Radio and optical properties of the LeMMINGs (Palomar) sample.}
	\label{tab1}
	\begin{tabular}{lC{2.5cm}ccccccccc}
	\hline	
\hline
Name & Hubble    & class  & $\sigma$   & log $M_{\rm BH}$ & log $L_{\rm [\ion{O}{iii}]}$ &  log Edd  &  det & morph &  log $L_{\rm core}$  & log $L_{\rm Total}$  \\
           &              &  BPT      &    km s$^{-1}$ &  M$_{\odot}$  & erg s$^{-1}$  & ratio &    &         &  erg s$^{-1}$ & erg s$^{-1}$\\     
(1)       & (2)         & (3)       & (4)             & (5)               & (6)                 & (7)       & (8)   &  (9)          &  (10)           &  (11)  \\
\hline
NGC~7817 & SAbc    & H          &66.7  & 6.21  &  39.29 & $-$1.51  &  U & $-$  & $<$35.64 & $-$ \\
IC~10 	  & IBm?      & H        &35.5  & 5.11  &   37.13 & $-$2.57 & U  & $-$  & $<$32.91 & $-$  \\
NGC~147   &  dE5 pec & ALG   &  22    & 4.28  &  $-$    &  $-$       &  U  & $-$     &  $<$32.39  & $-$ \\
NGC~185   & dE3 pec  &  L       &  19.9  & 4.10  & 34.63 &  $-$4.06 & U & $-$  &  $<$32.33 & $-$   \\
NGC~205   &  dE5 pec  & ALG  & 23.3    & 4.34$^{*}$ &  $-$    &  $-$  &U & $-$  &  $<$32.36 & $-$  \\
\hline
\end{tabular}
\begin{flushleft} Column description: (1) source name; (2)   morphological galaxy type taken from RC3
\citep{devaucouleurs91}; (3) optical spectroscopic classification based on BPT diagrams
and from the literature. H=H{\sc ii}, S=Seyfert, L=LINER, and ALG=Absorption line galaxy. `j\,H' marks the jetted H{\sc ii} galaxies and `RL' identifies the RL AGN; (4) stellar velocity dispersion $\sigma$ (\kms) from
\citet{ho09}; (5) logarithm of BH mass ($M_{\odot}$) determined
from $\sigma$ \citep{tremaine02} or from direct BH mass measurements
(galaxies marked with $^{*}$, \citealt{vandenbosch16}) ; (6) logarithm of [\ion{O}{iii}] luminosities from \citet{ho97a} or from the literature (non corrected for extinction, see Paper~II for references); (7) logarithm of Eddington ratio ($L_{\rm Bol}$/$L_{\rm Edd}$); (8) radio detection
status: `I' = detected and core identified; `U' =
undetected; `unI' = detected but core unidentified; `I+unI' =  detected and core identified with additional unknown source(s) in the field; (9) radio
morphological class: A = core/core--jet; B = one-sided jet; C =
triple; D = doubled-lobed ; E = complex; (10)--(11) logarithm of radio core and total luminosities at 1.5 GHz (erg s$^{-1}$).  To convert the radio luminosities in erg s$^{-1}$ to W Hz$^{-1}$ at 1.5 GHz, an amount of +16.18 should be subtracted from log $L_{\rm core}$ and log $L_{\rm Total}$. The full table is available as supplementary material.
\end{flushleft}
\end{table*}

Understanding the physics of the accretion/ejection of matter in low-luminosity regimes is important for a variety
of reasons. First, the steep local AGN luminosity function shows that
LLAGN outnumber the QSO population by a few orders of magnitudes at
$z<0.3$ \citep{heckman14,saikia18}. The large abundance of local LLAGN indicates
that such a population represents the common mode of SMBH accretion at low redshifts and gives a
snapshot of the ordinary relation between the SMBH and its host. Second, the identification of the LLAGN population would help to constrain the occupation
fraction of active SMBH in galaxies. Such a quantity is unknown at low stellar
masses $<$10$^{9-10}$ M$_{\odot}$ \citep{greene12,gallo19}, and is
crucial for establishing the BH mass density function and to calibrate
the prescriptions for SMBH--galaxy growth of semi-analytical and
numerical models \citep{shankar09,barausse17}. Third, the discovery of large cavities of hot gas that
have been formed in the intra-cluster medium typically by jets from LLAGN in
massive elliptical galaxies (e.g. \citealt{dunn06,werner19}), indicates
that jet-mode feedback is a crucial aspect of the low-luminosity stage of SMBH activity \citep{heinz07,fabian12}.

In the last decades, the preferred method for finding active SMBHs was by
X-ray selection, but even the deepest current observations 
cannot probe the heavily absorbed sources (e.g. the {\it Chandra} Deep Fields, \citealt{alexander03}). High-resolution radio observations (providing quantities such as, e.g., Spectral Energy Distribution [SED] or brightness temperature [$T_{\rm B}$]) offer complementary,  direct views of BH accretion even in dusty environments, and can detect
AGN at $\dot{m}$ below those detectable in other wave-bands

In active galaxies, apart from stellar processes, a plethora of radio-emitting mechanisms associated with SMBH accretion can compete (see \citealt{panessa19} for a review). Relativistic or sub-relativistic jets which accelerate particles \citep{padovani16,blandford19}, disc winds which shock and sweep the interstellar medium (ISM) \citep{zakamska14} and outflowing magnetically-active coronae \citep{laor08}, are the main astrophysical phenomena which can account for cm/mm-wavelength radiation in the pc-scale regions of AGN. The combination of radio and optical data can provide more robust diagnostics to separate SF  and AGN components (see e.g. \citealt{best05b,smolcic08,kauffmann08,best12,radcliffe18,muxlow20}) and break the degeneracy among all the possible radio-emitting physical processes: the goal of this work.

Jets\footnote{The term `jet’ indicates outflow of ejected plasma that becomes collimated (unlike a wind), which transports outwards mass, energy and angular momentum. The bulk speed can generally be  relativistic, leading to a radio-loud AGN, or sub-relativistic, corresponding to a radio-quiet one.}, generally seen in radio, but sometimes in other bands as
well (e.g. in the optical, as in M\,87, \citealt{perlman99}) are an unambiguous indicator of nuclear
activity seen in LLAGN (e.g. \citealt{koerding06}). The increasing evidence of finding jets associated with LLAGN compared to the high-luminosity AGN is supported by sparse radio studies
of LLAGN and QSOs
(e.g. \citealt{kukula99,nagar00,mezcua14,padovani16}), although a few studies  have found contradictory results (e.g. \citealt{jiang07,macfarlane21}). This result can be
generalised by the fact that the radio loudness, i.e. the ratio
between the radio and the optical emission associated with the active SMBH, increases at lower bolometric luminosities and at low Eddington
ratios \citep{ho02,ho08,koziel17,laor19}. The presence of a link between the SMBH
capability of emanating bright jets and their AGN multi-band nuclear properties (bolometric luminosities, Eddington ratios, etc.) has been interpreted in the light of different accretion modes. Among active galaxies, RIAF discs and large SMBHs ($>$10$^{8}$ M$_{\odot}$)
are generally more efficient at accelerating particles to relativistic velocities
than SAD and small SMBHs
\citep{meier01,nemmen07,begelman12,mckinney12}. All these
favourable conditions are generally met in LLAGN, which despite weak
optical nuclei, display a broad variety of radio properties, from
flat-spectrum cores to steep-spectrum kpc-scale jets \citep{nagar00}. 

Since jets are
observed across all types of active BHs (Galactic
BHs, X-ray binaries [XRBs], and AGN), it is still under debate whether
the accretion--jet symbiosis is scale-invariant, regardless of the
accretion mode in, e.g., the so-called `Fundamental Plane of BH
activity' \citep{merloni03,falcke04,plotkin12,bonchi13,gultekin19}. This 3D plane
(radio luminosity $L_{\rm radio}$, X-ray luminosity $L_{\rm X}$
and BH mass $M_{\rm BH}$) is believed to unify all active BHs at different mass
scales, indicating common physical mechanisms in accretion and jet
production among  all accreting compact objects.  Historically,
in analogy with the states in XRBs, LINERs are thought to be the hard-state equivalent of an XRB, where the
radiation is produced inefficiently and the jet is dominating over the
disc emission ($L_{\rm X}$ $\propto \dot{m}^{2}$, e.g. \citealt{esin97}), while  Seyfert galaxies are thought
to be the equivalent of high-$\dot{m}$ XRBs ($L_{\rm X}$ $\propto
\dot{m}$, e.g. \citealt{fender03}), where jets are never seen\footnote{Dark decaying jets have been rarely seen in particular transitional stages of the soft state (e.g. \citealt{rushton12,drappeau17}).}.  However, recently there is compelling evidence that the AGN optical classifications are not likely to map in a 1-to-1 manner to the spectral states observed from accreting stellar mass BHs and a growing list of disc observational constraints of XRBs challenge its application to AGN (see review from \citealt{davis20}).  For AGN, the $\dot{m}$ break between LINERs and Seyferts is typically at about $10^{-3}$, while for the XRBs hard states are typically seen below 2 per cent of $L_{\rm Edd}$ \citep{Maccarone_state_trans,Vahdat2019}, and soft states are typically seen above that level. A few Seyfert galaxies also show evidence of Fourier power spectra with similar characteristics to those of hard-state XRBs \citep{Markowitz2003,Vaughn_4395} and (low-luminosity) Seyferts often show jetted structures (e.g. \citealt{kukula95,thean00,giroletti09,kharb17}), differently from soft-state XRBs.

To have a less biased view of the properties of SMBH accretion and the SMBH--host link at low masses, an accurate census of the accretion and
jet properties is needed. This requirement is satisfied by the LeMMINGs (Legacy {\it e}-MERLIN Multi-band Imaging of Nearby Galaxy Sample) survey\footnote{\url{http://www.e-MERLIN.ac.uk/legacy/projects/lemmings.html}} \citep{beswick14}. It consists of 1.5-GHz observations (and upcoming 5-GHz observations) of 280 nearby galaxies from the Palomar sample, which is usually considered to be one of the best selected and most complete samples of nearby galaxies. The two data releases (\citealt{baldi18lem}, Paper~I, and \citealt{baldi21}, Paper~II) represent the deepest 1.5-GHz survey of local ($<$110 Mpc) active and inactive galaxies at milli-arcsecond resolution and $\upmu$Jy  sensitivity with the {\it e}-MERLIN array. Only partial conclusions on the optical-radio connection have been already drawn in Paper~I because of incompleteness of the sample. The full coverage of the [\ion{O}{iii}]-line optical and radio data in this work allows to exploit the statistical completeness of the sample to investigate  the origin of the radio cores in local galaxies.

In Section~\ref{sect2} we present the LeMMINGs project and the radio and optical [\ion{O}{iii}] properties of the sample. We show the optical--radio diagnostics to explore the nature of the radio emission for each optical class in Section~\ref{sect3}. In Section~\ref{sect4} we discuss the results and focus on each optical class. We revise the radio properties of the LLAGN population and draw our conclusions about disc--jet coupling in LLAGN and SMBH-host connection in nearby galaxies in Section~\ref{sect5}. A supplementary section \ref{app} focuses on the BH mass function of local galaxies.

\section{The Survey and the sample}
\label{sect2}

The LeMMINGs sample represents a subset of 280 galaxies from the Revised Shapley-Ames Catalog of Bright Galaxies and the Second Reference Catalogue of
 Bright Galaxies \citep{sandage81} (details in Papers~I and II). The sample is taken from the optical spectroscopic Palomar survey \citep{ho97a}, selecting only targets with $\delta$ $>$20$^{\circ}$, to be accessible to the {\it e-}MERLIN array. The sample is optically selected ($B_{\rm T} <$ 12.5 mag), so has no radio bias, and a median distance of 20 Mpc.  Based on the updated optical emission-line diagnostic diagrams (BPT, \citealt{baldwin81,kewley06,buttiglione10}), the sample has been classified into H{\sc ii}, Seyfert, LINER
 and Absorption Line Galaxies (ALG) (see Papers~I and II). The galaxies for which the active SMBH is the main photoionising source are Seyferts (18/280) and LINERs (94/280).  The inactive
galaxies are represented by H{\sc ii} galaxies (140/280) where star forming regions populated by massive
young stars mainly photoionise the surrounding gas, and ALG (28/280), which are optically inactive
galaxies and do not show evident emission lines and are typically
in early-type hosts.

{\it e-}MERLIN observations at L band (1.2--1.7\,GHz) of the 280 Palomar galaxies are presented in Papers~I and II. An angular resolution of $\lesssim$200 mas and high sensitivity ($1\sigma\sim 0.8\,$mJy\,beam$^{-1}$) enabled the detection of radio
emission at pc scale of 44.6 per cent (125/280) of the sample. We detected the radio cores\footnote{We define core as the unresolved  central component of  the  radio source, which pinpoints the location of a putative SMBH, and may represent  a jet base, an unresolved disc-driven emission or a nuclear  stellar cluster (NSC).} with typical radio sizes of $\lesssim$100\,pc and radio core luminosities, $L_{\rm core}$, in the range $\sim$10$^{34}$--10$^{40}$ erg\,s$^{-1}$ (Table~\ref{tab1}). For 106 of the 125 detected sources we identified the radio core within the structure, co-spatial with the optical galaxy centre: 56/94 LINERs, 12/18 Seyferts, 5/28 ALGs and 33/140 H{\sc ii} galaxies. Conversely for the remaining 19 `unidentified' sources, the detected radio emission was not associated with the central optical nucleus. We resolved parsec-scale radio structures with a broad variety of morphologies (Tab.~\ref{tab1}): core/core–jet (class A, the most common), one-sided jet (class B), triple sources (class C), double-lobed (class D), and complex shapes (class E) with extents of $\sim$3--6600\,pc. There are 31 sources with clear jets (class B, C and D), which are referred to as `jetted' galaxies.  LINERs and Seyferts are the most luminous sources, whereas H{\sc ii} galaxies are the least. LINERs show elongated core-brightened radio structures while Seyferts reveal the highest fraction of symmetric morphologies. The majority of the 33 radio-detected H{\sc ii} galaxies have single radio core or complex extended  structures, but seven of them show clear jets. ALGs exhibit on average the most luminous radio structures, similar to that of LINERs. 

When considering the galaxy morphological types, most of the sources
are late-type galaxies (LTGs, from Sa to Sd, $\sim$71 per cent), with a smaller
fraction of elliptical and lenticulars (E and S0, early-type galaxies [ETGs]). In terms of BH masses, we derive the values using the stellar velocity dispersions $\sigma$
measured from optical spectra (mostly from \citealt{ho09},
see Table~\ref{tab1}) and the empirical $M_{\rm BH}$--$\sigma$ relation from
\citet{tremaine02}, analogous to what was done in Paper~I. For 36 sources, we use
direct SMBH measurements (derived from stellar and gas dynamics,
mega-masers, or reverberation mapping) available from the $M_{\rm BH}$
compilation of \citet{vandenbosch16}. Our sample of 280 galaxies
harbours BHs with 10$^{4}$ $\lesssim$ $M_{\rm BH}$ $\lesssim$ 10$^{9}$ M$_{\odot}$. As a test, we also derive the BH masses using the $M_{\rm BH}$--$\sigma$ relation from \citet{vandenbosch16}, which is based on a larger sample of BH masses than the one from \citet{tremaine02}. The two relations  agree with each other within 0.3\,dex, for intermediate BH masses, $\sim$10$^{6}$--10$^{8}$ M$_{\odot}$,  but differ at  higher  BH  masses within  0.5\,dex  and  at  lower  BH  masses within  1\,dex.  This scatter defines the errors on the $M_{\rm BH}$ measurements. LINERs, Seyferts and ALGs typically host SMBHs with masses $\gtrsim$10$^{7}$ M$_{\odot}$, while H{\sc ii} galaxies have typically lower BH masses.

\subsection{[\ion{O}{iii}] emission line}
\label{o3}

The [\ion{O}{iii}] emission line is an optical forbidden $\lambda$5007\,\AA\, transition, produced by gas photoionisation by a strong radiation field either from an active SMBH or star-forming region. In the case of an AGN, because of its high energy ionisation level, [\ion{O}{iii}] emission is extended from several kpc on galaxy scale down to the innermost central region on pc scale, in the so-called `narrow-line region' (NLR) \citep{kewley19}. Despite being slightly dependent on the AGN orientation and obscuration (e.g. \citealt{risaliti11,bisogni17}), [\ion{O}{iii}] line
luminosity, $L_{\rm [\ion{O}{iii}]}$, is a good (but not ideal) indicator of the
bolometric AGN luminosity ($L_{\rm Bol}$ = 3500$\times L_{\rm
  [\ion{O}{iii}]}$, \citealt{heckman04} for LLAGN). Since the line emitting region can extend up to several kpc, the main caveat of the [\ion{O}{iii}] line  is the contribution from other ionising sources, such as shocks and purely stellar processes
\citep{binette94,dickson95,dopita:m87}, which could result in an overestimate of the SMBH accretion power. Conversely, if the kpc-scale NLR extends beyond the spectroscopic slit used by \citet{ho95} (1--2 arcsec, from a few pc to $\sim$1 kpc for our sample), this may lead to an underestimate of the AGN bolometric luminosity.

\begin{figure}
	\includegraphics[width=0.5\textwidth,angle=180]{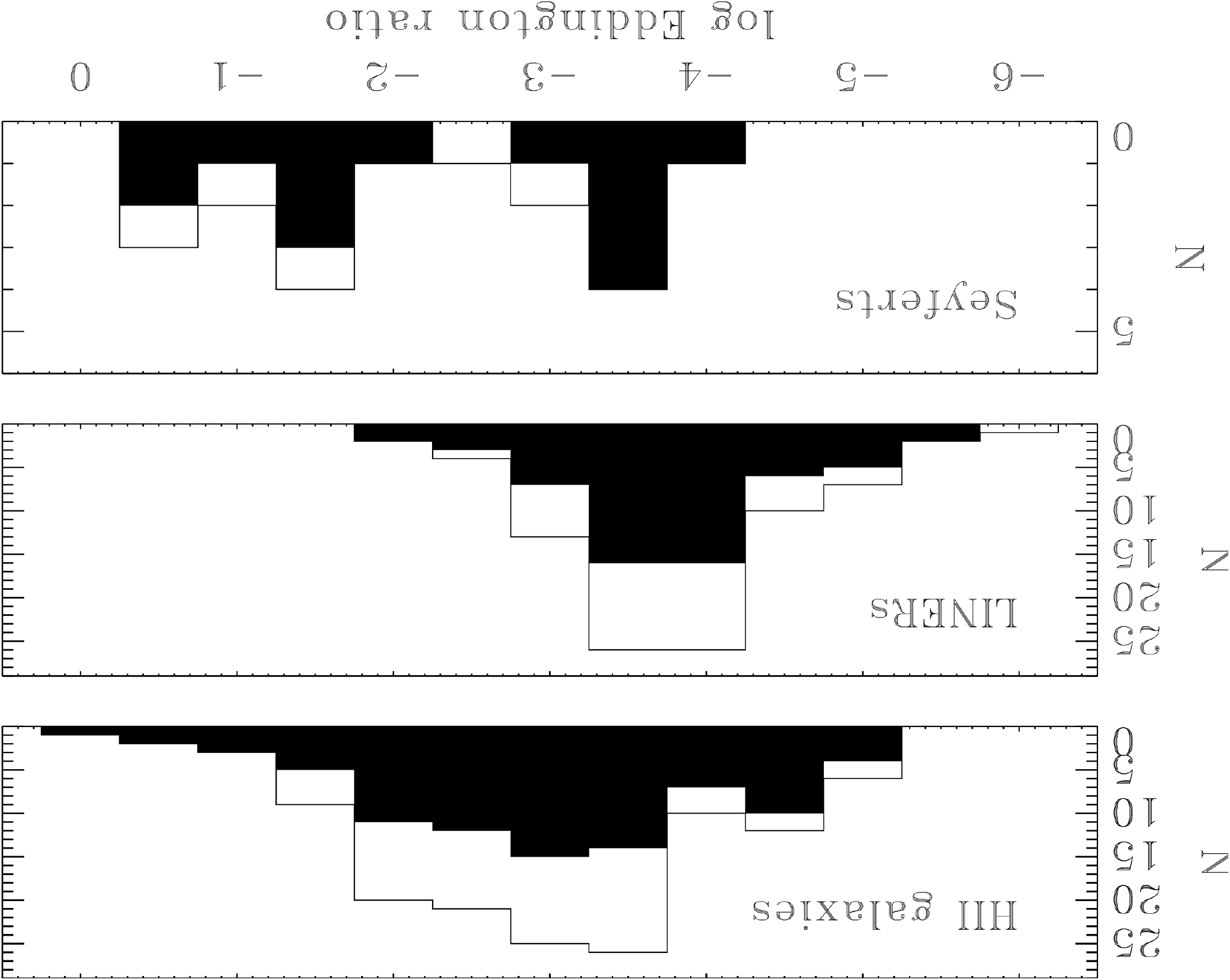}
        \caption[]{The Eddington ratio (ratio between the bolometric AGN luminosity, $L_{\rm Bol}$ = 3500\,$\times\,L_{\rm
  [\ion{O}{iii}]}$, \citealt{heckman04} for LLAGN and the Eddington luminosity) for the LeMMINGs sample with the [\ion{O}{iii}] line detected: H{\sc ii}, LINER, and Seyfert galaxies. The filled histogram identifies the radio detected sources.}
    \label{eddratio}
\end{figure}

For the LeMMINGs sample, we collect the  [\ion{O}{iii}] luminosities (Tab.~\ref{tab1}, not corrected for extinction) from  the Palomar optical survey \citep{ho97a}  and from the most recent surveys/observations (see Papers~I and II for more details). In our sample, Seyferts, LINERs and  H{\sc ii} galaxies
have mean [\ion{O}{iii}] luminosities of 5.0$\times$10$^{39}$, 3.5$\times$10$^{38}$, and 1.5$\times$10$^{38}$ erg
s$^{-1}$, respectively. For the vast majority of ALGs, upper limits on [\ion{O}{iii}] luminosities are not available.

Despite some caveats on $L_{\rm [\ion{O}{iii}]}$  as an indicator of AGN strength (e.g. \citealt{lamastra09,netzer09}, particularly at low luminosities), we can also calculate
the Eddington ratio, $L_{\rm Bol}$/$L_{\rm Edd}$, a gauge of the accretion rate, $\dot{m}$ 
(listed in Table~\ref{tab1} and shown in Figure~\ref{eddratio} for our sample). The
Eddington ratios of LINERs and Seyferts are generally below and above 10$^{-3}$, respectively, the typical threshold used to
separate between low and high $\dot{m}$ on to the SMBHs
\citep{best12}. For H{\sc ii} galaxies, in the conservative scenario that their line
emission is powered by the AGN, we note that, although less luminous
in line emission, they have intermediate Eddington ratios between
LINERs and Seyferts because their BH masses are typically lower than those of active galaxies.

Since optical emission lines (e.g. H$\alpha$ and [\ion{O}{iii}]) can be caused by hot, young  massive  stars  in  star-forming regions, we can also assume a stellar origin of the line emission: $L_{\rm [\ion{O}{iii}]}$ can thus be used as a SF rate (SFR) estimator, SFR (M$_{\odot}$/yr) = 7.9$\times$10$^{-42}$ $L_{\rm [\ion{O}{iii}]}$ (erg s$^{-1}$) following \citet{kennicutt98} and adopting log([\ion{O}{iii}]/H$\alpha$) = 0.0$\pm$0.5, \citealt{moustakas06,suzuki16}).

\subsection{Radio loudness}

The investigation of the nature of the radio emission in the LeMMINGs
galaxies would benefit from a separation between radio-loud (RL) and radio-quiet (RQ) AGN. For the former (i.e. radio galaxies, RGs) the relativistic jets are the main power source; for the latter different mechanisms (SF, sub-relativistic jets, and disc/corona winds) are likely to produce radio continuum \citep{panessa19}.  Accordingly, we have used the ratio
$L_{\rm core}/L_{\rm [\ion{O}{iii}]}$ defined as the spectroscopic radio loudness
parameter \citep{capetti06} based on local ellipticals. In general Seyfert galaxies and RQ AGN show a larger excess of line-emission (at a given radio-core luminosity) than those of RL objects, i.e.  $L_{\rm radio}$/$L_{\rm
  [\ion{O}{iii}]} <$ 10$^{-2}$ \citep{capetti07,ishibashi14}. Since the LeMMINGs sample contains a more heterogeneous mixture of early and late type galaxies than the sample
used by \citet{capetti06}, and RL AGN are known to be associated with the most massive BHs\footnote{More than 80 per cent of the local AGN with $M_{\rm BH}$ > 10$^{7.7}$ M$_{\odot}$ and with 150-MHz radio luminosities $>$ 10$^{21}$ W Hz$^{-1}$ are RL \citep{sabater19}.} \citep{chiaberge11}, we select the RL galaxies (black circles, Fig.~\ref{radioloudness}) based
on these two conditions: $L_{\rm core}$/$L_{\rm
  [\ion{O}{iii}]} >$ 10$^{-2}$  and $M_{\rm BH}$ > 10$^{7.7}$ M$_{\odot}$ (in agreement with local RL LLAGN, \citealt{baldi10b}). The selected RL AGN in the LeMMINGs sample are thus 18 LINERs and one jetted H{\sc ii} galaxy (see Tab.~\ref{tab1}). 

The radio loudness is known to increase with $M_{\rm BH}$  (Fig.~\ref{radioloudness}, see \citealt{nelson00,best05b}), and the presence of a bimodal, dichotomous or continuous distribution of radio loudness between RQ and RL AGN is still controversial (e.g. \citealt{kellerman89,cirasuolo03}). In fact, we note that other possible RL candidates (mostly LINERs) are close to our RL/RQ selection boundary. A clean RQ/RL separation simply based on the radio-loudness parameter and BH mass is not possible, as a detailed radio study of the jet with very-long baseline interferometers (VLBI)  would be needed (see \citealt{giovannini03,giovannini04}). Therefore, our criteria select the most bona fide RL AGN of the sample. Furthermore, the radio loudness is also known to inversely correlate with the Eddington ratio \citep{ho08,yang20}. However, since in our case both quantities are derived from the [\ion{O}{iii}] luminosities (i.e. the spectroscopic radio loudness $\propto L_{\rm [\ion{O}{iii}]}^{-1}$ and Eddington ratio $\propto  L_{\rm [\ion{O}{iii}]}$), an inverse correlation between these two quantities is hence expected.

\section{The optical--radio connection and the origin of the radio emission}
\label{sect3}

BPT diagrams attribute the optical emission lines of LINERs and Seyferts to photoionisation from AGN activity, whereas for H{\sc ii} galaxies the line ionisation is most likely due to SF processes. The paucity of studies on the nuclear emission of ALGs casts doubts on the nature  of this class of sources. However, the origin of the radio emission associated with an optical nucleus for each class is controversial, since several processes compete \citep{panessa19}. The [\ion{O}{iii}] line emission taken from the Palomar spectroscopic survey is extracted from a more extended nuclear region than the sub-arcsec radio emission detected by our {\it e}-MERLIN survey. A direct spatial comparison between the optical and radio emitting regions is therefore not possible in this work.  Consequently, a possible mismatch between the spatial distributions of the radio and optical regions could lead to a misinterpretation of the origin of the radio emission, e.g. genuine AGN-driven nucleus in a H{\sc ii} galaxy. Fortunately, forthcoming studies with integral-field spectroscopy (e.g. from the MaNGA survey, \citealt{bundy15}) on local radio-emitting galaxies (e.g. \citealt{roy21}, Mulcahey et al in prep.) and with {\it HST} photometry on the LeMMINGs sample itself (Dullo et al. in prep.) will clarify this point.

In analogy to Paper~I for a sub-sample (103 objects), we here carefully study the connection between the radio emission down to $\sim$50 pc and the sub-kpc [\ion{O}{iii}] emission, weighing the role of $M_{\rm BH}$,
for the entire LeMMINGs sample. Specifically, we explore the $L_{\rm core}$--$L_{\rm [\ion{O}{iii}]}$--$M_{\rm BH}$ relation to investigate the origin of nuclear radio emission in active and inactive galaxies.

\subsection{Radio properties versus BH mass}
\label{sec:bhmass}

Hints to possibly identify the genuine AGN origin of the nuclear
emission come from two quantities: the BH mass and the radio
luminosity. The former is often used as an indicator of BH activity as
active nuclei are preferentially associated with massive systems
(e.g., \citealt{chiaberge11}). The latter roughly establishes the likelihood of the source being radio-jet dominated, a sign of an
active SMBH \citep{cattaneo09}. Both quantities are connected in
active nuclei, as AGN tend to become more radio powerful
(e.g. radio louder) at larger $M_{\rm BH}$
(e.g. \citealt{best05b}).

\begin{figure}
	\includegraphics[angle=90,width=0.48\textwidth]{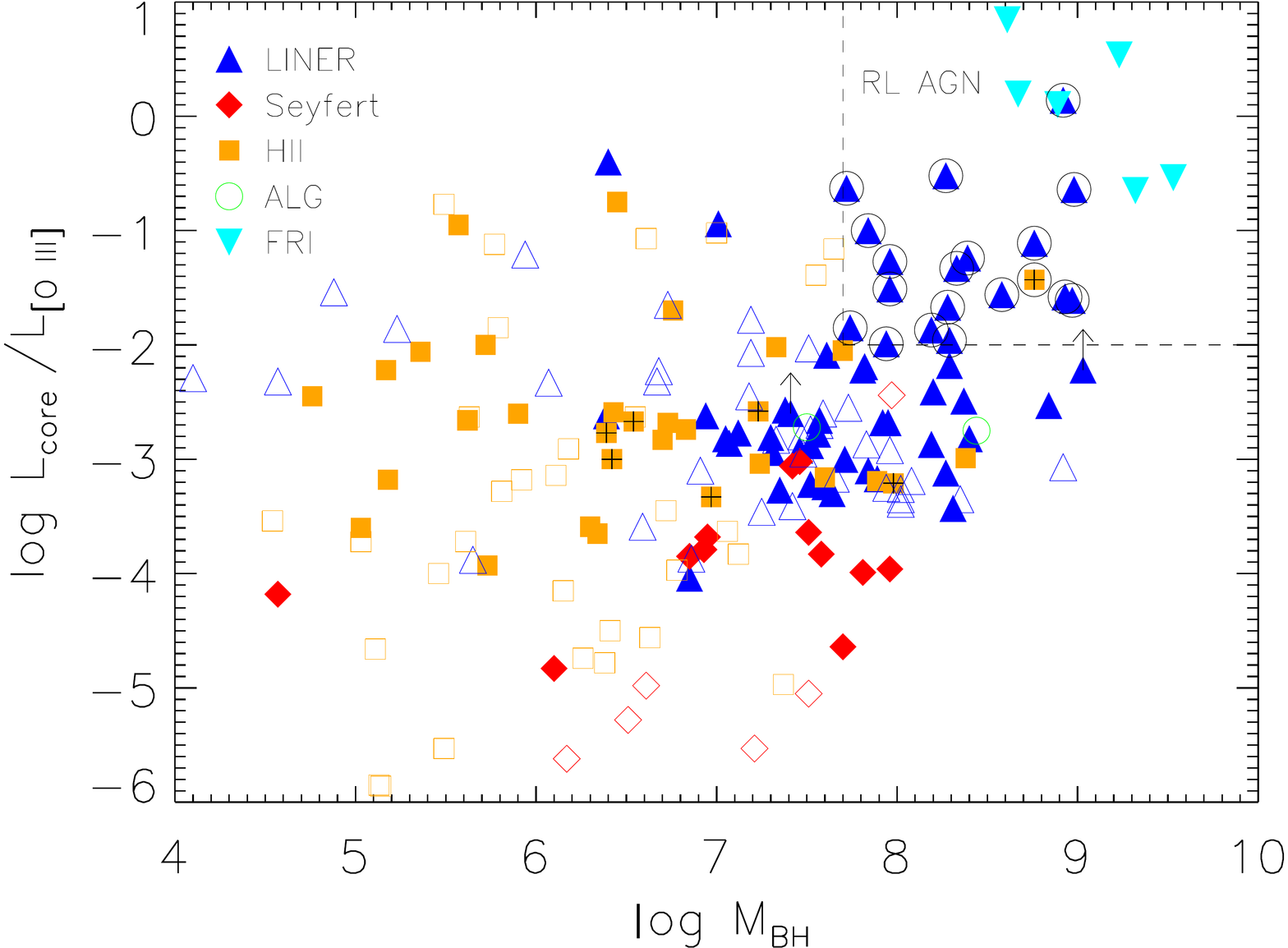}
	\vspace{-0.5cm}
    \caption[]{The spectroscopic radio-loudness parameter $L_{\rm
        core}/L_{\rm [\ion{O}{iii}]}$ as a function of $M_{\rm BH}$ (in M$_{\odot}$) for our sample. The area limited by the dashed line encloses the RL AGN (black circles), selected to have  $L_{\rm core}$/$L_{\rm [\ion{O}{iii}]} >$ 10$^{-2}$ and $M_{\rm BH} >$ 10$^{7.7}$ M$_{\odot}$. The symbols and colour codes
      are described in the legend: LINERs as blue up-pointing triangles, Seyferts as red diamonds,
and  H{\sc ii} galaxies as orange squares. The filled (empty) symbols are the radio detected  (non-detected with 3$\sigma$ upper limits) radio sources. The plus signs identify the jetted H{\sc ii} galaxies. We also include six 3C/FR\,Is (pale-blue down-pointing triangles) to highlight the locus of RL AGN.}
    \label{radioloudness}
\end{figure}

\begin{figure*}
	\includegraphics[width=0.88\textwidth]{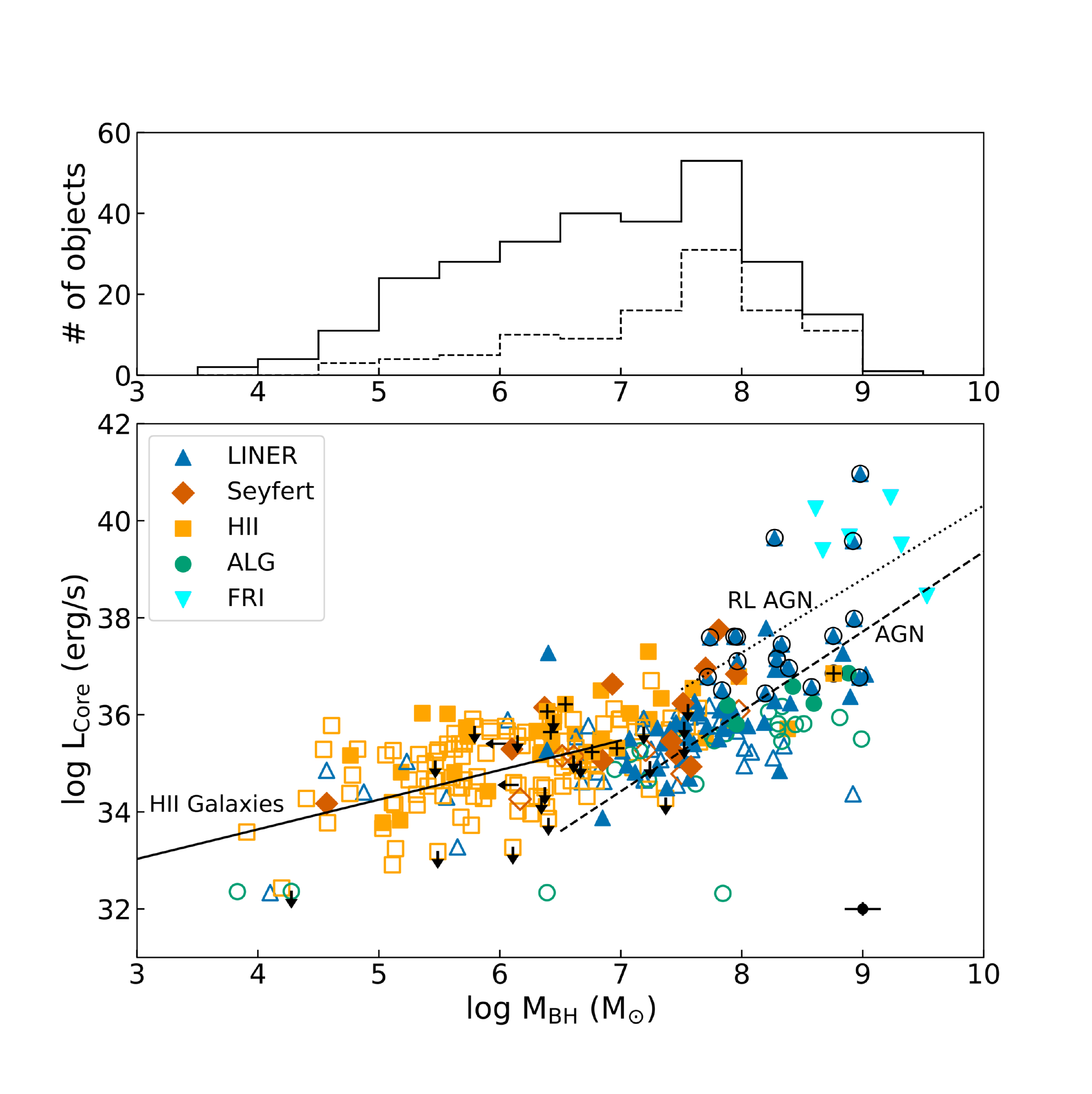}
	\vspace{-1.2cm}
        \caption[]{In the upper plot, the histograms of the entire sample (solid line) and of radio-detected sources (dashed line) in bins of BH mass. In the lower panel, 1.5-GHz core luminosities ($L_{\rm core}$
          in erg s$^{-1}$) as a function of the BH masses
          (in M$_{\odot}$) for the sample, divided per optical class
          (symbol and colour coded as in the legend, analogous to Fig.\ref{radioloudness}). The filled symbols
          refer to the detected radio sources, while the empty symbols
          refer to non-detected radio sources  (3$\sigma$ upper limits). The unidentified sources (detected but the radio core has not been identified)
          are the empty symbols with radio 3$\sigma$ upper limits. The jetted H{\sc ii}
          galaxies show an additional plus sign. The dotted line 
          represents the linear correlation found for RL AGN (marked by black circles) and 3C/FR~Is, the dashed line for active and jetted (ALG and H{\sc ii}) galaxies for $M_{\rm BH}$
          $\gtrsim$ 10$^{6.5}$ M$_{\odot}$ and the solid line for the non-jetted H{\sc ii} galaxies for $M_{\rm BH}$ $\lesssim$10$^{6.5}$ M$_{\odot}$ (see Sect.~\ref{sec:bhmass} for details and Table~\ref{statistics} for the best-fit parameters using censored-data statistics). In the bottom-right corner
          we show the typical error bars associated with the data points.}
\label{core_mbh}
\end{figure*}

\begin{figure}
	\includegraphics[width=0.47\textwidth,angle=180]{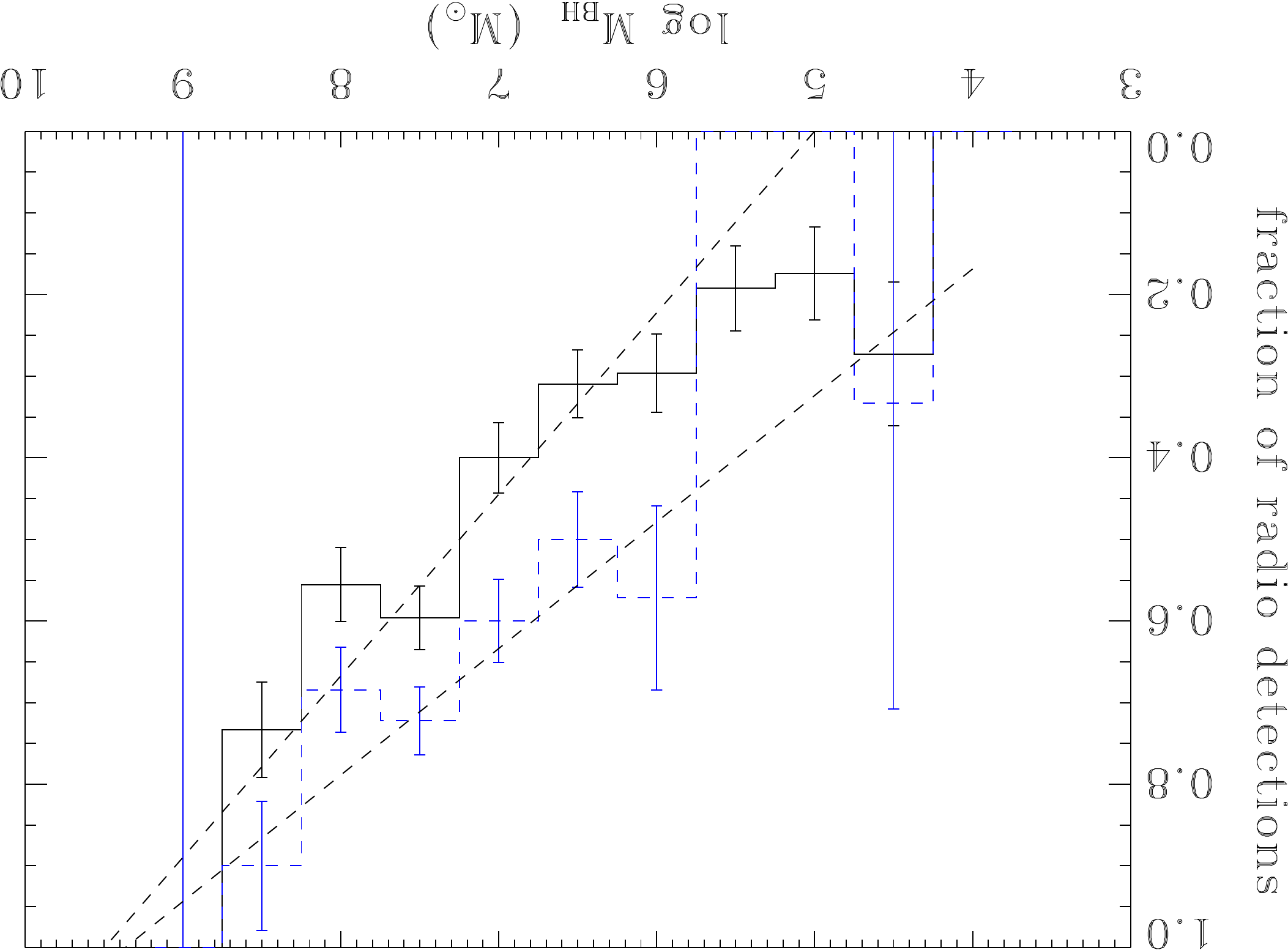}
        \caption[]{Histograms of the fraction of radio detected galaxies in the LeMMINGs sample in $M_{\rm BH}$ bins. The solid-line histogram considers the entire sample, while the blue dashed line is the distribution of only the  active galaxies (LINER, Seyferts and jetted ALG and H{\sc ii} galaxies). The two dashed lines represent the best fit for two distributions (Tab.~\ref{statistics} for the best fit parameters).}
\label{f_d}
\end{figure}

\begin{figure}
        \includegraphics[width=0.47\textwidth]{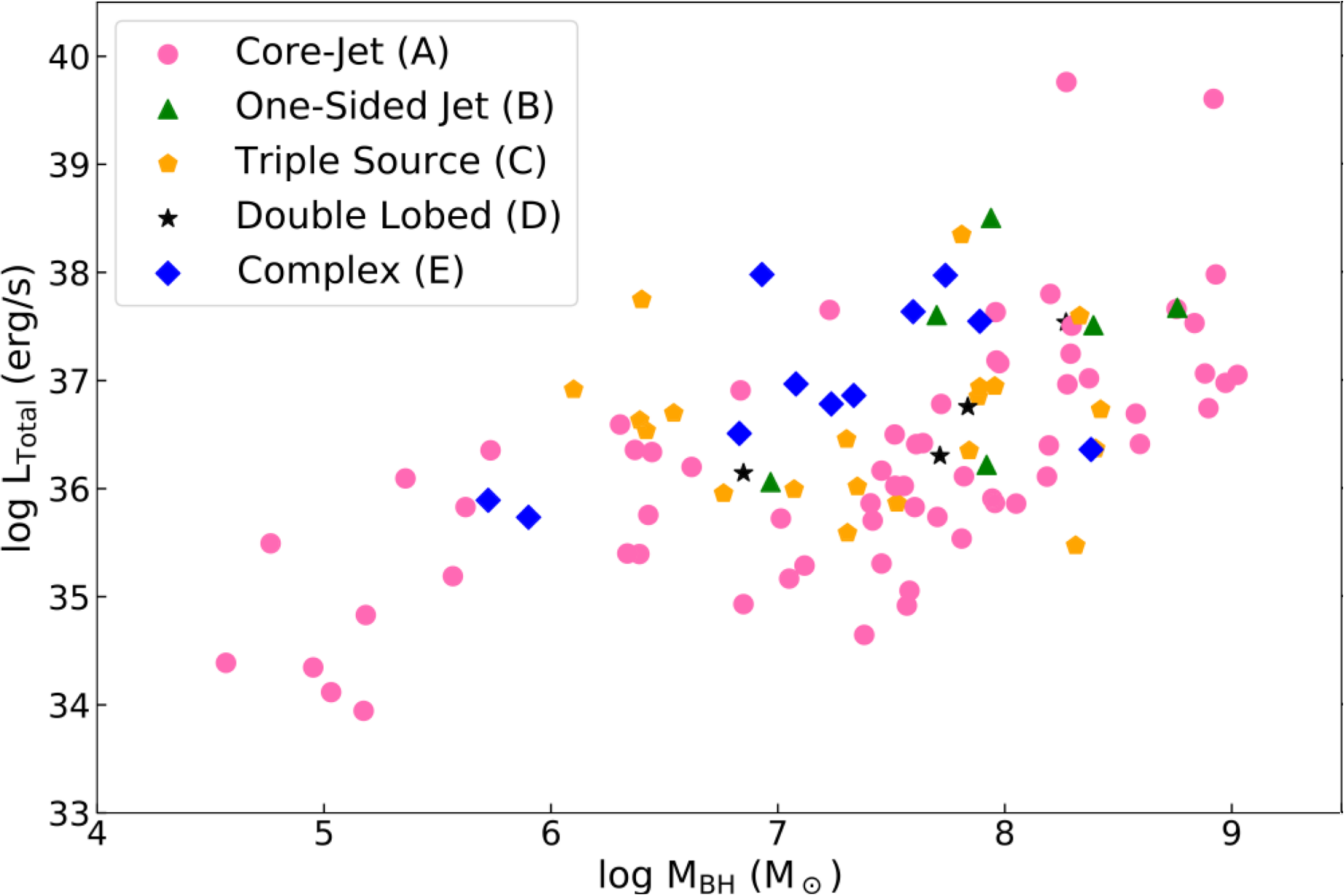}
\caption[]{The total 1.5-GHz luminosity
  ($L_{\rm Total}$ in erg s$^{-1}$) as a function of the BH masses
  (M$_{\odot}$) for the LeMMINGs sample, divided per radio morphological class
  A, B, C, D, and E (symbol and colour coded according to the legend).}
    \label{tot_mbh}
\end{figure}

\setcounter{table}{1}
\begin{table*}
\begin{center}
\caption{Statistical censored analysis of radio-optical correlations of the full LeMMINGs sample presented as a function of galaxy type.}
\begin{tabular}{lllc|cccc|cc} 
\hline
    $X$  & $Y$ & Fig. & sub-sample  & Stat & $\rho_{XY}$ & $P_{\rho_{XY}}$ & $\rho_{XY,D}$ &  Slope  & Intercept  \\
    (1)   &  (2)   & (3)  & (4) & (5) & (6) & (7) & (8) & (9) & (10)\\
\hline
    $\log M_{\rm BH}$  & $\log L_{\rm core}$     & \ref{core_mbh}     & active galaxies & S & 0.549 & $<$0.0001 & 0.526 & 1.65$\pm$0.25  &  23.0$\pm$8.1  \\
    $\log M_{\rm BH}$  & $\log L_{\rm core}$     & \ref{core_mbh}      &  non-jetted H{\sc ii}  & S & 0.575 & 0.0003  & 0.541  & 0.61$\pm$0.33  &  31.2$\pm$8.7  \\
    $\log M_{\rm BH}$  & $\log L_{\rm core}$  & \ref{core_mbh}         &  RL AGN + 3C/FR\,I          &    P & 0.526  &  0.00069 & 0.496    &   1.52$\pm$0.31   &  25.1$\pm$4.1 \\
    \hline
    $\log$ $M_{\rm BH}$  & $f_{d}$ &  \ref{f_d} & all  ($M_{\rm BH} > 10^{6.5}$ M$_{\odot}$)  & P & 0.951 &  0.00098 & $-$  & 0.22$\pm$0.03 &  -1.11$\pm$0.24 \\
    $\log$ $M_{\rm BH}$  & $f_{d}$  & \ref{f_d} &  active gal. ($M_{\rm BH} > 10^{6.5}$ M$_{\odot}$)  & P  & 0.925 &  0.0028 & $-$  & 0.15$\pm$0.03 &  -0.45$\pm$0.21 \\
    \hline
    $\log L_{\rm [\ion{O}{iii}]}$  & $\log L_{\rm core}$  & \ref{coreo3} & Seyferts          & K  & 1.190    &  0.0002     & 0.920   &   0.99$\pm$0.20  &  -3.5$\pm$5.5 \\
  $\log L_{\rm [\ion{O}{iii}]}$  & $\log L_{\rm core}$  & \ref{coreo3} &  RL AGN + 3C/FR\,I          &  P &  0.876    &  9.4$\times$10$^{-9}$  & 0.840 &  1.31$\pm$0.18  &  -13.1$\pm$7.0 \\   
  $\log L_{\rm [\ion{O}{iii}]}$  & $\log L_{\rm core}$  &  \ref{coreo3} & RQ LINERs  & S        &   0.579     &  $<$0.0001   & 0.542 &     1.21$\pm$0.28  &  -11.1$\pm$6.4            \\
    $\log L_{\rm [\ion{O}{iii}]}$  & $\log L_{\rm core}$ & \ref{coreo3} &   non-jetted H{\sc ii}   & S  & 0.347       &  $<$0.0001   & 0.323 &   0.79$\pm$0.30  &    5.0$\pm$10.2            \\
\hline
$\log$ Eddington ratio &  $\log L_{\rm core}$ & \ref{core-edd}   & Seyfert  &  K   &  0.591  &  0.0149 &  0.582 &0.98$\pm$0.36 & 37.6$\pm$13.4 \\
\hline 
    0.83 $\log L_{\rm [\ion{O}{iii}]}$ + 0.82 $M_{\rm BH}$ & $\log L_{\rm core}$        & \ref{fbhpo3}          &  Seyferts    &   K    &   1.072     &   0.0007   & $-$     &   0.84$\pm$0.21  &   3.3$\pm$10.2  \\
    0.83 $\log L_{\rm [\ion{O}{iii}]}$ + 0.82 $M_{\rm BH}$ & $\log L_{\rm core}$       & \ref{fbhpo3}       &   RL AGN  + 3C/FR\,I     &  P  &     0.886   & 3.8$\times$10$^{-9}$  & $-$   &  1.20$\pm$0.15  &   -9.1$\pm$6.1        \\
    0.83 $\log L_{\rm [\ion{O}{iii}]}$ + 0.82 $M_{\rm BH}$ & $\log L_{\rm core}$  & \ref{fbhpo3}     &  RQ LINERs    &    S  &   0.543     &   $<$0.0001  & $-$    &   0.83$\pm$0.19  &  4.0$\pm$9.5     \\
    0.83 $\log L_{\rm [\ion{O}{iii}]}$ + 0.82 $M_{\rm BH}$ & $\log L_{\rm core}$          & \ref{fbhpo3}         &   non-jetted H{\sc ii} & S & 0.426      &  $<$0.0001  & $-$ &   0.60$\pm$0.25  &   13.0$\pm$12.9       \\
    \hline
\end{tabular} 
\label{statistics} 
\end{center} 
\begin{flushleft}
Column description: (1)-(2) the two variables of the considered
relation; (3) Figure; (4) the sub-sample of galaxies for the tested correlation. `Active galaxies' include AGN (LINER and Seyferts) and jetted (ALG and H{\sc ii}) galaxies; (5)--(6)--(7) the statistical analysis used for the given sub-sample to calculate the associated linear regression coefficient  $\rho_{\rm XY}$ and the probability that there is no
correlation $P_{\rho_{XY}}$: $S$ for the censored generalised Spearman's
correlation coefficient (for $>$30 objects), $K$ for the censored generalised Kendall's $\tau$ correlation
coefficient (for $<$30
objects)  and $P$ for the Pearson correlation coefficient for fully a detected data set; (8) censored partial rank correlation coefficient  between X and Y adjusting for the target distance D; (9)-(10) the slope and the intercept of the best fits with their 1-$\sigma$ errors. 
\end{flushleft}
\end{table*}

The distribution of radio-detected galaxies as a function of
BH mass in the LeMMINGs sample is shown in Figure~\ref{core_mbh}
(upper panel) and it is clear that the detection fraction increases
with $M_{\rm BH}$ (a closer look in Fig.~\ref{f_d}). Similarly, a positive trend between the radio core luminosities, $L_{\rm core}$,
and $M_{\rm BH}$ is also observed (lower panel, Fig.~\ref{core_mbh}). In analogy to Paper~I, for the objects with $M_{\rm BH}\gtrsim 10^{6.5}$ M$_{\odot}$, a clear sequence includes all active and jetted galaxies, despite the large scatter. However, below 10$^{6.5}$ M$_{\odot}$, which mostly includes H{\sc ii} galaxies, a flatter correlation emerges. Interestingly, the small sub-group of jetted H{\sc ii} galaxies  have $M_{\rm BH} \gtrsim 10^{6.5}$ M$_{\odot}$ and generally
follow the sequence of the active galaxies. 

In order to assess the presence of correlations, we performed a
statistical censored analysis (\textsc{ASURV} package;
\citealt{feigelson85,lavalley92}) which takes into account upper
  limits. We used the \verb'schmittbin' task \citep{schmitt85} to
  calculate the associated linear regression coefficients for two sets
  of variables. The best fit is represented by the bisector of the two
  regression lines obtained by switching the $x$ and $y$ axes as
  dependent and independent variables.  In order to estimate the
  quality of the linear regression, for small data sets ($N<30$) we
  also derived the generalised Kendall’s $\tau$ correlation coefficient \citep{kendall38}
  between the two variables, using the \verb'bhkmethod'
  task. Otherwise for larger samples, we used the Spearman’s rank order
  correlation coefficient, using the \verb'spearman' task
  \citep{akritas89}. To measure the linear correlation between two fully detected sets of data, we estimate the Pearson correlation coefficient $r$ \citep{pearson95}. We also test the possible influence of the sample distance D in driving the established correlations, estimating the censored partial rank coefficient $\rho_{\rm XY,D}$ \citep{akritas96}. Table~\ref{statistics} reports the parameters of
  the statistical analysis of the correlations we analyse hereafter in this work.

We fit the linear (in a $\log$-$\log$ plot, hereafter) radio--$M_{\rm BH}$ correlation by including
all the active galaxies, i.e.  those which show characteristics of AGN activity in optical (emission line ratios) or radio (presence of jets) bands: RQ/RL LINERs, Seyferts and jetted ALG and H{\sc ii}
galaxies. For BH masses $\gtrsim$10$^{6.5}$
M$_{\odot}$, we find $L_{\rm
  core} \propto M_{\rm BH}^{1.65 \pm 0.25}$ with a Spearman's
correlation coefficient of 0.549. This value shows the likelihood of
the two quantities not correlating as less than 1$\times$10$^{-4}$
(Tab.~\ref{statistics}).

 In contrast, the non-jetted H{\sc ii} galaxies tend to fall on another radio--$M_{\rm BH}$ sequence with $M_{\rm BH}\lesssim$10$^{6.5}$ M$_{\odot}$. Indeed, their relation
 clearly flattens with respect to that of active galaxies as $L_{\rm core}\propto M_{\rm BH}^{0.61\pm
   0.33}$ with a chance (Spearman's $\rho$ = 0.575) probability of correlation of
 3$\times$10$^{-4}$ (Tab.~\ref{statistics}). The two incident radio--$M_{\rm BH}$ relations valid for star-forming and active galaxies strengthen the scenario of a different origin of the radio emission: SF-driven for non-jetted H{\sc ii} galaxies and AGN-driven for active galaxies.

\begin{figure*}
	\includegraphics[width=0.87\textwidth,angle=180]{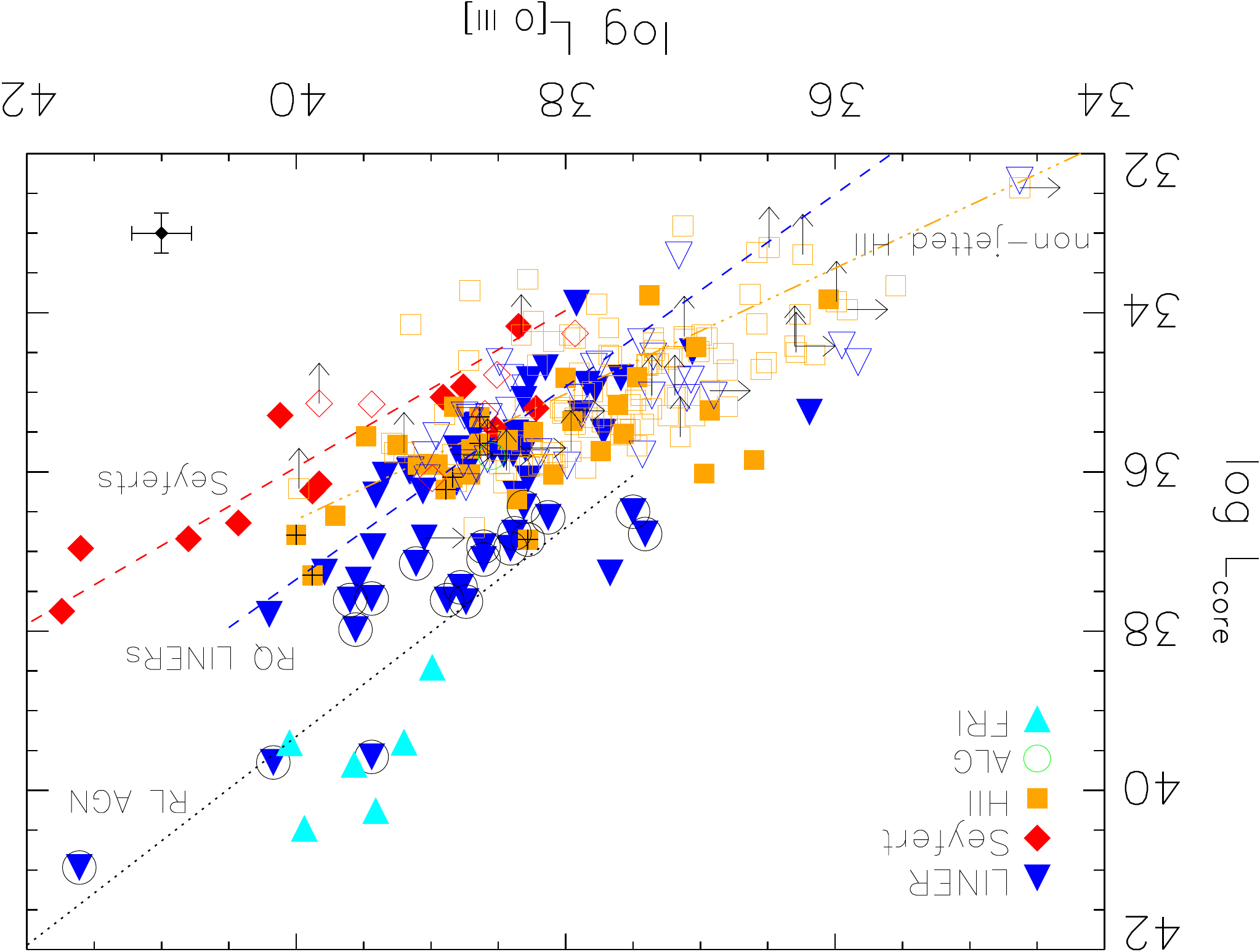}
        \caption[aaa]{[\ion{O}{iii}] line luminosity ($L_{\rm [\ion{O}{iii}]}$ in
          erg s$^{-1}$) vs 1.5-GHz core luminosities ($L_{\rm core}$ in
          erg s$^{-1}$) for the LeMMINGs sample. The different optical
          classes are coded (symbol and colour)  according
          to the legend of Fig~\ref{radioloudness}. The jetted H{\sc ii} galaxies show an
          additional plus sign. The filled symbols refer to the
          detected radio sources, while the empty symbols refer to
          non-detected radio sources  (3$\sigma$ upper limits). The unidentified sources (detected but the radio core has not been identified) are the
          empty symbols with radio 3$\sigma$ upper limits. The dotted line
          represents the linear correlation by fitting only the
          RL AGN (black circles) together with the 3C/FR~Is, the blue dashed line represents the fit by
          including RQ LINERs, the red dashed line for Seyferts and the yellow triple-dot-dashed line for non-jetted H{\sc ii} galaxies (see Table~\ref{statistics} for best-fit parameters). The
          black bars in the bottom-right corner indicate the typical errors
          for the data-points.}
    \label{coreo3}
\end{figure*}

Paper~I suggested that LINERs and low-power RGs are powered by a common
central engine, e.g. a RIAF disc with a coupled jet
\citep{falcke04,nemmen14}. Paper~I compiled a sample of six \citeauthor{fanaroff74} type~I (FR~I) RGs (i.e. 3C~66B, 3C~264, 3C~78, 3C~338,
  3C274, and 3C~189, from the Revised Third Cambridge Catalogue, 3C, \citealt{bennett62a}) at $z<$0.05 observed with
MERLIN at 1.5\,GHz (see Table 4 in Paper~I). In Figures~\ref{radioloudness} and \ref{core_mbh}, FR~Is
stand out from the rest of the sources for their large radio
luminosities and BH masses, and thus higher radio loudness, and clearly extend the LINER population to higher values. Therefore, we
fit the mass-radio relation valid for the whole RL AGN group (RL Palomar + FR~I galaxies, dotted line in Fig.~\ref{core_mbh}). We
find $L_{\rm core}\propto M_{\rm BH}^{1.52 \pm 0.31}$ with a
Pearson correlation coefficient of 0.526 and a two-sided
no-correlation probability significance of 6.9$\times$10$^{-4}$.

Two different radio-$M_{\rm BH}$ sequences below and above $\sim$10$^{6.5}$ M$_{\odot}$, respectively, for star-forming and active galaxies have been thus validated, also by excluding the distance effect in driving the correlations (as the censored partial rank coefficients only marginally decrease, see Tab~\ref{statistics}). Therefore, now that the presence of a break at $M_{\rm BH} \sim$10$^{6.5}$ M$_{\odot}$ has been highlighted, we can focus on the fraction of radio detections, $f_{d}$, as a function of $M_{\rm BH}$ for the entire LeMMINGs sample (Fig.~\ref{f_d}). For $M_{\rm BH} >$10$^{7.5}$ M$_{\odot}$, this fraction is 50 per cent, reaching $>$75 per cent in the last bins, $>$10$^{8.5}$ M$_{\odot}$. Conversely, the fraction flattens at $\sim$20 per cent below $\sim$10$^{6.5}$ M$_{\odot}$, where the SF was found to largely contribute in our sample (Fig.~\ref{core_mbh}). By considering only the sub-sample of active galaxies, the fraction of radio detections is unexpectedly higher than that of the entire sample in each BH mass bin. We also note that the $f_{d}$ reaches 100 per cent for active galaxies with $L_{\rm core}$ $\gtrsim$10$^{36}$ erg s$^{-1}$ ($\gtrsim 10^{19.8}$ W Hz$^{-1}$) and $M_{\rm BH}$ $\gtrsim$ 10$^{7}$ M$_{\odot}$. We fit the two detection fraction distributions as a function of BH mass for $M_{\rm BH} >$10$^{6.5}$ M$_{\odot}$ to avoid both the SF flattening and a single AGN case (Seyfert NGC~4395\footnote{The active galaxy with the lowest $M_{\rm BH}$ of the sample, 3.7$\times$10$^{4}$ M$_{\odot}$}). The best fits (see statistical parameters in Tab.~\ref{statistics}) are $f_{d}$ $\propto M_{\rm BH}^{0.22\pm0.03}$ and $\propto M_{\rm BH}^{0.15\pm0.03}$, respectively, for the whole LeMMINGs sample and for the active galaxies.

All of the radio morphological classes are represented along the
radio (both $L_{\rm core}$ and $f_{d}$) $-$ $M_{\rm BH}$ sequence. The total integrated radio luminosity, $L_{\rm Total}$, also broadly increases with the BH mass
(Fig.~\ref{tot_mbh}). There is no overall trend which links a specific
radio morphological classification to a given range of radio
luminosities.  The fraction of targets assigned to class B, C, D and E (e.g. clear jetted structures or complex) typically increases with $L_{\rm Total}$/$M_{\rm BH}$ because of their large extended, and sometimes diffuse, radio emission.  Core/core--jet structures (class A) are observed
across a slightly broader  $M_{\rm BH}$ and $L_{\rm Total}$ range
than the other classes. A similar effect has been also noted in the $L_{\rm core}$--$M_{\rm BH}$ plot. This could be the consequence of two effects: the preference for faint sources to appear as single cores and a Doppler boosted flux in the case of aligned sources. However, the latter scenario, i.e. a core boosting, is expected to marginally affect our results at 1.5\,GHz in the sub-mJy regime (only for RL AGN, see e.g. blazar-like heart of NGC~1275, \citealt{walker94}).

\subsection{Radio properties versus [\ion{O}{iii}] luminosity}
\label{sec:radioo3}

The comparison between radio and optical properties of galaxies is a powerful tool to explore the nature of their nuclear emission. The [\ion{O}{iii}] luminosity represents a robust upper limit (unless there is strong dust extinction) to any bolometric emission from an accretion disc in the case of an AGN \citep{lamastra09} (or from stellar emission in the case of an H{\sc ii} galaxy, \citealt{moustakas06}). The radio emission efficiency, i.e. the fraction of the radio emission produced with respect to the AGN (or SF) bolometric luminosity, offers a good diagnostic to investigate the nature of the nuclei.

The radio core luminosity as a function of [\ion{O}{iii}] luminosity is
shown in Figure~\ref{coreo3} for the LeMMINGs sample (without ALGs which lack optical counterparts, but including NGC\,5982 with a $L_{\rm [\ion{O}{iii}]}$ upper limit). The three optical classifications (LINER, Seyfert, H{\sc ii} galaxy) tend to cluster in
different regions of the [\ion{O}{iii}]--radio diagram, despite a large overlap, similarly to what has been noted in the analogous plot for a smaller sub-sample in Paper~I. Seyferts are up to 100 times more luminous in the [\ion{O}{iii}] emission line compared with LINERs for a given $L_{\rm core}$, or equivalently have 100 times less luminous radio cores. The H{\sc ii} galaxies have the lowest radio and [\ion{O}{iii}] luminosities, whilst the
LINERs have intermediate values. Analogously to Paper~I, we
report that the LINERs appear to broadly follow a linear trend in the
[\ion{O}{iii}]-radio plot. Similarly to Fig.~\ref{core_mbh}, the relation formed by H{\sc ii} galaxies shows a  break with respect to that valid for active galaxies at lower luminosities. Compared to the results from Paper~I, we note that Seyferts show a distinct relation with respect to the other classes. The current complete sample allows us to derive more robust relations by using the censored data analysis (see Table~\ref{statistics}) than those obtained in Paper~I.

First, Seyfert galaxies clearly show a striking linear correlation in the
form $L_{\rm core}$$ \propto L_{\rm [\ion{O}{iii}]}$$^{0.99 \pm 0.20}$ with a
(Kendall's $\rho$ = 1.190) probability of 0.0002 of being caused fortuitously.  At higher radio
luminosities, RL AGN follow the 3C/FR\,Is on a steeper linear fit, $L_{\rm
  core}$$\propto L_{\rm [\ion{O}{iii}]}$$^{1.31 \pm 0.18}$ (Pearson's P value = 9.4$\times$10$^{-9}$). RQ LINERs fill
the gap between RL AGN and Seyferts with a correlation almost parallel to
the one found for RGs, i.e. $L_{\rm core}$$\propto L_{\rm [\ion{O}{iii}]}$$^{1.21 \pm 0.28}$ (Spearman's P value $<$ 0.0001). At low powers, the sequence of non-jetted H{\sc ii} galaxies clearly flattens with respect to that formed by active galaxies, but with a broader scatter,
i.e. $L_{\rm [\ion{O}{iii}]}$$\propto L_{\rm core}$$^{0.79 \pm 0.30}$ (Spearman's P value
$<$ 0.0001). Once the distance dependence has been considered, these optical-radio relations are still statistically valid, because the censored partial rank coefficients are only slightly smaller than the generalised correlation coefficients (see Tab.~\ref{statistics}).

\begin{figure}
        \includegraphics[width=0.48\textwidth,angle=180]{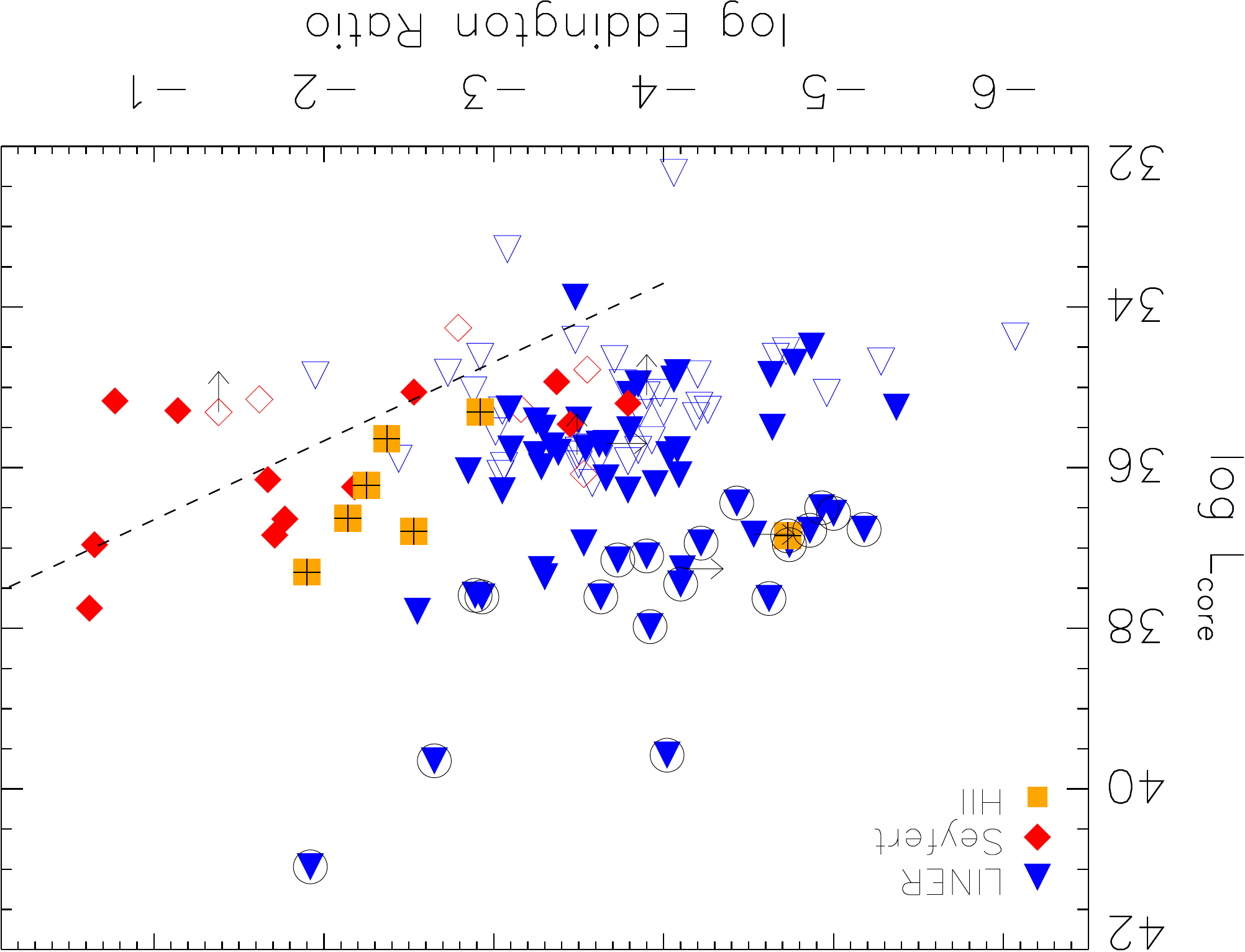}
        \caption[aa]{Eddington ratio (whose bolometric luminosity is estimated from $L_{\rm [\ion{O}{iii}]}$, see Sect.~\ref{o3}) vs 1.5-GHz core luminosity (erg s$^{-1}$) only for active galaxies (LINER, Seyferts and jetted H{\sc ii}
          galaxies, and no ALGs for the lack of their optical counterparts). Symbols and colours, described in the legend, are coded analogously to Fig.\ref{radioloudness}. The dashed line represents
          the best fit for Seyferts (by using censored-data statistics, see Table~\ref{statistics}).}
    \label{core-edd}
\end{figure}

Beyond the optical classification, the
sources with a core/core--jet or triple structure tend to follow the
correlation of the active galaxies, as they are mostly classified as
LINERs and Seyferts. Similarly, most of the ETGs also follow the correlation of active galaxies. Conversely, we do not find any statistically significant correlation between the Eddington ratios and $L_{\rm core}$ for our sample (see Fig.\ref{core-edd}, probability of fortuitous correlations $>$0.04). Nevertheless, we note for Seyferts: i) a mild positive trend  where $L_{\rm core}$ and  Eddington ratio linearly increases with a slope 0.98$\pm$0.36 with a large scatter $\sim$1 dex (Kendall's $\rho$ = 0.591 [ essentially unchanged correlation coefficient, if considering the distance dependence] and probability of a null correlation 0.0149, Tab.~\ref{statistics}]) and ii) the jetted morphologies correspond to objects with higher Eddington rates.

\subsection{The optical Fundamental plane of BH activity}

Observational attempts at explaining the similarities between the
X-ray and radio properties of SMBHs and XRBs eventually
culminated in the discovery of the so-called `Fundamental Plane of Black Hole
Activity' \citep[hereafter `FPBHA'][]{merloni03,heinz03,falcke04}. The
FPBHA is a 3D hyper-plane that correlates the radio luminosity with the
X-ray luminosity, scaled by the BH mass, and seems to hold for
XRBs and active SMBHs. By using a similar approach, \citet{saikia15}
introduced a new version of the FPBHA using the [\ion{O}{iii}] luminosity as
a tracer of the accretion, instead of the X-ray luminosity and found
an analogous correlation for active BHs. The advantage of  using the [\ion{O}{iii}] line instead
of the X-ray data is the accessibility from the ground with reasonable resolution. Full coverage of the optical and radio data for the entire LeMMINGs sample allows a complete analysis of the Fundamental Plane (FP). New X-ray {\it Chandra} observations and the X-ray-based FP of the LeMMINGs sample will be addressed in forthcoming papers (Williams et al., in prep.)

\begin{figure*}
        \includegraphics[width=0.87\textwidth,angle=180]{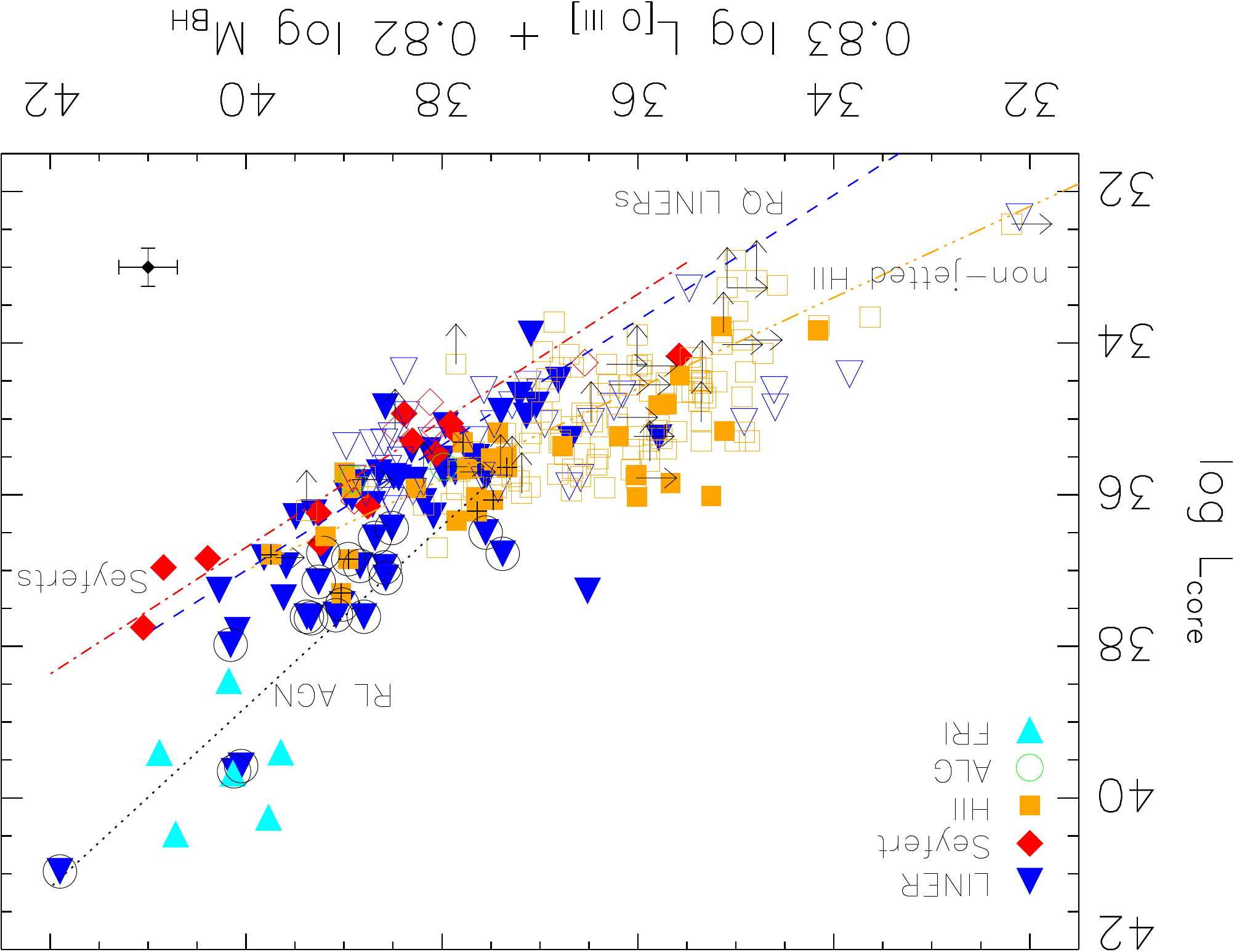}
        \caption[aa]{The fundamental plane of BH activity
          (FPBHA) in the optical band for the LeMMINGs sample with
          {\it e}-MERLIN data, i.e. $L_{\rm core}$ vs 0.83$\times$log
          $L_{\rm [\ion{O}{iii}]}$ + 0.82$\times$$\log$$M_{\rm BH}$ (1.5-GHz core and [\ion{O}{iii}]
          luminosities in erg s$^{-1}$ and BH masses in
          M$_{\odot}$) parameterised as expressed by
          \citet{saikia15}. Colour and symbols are explained in the legend. The
          filled symbols refer to the detected radio sources, while
          the empty symbols refer to non-detected radio sources (3$\sigma$ upper limits). The
          unidentified sources are the empty symbols with radio 3$\sigma$ upper
          limits. The dotted line
          represents the best fit for RL AGN (large black circles) and 3C/FR~Is, the blue dashed line represents the fit for RQ LINERs, the red dot-dashed line for Seyferts and the yellow triple-dot-dashed line for non-jetted H{\sc ii} galaxies (see Table~\ref{statistics} for best fit parameters using censored-data statistics). The
          black bars in the bottom-right corner indicate the typical errors
          for the data-points.}
    \label{fbhpo3}
\end{figure*}

Figure~\ref{fbhpo3} depicts the optical FPBHA of the LeMMINGs
galaxies, by using the same
parametrisation expressed by \citet{saikia15}. Note that these authors
used 15-GHz {\it VLA} observations at a resolution of $\sim$0$\farcs$13,
comparable with that of our {\it e-}MERLIN data.  Yet, their observations are possibly less contaminated by SF and more sensitive to the
nuclear optically-thick radio emission than our 1.5-GHz observations because of the higher radio frequency.

In Paper~I, despite a limited number of sources, we noted that the
active galaxies appear to broadly follow a common correlation in the optical FP,
stretching their luminosities up to FR~I RGs. Nevertheless,
Paper~I also noted the three classes are slightly stratified within the FP scatter
across the radio luminosities, possibly reminiscent of the different
accretion states for each optical class. Here, thanks to the sample completeness, we are able to recognise that different
optical classes move separately along the plane. LINERs still appear to follow the FR~Is at lower luminosities. The RL (FR~I, \citealt{parma86}) jetted H{\sc ii}, NGC 3665, lies on the RL AGN relation and the rest of the jetted H{\sc ii} galaxies agrees with the FPBHA of RQ LINERs. Seyferts, on the other hand, seem to cover a different sequence in the plane because of their lower radio core luminosities. Non-jetted star forming galaxies stand out from the rest of the sample because of their smaller BH masses and lower
[\ion{O}{iii}] and radio luminosities. Therefore, similar to the approach
taken in Paper~I, we prefer to
fit the FPBHA for the different classes, by using the parametrisation introduced by \citet{saikia15} (in units of erg s$^{-1}$
for luminosities and M$_{\odot}$ for BH masses)\footnote{We refrain from providing a new optical FPBHA parametrisation independent of that of \citeauthor{saikia18} because of the presence of a large fraction of upper limits in radio and [\ion{O}{iii}] data and the fact that [\ion{O}{iii}] luminosity is not an absolute estimator of the AGN activity. We will dedicate an accurate analysis on the X-ray-based FPBHA,  by using the state-of-the-art censored-data statistics (Williams et al., in prep.).}:

\vspace{-0.5cm}
$$\log L_{\rm core} = (0.83 \,\ \log L_{\rm [\ion{O}{iii}]} + 0.82 \,\, \log M_{\rm BH}) \, m + q $$

\noindent
where \citeauthor{saikia15} found $m$ = 1.0 and $q$ = $-$3.08 for their
sample. Table~\ref{statistics} collects all the statistical results
obtained from fitting the FP for different classes. Seyferts exhibit a tight
linear correlation even evident to the naked eye, with a slope of $m$=0.84$\pm$0.21. The generalised Kendall's $\tau$ coefficient of the Seyfert FP is 1.072
which indicates that two axis quantities do not correlate with a
probability of 0.0007. The whole LINER population shows a large scatter in the FP. A better understanding of this population is possible by separating RL and RQ LINERs, which follow
different tracks in Fig.~\ref{fbhpo3}: RL LINERs extend the FR~Is to lower regimes, whereas RQ in general tend towards the locus of Seyferts. We search for linear correlations
separately for the two LINER classes. For RL AGN (RL LINERs + FR~Is), the fit clearly
appears steeper than the Seyfert plane, with a slope of $m$=1.20$\pm$0.15
(Pearson's P value = 3.8$\times$10$^{-9}$). For the RQ LINERs, the plane is statistically consistent
with the one defined by Seyfert galaxies but it extends to lower
luminosities with a larger scatter, bridging the gap between RL LINERs and Seyferts: $m$ = 0.83$\pm$0.19 with a (Spearman's) P value $<$0.0001. The
non-jetted H{\sc ii} galaxies break the generic FP of the active galaxies with a flatter slope ($m$ =
0.60$\pm$0.25, Spearman's P value $<$0.0001) extending to the lowest luminosities
and BH masses of the sample.

Although  $L_{\rm core}$ has been found to correlate with both $M_{\rm BH}$ and $L_{\rm [\ion{O}{iii}]}$ for the LeMMINGs sample (Sect.~\ref{sec:bhmass} and \ref{sec:radioo3}), a mutual dependence could stretch the optical FPs we observe. The scatters of the FPs of the optical classes (FWHM $<$0.8 dex) are typically smaller than the scatters of the relations
defined for each class in the radio-$M_{\rm BH}$ and the radio-[\ion{O}{iii}] planes
(FWHM $<$1 dex).  This suggests that the observed FPs are not
stretched  by any particular dependence on one of the three quantities ($L_{\rm core}$, $M_{\rm BH}$, $L_{\rm [\ion{O}{iii}]}$)
which establish the plane, as it has been also confirmed by other studies (e.g. \citealt{saikia18a}). Nevertheless, the large number of upper limits in the radio and line measurements may undermine the reliability of the fitted FPs. To test the accuracy of the established FPs, in the cases of non-detected Seyfert and LINER galaxies in the radio band, if we assign a core
luminosity extrapolated by the $M_{\rm BH}$-$L_{\rm core}$ relation found
for active galaxies, the corresponding data-points would lie on the plane of their
corresponding class. Analogously, non-detected ALGs (in
[\ion{O}{iii}], but detected in radio) would sit between RL and RQ LINERs. Conversely,
non-jetted H{\sc ii} galaxies would follow the plane of the RQ LINERs if they had the core luminosities assigned by the $M_{\rm BH}$-$L_{\rm core}$ relation valid for AGN at their BH masses. This result reinforces the discrepancies of the observed FPs and their physical meanings for the different optical classes. In summary, star forming and active galaxies follow different FPs, as also evident in the radio-[\ion{O}{iii}] plane (Fig.~\ref{coreo3}), which are likely a consequence of different origins of their radio emission in relation to their main source of energy (stellar or BH accretion). 

\vspace{-0.5cm}
\section{Discussion}
\label{sect4}

The LeMMINGs survey has unveiled radio cores at the centre in around half of our sample of local Palomar galaxies. Here we investigate the origin of the nuclear 1.5-GHz emission and discuss the optical--radio association in the context of physical models of disc--jet coupling\footnote{Disc--jet coupling refers to the connection between the inflowing accretion mode and the outflowing mode, in the form of e.g., jets, winds, or slow outflows \citep{panessa19}.} and SF for each optical class (see Table~\ref{sketch} for a simple summarising sketch, see also \citealt{panessa19}).

\subsection{LINERs}
\label{liner}

LINERs are amongst the most frequently detected and most luminous sources in the LeMMINGs sample. They are mostly detected in ETGs and feature bright radio cores and jets.

Near-infrared, optical and X-ray nuclear emission in LINERs has generally been interpreted as related
to an active SMBH
\citep{terashima02,balmaverde06a,gonzalez06,flohic06,kharb12}, although reservations have been raised  in the past (e.g., stellar-dominated optical nuclei,  \citealt{capetti11c}). In contrast, several pieces of evidence convincingly argue in favour of an AGN origin of the radio emission
in LINERs. First, their high brightness temperatures  measured with {\it e}-MERLIN and VLBI
($T_{\rm B}>$10$^{7}$ K, Papers I and II, \citealt{falcke00}) suggest synchrotron emission from (mildly to highly) relativistic jets. Second, core luminosities for LINERs
correlate with their BH masses up to the classical RL regime with $M_{\rm BH} \sim 10^9 M_{\odot}$, suggesting that the mass of the SMBH plays a major
role in the production of radio emission. Third, LINERs are
often associated with either symmetric or asymmetric pc/kpc-scale jets with luminosities of $\sim$10$^{36}$ erg
s$^{-1}$, similar to low-power RGs \citep{fanti87}. Such morphologies and powers cannot easily be
explained by SF, as the latter resides in the sub-mJy regime below 10$^{34}$ erg
s$^{-1}$ \citep{bonzini13,mascoop21}. Fourth, the jets tend to be
core-brightened, similar to small FR~Is \citep{capetti17a}, suggesting that the jets are
probably collimated and launched relativistically on parsec scales, but rapidly slow down along
the jet propagation axis on kpc scales. These characteristics are similar to what is
seen and modelled in nearby low-luminosity RGs
\citep{morganti87,giovannini05,massaglia16,rossi20}.

The analysis of LINERs in this work has led to rather different results
compared to what we found in Paper~I, probably due to the inhomogeneous population of LINERs in the latter. Here we identify two sub-classes, RL and RQ.  A group of eighteen RL LINERs have the lowest Eddington ratios
($\lesssim$10$^{-4}$) and are the brightest sources of the sample. They typically
reveal core--jet and twin jet morphologies, reside in ETGs and extend the
relations of 3C/FR~Is to low radio and [\ion{O}{iii}] luminosities.  The complementary RQ group of LINERs have slightly
higher accretion rates than the RLs (Eddington ratios mostly $\sim$10$^{-4}$--10$^{-3}$) and form an intermediate optical--radio relation between RL AGN and
Seyferts.

Similarities in the correlations of RL LINERs and FR~Is
support the idea that they likely share the same central engine, capable of launching relativistic jets.  The correlation between radio and optical (continuum) emission found for FR~Is is best explained as the result of a single emission process in the two bands (i.e. non-thermal synchrotron emission from the base of the relativistic jet, \citealt{chiaberge:ccc}). Analogously, an [\ion{O}{iii}]-radio correlation found for FR~Is and low-luminosity RL LINERs \citep{balmaverde06b} suggests a similar ionising central source. 
Therefore, this interpretation is plausibly also valid for the RL LINERs in our sample,  indicating that they represent the scaled-down version of FR~Is with lower luminosities and accretion rates, as also reported in previous studies \citep{falcke04,allen06,balmaverde08}. Multiple theoretical
and analytical studies  of FR~Is (e.g.
\citealt{meier01,begelman12,mckinney12}) suggest that they are powered by a RIAF
disc, usually as an advection-dominated accretion flow (ADAF, see reviews by
\citealt{narayan94,narayan08}), which can
efficiently produce jets. As LINERs are usually radio-louder
than the other optical classes and similar to low-power RGs \citep{capetti06,kharb12}, a model of an ADAF disc with a low-power jet (jet-dominated accretion flows, JDAF, \citealt{falcke04}) has been used as well to describe the disc--jet coupling of LINER-like AGN.

Conversely, the RQ LINERs appear to deviate from the RL AGN
correlations, pointing to a different interpretation for their central engine. A possible truncated disc with an
optically thick disc at larger radii
\citep{chen89,maoz07,narayan08,nemmen14} could represent an
intermediate regime between the low-$\dot{m}$ JDAF model for RL LINERs and high-$\dot{m}$ SAD models for luminous Seyferts. The accretion flow may
begin as a thin SAD but, at a certain transition
radius, gradually switches from a cold to a hot ADAF mode, resulting
in composite characteristics with weaker sub-relativistic jets and slightly higher
accretion rates. This result is broadly consistent with what has been found for a general dual population of RL and RQ LINERs in the local Universe (e.g.,
\citealt{chiaberge05,balmaverde06b}).

\begin{table}
\begin{center}
\caption{Simplistic sketch of the origin of radio emission in relation to the accretion modes in local galaxies.}
\begin{tabular}{c|C{0.01cm}cC{1cm}C{2.3cm}} 
class  &  & radio    &  $\dot{m}$  &    disc  \\
\hline
{\scriptsize RL LINER}   & &  {\scriptsize relativistic jets}   & $\lesssim 10^{-3}$ &  {\scriptsize RIAF} \\
{\scriptsize RQ LINER}  &   & {\scriptsize sub-relativistic jets} &  $\lesssim 10^{-3}$  &  {\scriptsize truncated thick disc} \\
\multirow{2}{*}{{\scriptsize (RQ) Seyfert}}   & \rdelim\{{2}{20pt} & {\scriptsize sub-relativistic jets} & $\lesssim 10^{-2}$  &   {\scriptsize JED/truncated slim disc} \\
                       &  & {\scriptsize disc/corona wind}  & $\gtrsim 10^{-2}$ &  {\scriptsize SAD} \\
{\scriptsize ALG}    &  &  {\scriptsize (sub-)relativistic jets}  &  $\lesssim 10^{-3}$ &  {\scriptsize recurrent/starving RIAF} \\
{\scriptsize jetted H~II} &  \multicolumn{2}{c}{{\scriptsize sub-relativistic jets and SF}} &  $\lesssim 10^{-3}$ &  {\scriptsize RIAF?}   \\
{\scriptsize non-jetted H~II} &  & {\scriptsize SF}                    &               &    \\
\hline
\end{tabular} 
\label{sketch} 
\end{center} 
\begin{flushleft}
Column description: (1) optical class;  (2) radio properties: sub-/relativistic jets, disc or magnetically-active corona wind, and SF; (3) Eddington-scaled accretion rate $\dot{m}$; (4) disc mode (see Sect. \ref{liner}--\ref{hii} for details).
\end{flushleft}
\end{table}

\subsection{Seyferts}
\label{seyfert}

Seyfert galaxies show the highest radio detection fraction in the LeMMINGs sample. Two thirds of the detected Seyferts are found in spiral galaxies and one third in ETGs. The Seyferts with the highest Eddington ratios and BH masses are the most luminous  at 1.5\,GHz, featuring edge-brightened radio morphologies.

Seyfert nuclei lie in a different
region of the optical--radio plane with respect to RL LINERs and FR~Is: for a similar radio luminosity, they show a
significant optical excess (2--3 dex). On the one hand, their
optical line excess is
qualitatively similar to that observed in luminous type-I QSO, where their
high-energy output is interpreted as a result of thermal
emission from a radiatively efficient accretion disc (SAD,
\citealt{panessa06,cappi06,singh11}). On the other hand,  a large variety of radio characteristics observed in local Seyferts (e.g. morphologies, radio SED in relation to accretion properties, \citealt{laor19}) prevents a complete comprehension of the origin of their radio emission \citep{panessa19,silpa20}, although a non-thermal synchrotron origin for compact cores in Seyferts has been clearly established \citep{kukula99,mundell00,nagar00,ho01a}.

As a first approximation, the radio {\it and} optical properties observed in our sample of Seyferts are consistent with the general properties of SADs: higher accretion rates ($>$10$^{-3}$ Eddington rate) and lower efficiency at launching and collimating jets than ADAF-dominated LINERs. In fact, diffuse lobe-like radio structures are observed more frequently
in Seyferts than in LINERs in LeMMINGs and in other samples of local Seyferts (e.g. \citealt{kukula93,kukula95,morganti99,kharb06,gallimore06}). These lobes\footnote{We cannot fully rule out SF-driven super-winds as the cause of the observed lobes in Seyferts \citep{pedlar85,heckman93}. However, such a scenario is unlikely because these structures are primarily observed in star forming galaxies at low radio luminosities ($<$10$^{34}$ erg
s$^{-1}$), where other than radio bubbles, extended SF in the galactic disc is observed.} have been interpreted as sub-relativistic outflows at the core, which inflate bubbles and produce bow shocks in the ISM \citep{middleberg04,yuan14}.

To complete the puzzle, Seyferts are clearly
radio-quieter than RQ LINERs, suggesting a different disc--jet coupling.  Their strikingly distinct correlation between
the radio core and disc luminosity, and the lack of
continuity between Seyferts and other AGN classes, rule out
the possibility that Seyfert jets are just the scaled-down versions of
RG relativistic flows \citep{talbot21}. Our study also reveals a moderate increase of the radio luminosity and jetted structure fraction at higher Eddington ratios, which can be interpreted as increasing jet luminosity with higher $\dot{m}$ ($P_{\rm jet}$ = $\eta_{\rm jet}\dot{m}c^{2}$ where $\eta_{\rm jet}$ is the jet production efficiency). 
This is different from the literature where there are reports of a tendency of an increased jet production ($\eta_{\rm jet}$) as the system goes fainter from QSOs to low-luminosity Seyferts (e.g. \citealt{trippe14,wojtowicz20}), which is consistent with the anti-correlation between radio loudness and bolometric luminosity in AGN (e.g. \citealt{terashima03,panessa07,ho08}) and in XRB \citep{fender09}.  Such an effect could be caused by an incomplete separation between RQ and RL Seyferts and a lack of angular resolution to resolve jets. Previous studies were based on more heterogeneous samples of radio-emitting Seyferts/QSOs than our sample which consists of a complete sample of resolved jetted RQ Seyferts. In fact, recent studies (\citealt{rusinek20,chang21}) which separate RQ and RL AGN, show a slight $\eta_{\rm jet}$ increment to small Eddington ratios for RQ Seyferts, an effect that is enhanced by including RL Seyferts. Therefore, considering a differential dependence on $\dot{m}$ for each radio-emitting physical mechanism competing in the {\it total} radio production $\eta_{\rm radio}$ (= $\eta_{\rm jet}$ + $\eta_{\rm disc-wind}$ + $\eta_{\rm corona-wind}$) could solve the tension \citep{laor19}. At low $\dot{m}$, a (core-brightened) sub-relativistic jet ($\eta_{\rm jet}$  $>$ $\eta_{\rm disc-wind}$ + $\eta_{\rm corona-wind}$) might dominate over a (edge-brightened) radio-emitting disc/corona wind, which, in turn, becomes important at high $\dot{m}$ ($\eta_{\rm disc-wind}$ or $\eta_{\rm corona-wind}$ $>$ $\eta_{\rm jet}$) (see Sect.~\ref{disc-jet} for discussion and comparison with XRBs). Such scenario would account for both the heterogeneous radio properties of Seyferts and the general tendency of low-luminosity Seyferts to expel more compact jets than QSOs over a large range of Eddington ratios, covering the gap between two extreme classes: almost-radio-silent  high-$\dot{m}$ QSOs and jetted low-$\dot{m}$ LINERs. In addition, the role of the BH parameters (e.g. mass, spin) is to sustain and reinforce the jet/wind launch \citep{yang21}.

To accommodate the disc and radio similarities respectively with QSOs and LINERs, several scenarios have been proposed within the large breadth of SAD models. At small Eddington ratios, a solution could be the model of Jet Emitting Discs (JED,
\citealt{ferreira10}) which are much less dense than an optically-thick disc and arise at
larger magnetisation, where magneto-centrifugally driven jets are
launched \citep{ferreira97}. Another scenario is the
presence of a truncated slim disc whose inner radius, where the jet is
anchored, shrinks, by dragging higher poloidal magnetic flux strength
closer to a spinning SMBH. This condition favours the launch of strong
radio outflows \citep{tchekhovskoy11}, and leads to an increase of the jet Lorentz factor $\Gamma_{\rm jet}$, similar to  what is seen in the hard-state XRBs
(e.g. \citealt{fender04}). At higher Eddington ratios, a quasar-driven uncollimated wind is favoured, able to shock the ISM and accelerate  relativistic electrons producing  synchrotron  radio  emission  \citep{faucher12,zakamska14}. An alternative scenario is an outflow of coronal plasma above the accretion disc, which
slightly collimates within a narrow nozzle near the SMBH and then 
fans out to form a diffuse radio morphology
\citep{donea02,markoff05,king11,raginski16}.

\subsection{Absorption line galaxies}

The ALGs lack evidence of BH-accretion activity in the optical band but may still conceal a weak AGN at their centres, since they show bright radio cores and jets.

Apart from a few irregular/dwarf ALGs which are probably powered by stellar processes \citep{paudel20}, when detected in radio, ALGs
appear indistinguishable from LINERs: massive ETGs hosting SMBHs with masses $\sim$10$^{8}$ M$_{\odot}$ and core-brightened radio morphologies. In addition, the radio-detected ALGs are potential RL AGN, since their optical counterparts are not detected. This scenario is supported by a multi-band study of local line-less RL AGN whose nuclei and hosts are similar to those of LINER-like RGs \citep{best05b,baldi10b},
although at slightly higher luminosities than those of LeMMINGs sample
(10$^{39}$--10$^{41}$ erg s$^{-1}$). The radio detection of ALGs in this survey, and in previous studies, argue for the
presence of an active SMBH in at least a quarter of ALGs. In this scenario, the optically weak characteristics of ALGs could also be reconciled with the picture of low-$\dot{m}$ RGs, where nuclear conditions (i.e. low ionising source, poor gas availability) do not favour optical emission from a compact NLR.  This picture has been observed in particularly faint FR~Is \citep{buttiglione09} and relic RGs \citep{capetti13}.

Even under favourable conditions, local SMBHs are known to exhibit levels
of activity much lower than those expected from gas supplying rates
onto the galactic nuclei, and only a small fraction of silent SMBHs
can turn into AGN. Dynamic stability of gas reservoirs and particular conditions of magnetic field loops, which trap stars and gas in orbits, could lead to
a fuel-starved SMBH \citep{inayoshi20}, thus resulting in
ALGs. The problem of dormant massive BHs has much been discussed in
the last decades (e.g \citealt{fabian88,kormendy01,herpich18}) and the
inactivity of ALGs falls in this investigation.

Another possible interpretation of the apparent absence of nuclear
activity in ALGs is intermittent
accretion  \citep{czerny09}. For example, the currently
quiescent Sgr~A* was recently found to have undergone a period of activity a few million years ago, which created a radio outflow as a consequence of an accretion event \citep{heywood19}. Around 30 per cent of fading AGN have been found to feature jets \citep{esparza20}. The large fraction of non-detected ALGs suggests that if they have intermittent periods of accretion activity, they are short-lived (10$^{4}$--10$^{5}$ years, \citealt{reynolds97a}). This is supported by estimates based on their radio jet lengths (Papers I and II). Any intermittent activity could be caused by occasional
accretion events lasting a few Myr, where X-ray emitting hot-gas
atmosphere, typical of ETGs \citep{forman85}, feeds the strangulated SMBH at a very low $\dot{m}$, and establishes a quasi-spherical
accretion regime which supports the jet launching \citep{ho02,allen06}. 

In summary, unfavourable nuclear conditions, SMBH dormancy and nuclear recurrence can account for the optical and radio detections (or lack thereof) in ALGs. However, once the accretion is set, the activity phases visible in the two bands are not necessarily synchronised: the jet production can possibly lag the disc activity by a few 10$^{2}$-10$^{4}$ yr  (depending on $M_{\rm BH}$ and jet length and power) as expected by disc--jet evolution models  (e.g. \citealt{czerny09}). 

\subsection{H{\sc ii} galaxies}
\label{hii}

The H{\sc ii} galaxies are generally interpreted as SF-dominated nuclei, based
on the emission line ratios in BPT diagrams, but this does not
preclude them from having a weak AGN. H{\sc ii} galaxies show mostly compact single cores and extended complex structures.

The analysis of the full LeMMINGs sample confirm the results from Paper~I: the presence of a dual population of H{\sc ii} galaxies, i.e. jetted and
non-jetted, which are probably related to different origins of the
core emission. In general, jetted H{\sc ii} galaxies are found to have larger $M_{\rm BH}$ ($\gtrsim$10$^{6.5}$ M$_{\odot}$) than
their non-jetted companions.

The vast majority of radio-detected H{\sc ii} galaxies are non-jetted. They are less [\ion{O}{iii}] luminous by a factor $\sim$30 than active galaxies, and typically have a large radio excess with respect to the [\ion{O}{iii}]--radio correlation found for LINERs at low $L_{\rm [\ion{O}{iii}]}$.  All these characteristics are consistent
with a scenario of a nuclear
starburst on a scale of $<$1 kpc, producing stellar emission at 1.5\,GHz. To corroborate this picture of a stellar
origin, Paper~I concluded that  thermal/non-thermal and free-free radio emission predicted from supernova (SN) progenitors, SN explosions and  SN remnants and H{\sc ii} regions \citep{ulvestad81,condon92} are sufficient to account for radio luminosities of non-jetted H{\sc ii} galaxies. In addition, a population of SNe expected from high SFR ($\sim$30--50 M$_{\odot}$ yr$^{-1}$) might be able to
blow bubble-like super-winds \citep{weaver77,heckman15}, which would match the extended irregular morphologies (E class) observed in some sources.

In contrast, three aspects of the (seven) jetted H{\sc ii} sources point to the presence of an active SMBH, powering their radio emission: i) larger BH masses than the non-jetted companions; ii) core-brightened and elongated radio morphologies, ii) similar line-radio correlation and FPBHA to those found for RQ/RL LINERs. As an example, the well-known FR~I  (NGC~3665,
\citealt{parma86}) belongs to this sub-sample. We conclude that it is likely that this population of jetted H{\sc ii} galaxies does
 host weak AGN, whose output is too dim to
significantly contribute to the optical emission,
dominated by nuclear SF. The possible LLAGN signatures found in jetted H{\sc
  ii} galaxies are similar to those of sub-Eddington LINERs (likely powered by a RIAF), rather than
the efficiently accreting Seyferts, and could represent the tip of the iceberg of a large population of weakly-active sub-mJy star forming galaxies at low powers \citep{padovani16, muxlow20}. Another scenario to account for their radio-optical properties is the
combination of both SF and an active BH in the event of a jet shocking a dense ISM and triggering in-situ SF \citep{gaibler12,dugan14}.  The SF--AGN co-existence underlines the symbiotic relationship between these two types of activity, particularly in LTGs (e.g. \citealt{santini12}).

\subsection{Low-luminosity AGN and disc--jet coupling}
\label{disc-jet}

Almost half ($\sim$45 per cent) of the LeMMINGs galaxies are not radio silent, but are characterized by low radio
powers ($>$ 10$^{17.6}$ W Hz$^{-1}$).
Our survey has revealed that local galaxies can show evidence of weak AGN-driven activity, in the form of sub-galactic jets (B, C, D classes) or optical line ratios (LINER or Seyfert) down to $L_{\rm core}$ $\gtrsim$10$^{18.7}$ W Hz$^{-1}$. By binning the sample by radio luminosity and BH mass, the fraction of radio detections reaches 100 per cent for galaxies with $L_{\rm core}$ $\gtrsim$10$^{19.8}$ W Hz$^{-1}$ (twice than the survey sensitivity limit)  and with $M_{\rm BH}$ $\gtrsim$ 10$^{7}$ M$_{\odot}$, regardless of optical type. This limit represents the lowest luminosity for which the most massive BHs hosted by local (Palomar) galaxies are `always' switched on. This limit is lower than what  has been obtained from previous radio surveys, $\sim$10$^{20}$--10$^{21}$  W Hz$^{-1}$ for the most massive galaxies \citep{mauch07,cattaneo09,sabater19}.

Assuming that some (if not all) of the radio
emission in LLAGN comes from jets, the jet fraction is expected to increase with decreasing
Eddington ratios, and consequently the radio loudness  should increase with
increasing BH mass \citep{ho08}. However, the jet/radio properties of nearby galaxies have been shown to be more complicated than a simple bimodality (presence or lack of jet in relation of low- or high-$\dot{m}$ disc), suggesting specific modes of accretion and jet launching mechanisms for different optical AGN classes \citep{best12,hardcastle18}. The high sensitivity and resolution of our survey casts light on the entanglement between  radio/jet and optical/disc properties in LLAGN. Further help in disentangling 
this complexity comes from the comparison with the analogous disc--jet coupling observed in XRBs \citep{remillard06,ontiveros21}. Within the low/hard XRB state, there
are indications that the transition radius between an inner ADAF and
an outer thin disc decreases with increasing luminosity
\citep{narayan05} and a short-lived strong radio outburst occurs
where highly relativistic jets are launched (e.g.,
\citealt{fender04}). Conversely, clear jets are never seen in
high-soft XRBs, while our survey and other studies demonstrate that their {\it putative} AGN-equivalent (low-luminosity) Seyferts can produce jets as much as LINERs do. While jetted AGN and low/hard XRBs clearly lie on the same FPBHA \citep{merloni03,falcke04}, \citet{gultekin19}, by including radio-active high/soft state XRBs, cannot rule out that the latter and Seyferts are inconsistent with the FP made up of low/hard state XRBs and LINERs. Furthermore, \citet{fischer21} has recently shown that,  Seyferts, once resolved with the VLBI on sub-parsec scales, can have corresponding $L_{\rm core}$ upper limits that are systematically below the predictions from the FPBHA. The tension with the analogies between stellar mass BHs and active SMBHs probably rests on attempting to use the optical class to set the association. The Eddington-ratio argument could partially resolve this tension. For XRBs, sources are almost entirely in hard states below 2 per cent of  $L_{\rm Edd}$ \citep{Maccarone_state_trans,Vahdat2019}, while above that value both states are present. For AGN classes, LINER and Seyfert distributions usually roughly break up at 0.1 per cent of $L_{\rm Edd}$. Therefore, the global properties of LINERs (e.g. low Eddington ratios; weak or strong, persistent radio jets; hard X-ray spectra) make this class similar to hard-state XRBs. Conversely, Seyferts show more heterogeneous radio and disc properties and consists of a mixture of hard and soft states \citep{ontiveros21}. Low-$\dot{m}$ Seyferts (between 0.1 and 2 per cent of $L_{\rm Edd}$) could be more similar to hard XRBs, able to launch compact jets, while the high-$\dot{m}$ Seyferts  are more similar to soft XRBs. However, as disclosed in Sect.~\ref{seyfert}, the current status of results on Seyferts based on radio observations reveals a higher level of complexity than this scenario. In fact,
we would expect a larger fraction of jetted structures at low $\dot{m}$, similar to hard XRBs, which is opposite to what we observe in our sample, but in agreement if we consider a larger population which includes luminous Seyferts and QSOs. A possible explanation to this apparent contradiction is that different physical mechanisms of radio production co-exist in Seyferts, which makes the comparison with XRBs even more challenging. At higher accretion rates ($\dot{m} \gtrsim 10^{-2}$), disc and corona wind  are expected to play an important role in the radio emission \citep{laor19} and could account for the observed tendency to display edge-brightened structures. At lower $\dot{m} \lesssim 10^{-2}$, a sub-relativistic compact jet, more similar to RQ LINERs, could dominate over the other radio-emitting physical processes (Table~\ref{sketch}).  

Although (simultaneous) high-frequency radio (e.g. mm-band) and X-ray observations would be ideal to isolate the pc-scale emission and its link to the disc (see \citealt{bell11,behar20}), in our 1.5-GHz survey, we observed a significant  dependence of radio-optical properties (i.e. disc--jet coupling) with $M_{\rm BH}$. The small range of stellar BH masses of XRBs does not enable a straightforward comparison of radio models valid for XRBs with the results from the LeMMINGs survey. Instead, this is possible for AGN in general. In fact, AGN disc models predict different mass dependencies for the radio emission. The similar slopes found for RL and RQ AGN ($L_{\rm core}\propto M_{\rm BH}^{1.5-1.65}$) despite the large scatter, can be interpreted as a single strong connection between the radio output and the BH mass, regardless of the type of the radio product (i.e. relativistic or sub-relativistic jets, winds, etc). In the observed range of $L_{\rm core}$--$M_{\rm BH}$ slopes, a degeneracy of models exists: for jet-dominated sources (e.g., JDAF, \citealt{falcke04}) $L_{\rm radio} \propto \bar{a}^{2} M_{\rm BH}$ (where $\bar{a}$ is the BH spin), ADAF models which predict radio emission $L_{\rm radio}$ $\propto$ $M_{\rm BH}^{8/5}$ $\dot{m}^{6/5}$ \citep{yi99}, while relativistic jets can be described by the Blandford--Znajek (BZ, \citealt{blandford77}) process as  $L_{\rm radio} \propto \bar{a}^{2} M_{\rm BH}^{2} B^{2}$ (where $B$ is the magnetic field). Across the different AGN classes, a large range of BH spins, accretion rates and magnetic field strengths could account for the significant but scattered relations we observe. The transition from RQ AGN (jetted H{\sc ii} galaxies, RQ LINERs and Seyferts) to RL AGN (RL LINERs and FR~Is) could be caused by an increment of BH or disc parameter values, which result in a boost of the radio emission \citep{blandford19,chen21}.

The radio emission production in XRBs is related to accretion
rate and high-energy-band luminosity (from optical to X-ray, produced by physical processes in disc and/or corona). Their jet contribution  to the observed high-energy flux is largely debatable (e.g. \citealt{fender01,markoff03}). Despite the problematic comparison between states of SMBHs and stellar BHs, radio and disc emission models of XRBs could still help to understand the physical processes involved in active SMBHs. For the jet-mode inefficient discs in XRBs, $L_{\rm radio}$ $\propto$ $L_{\rm
  X}^{0.5-0.7}$, whereas for disc-dominated sources $L_{\rm radio}$
$\propto$ $L_{\rm X}^{1.4}$ (see \citealt{coriat11}). In our LeMMINGs sample, assuming that [\ion{O}{iii}] is a good
indicator of the bolometric AGN power which scales with the X-ray
output, LINERs show a steeper radio-disc luminosity dependence
($L_{\rm core} \propto L_{\rm [\ion{O}{iii}]}^{1.2-1.3}$ for RL and RQ LINERs) than Seyferts ($L_{\rm core}$ $\propto L_{\rm [\ion{O}{iii}]}$). This is opposite to the equivalent relationship observed in hard/soft XRBs.  Furthermore, the absence of a clear dependence of the radio luminosity on Eddington ratios in LINERs casts new light on a possible disparity between jetted accreting compact objects, since hard-state XRBs show $L_{\rm radio} \propto
\dot{m}^{1.4}$ (e.g. \citealt{koerding06}), unless $L_{\rm [\ion{O}{iii}]}$ scales as $\dot{m}$ for LINERs \citep{netzer09}.

In the context of AGN models, for SAD-dominated sources the ratio $L_{\rm radio}$/$L_{\rm X} \propto \eta_{\rm jet} \, \bar{a}^{2}$  \citep{yi99}  is expected to be constant and limited to a narrow range of values, as we find for our Seyferts, if the $\bar{a}$ range is finite. Conversely, for LINERs, that are powered by RIAF discs, the relation between bolometric AGN and jet power is difficult to model because jet emission can dominate their entire SED \citep{koerding08} and [\ion{O}{iii}]-line contamination  from jet photoionising shocks can overpredict the bolometric AGN power \citep{capetti:cccriga,netzer09}. However, in a simplistic scenario of a  disc origin of the radio emission, ADAF discs would predict much shallower relations ($L_{\rm core} \propto L_{\rm X}^{\alpha}$  with $\alpha < $0.6 depending on bremsstrahlung-dominated and multiple Compton scattering regimes, \citealt{yi98}) than the observed slopes for LINERs (1.2-1.3). Furthermore, the observed radio-[\ion{O}{iii}] relation found for LINERs is even steeper (despite the large scatter) than any previous relations between the kinetic jet power and the bolometric AGN luminosities found for LINER-like RL AGN (slopes $<$1, e.g. \citealt{capetti06,merloni07,baldi19b}). Therefore, assuming that RL LINERs are the scaled-down version of RGs and normalising the $L_{\rm radio}$/$L_{\rm X}$ comparison with previous studies based on this assumption, the steeper slopes might be the consequence of a closer view of jet launching mechanism even in RQ LINERs. The high sensitivity and sub-arcsec resolution (crucially intermediate between {\it VLA} and VLBI) provided by our survey allowed us to probe the parsec scale region near the core where the jet is launched and reveal the relevant role of $\Gamma_{\rm jet}$ and BH spin/mass in the jet production. These parameters could eventually induce much steeper relations than those established with shallower radio observations.

Although, the reliability of the [\ion{O}{iii}]--X-ray conversion could affect this comparison, the difference between AGN and XRBs is stark. In conclusion,  the apparent disparity between AGN and XRB states suggests that LLAGN show different disc--jet couplings and more complicated mechanisms of radio production than the XRBs (without considering transition phases). However, a more detailed (radio/X-ray) study and comparison between LLAGN and XRBs is needed and is not the goal of this work.

Another variable which can increase the scatter of the observed radio-optical relations and complicate our comprehension of the
disc--jet connection, is the core variability.  Radio flickering and transient events have been documented in the nuclei of some LLAGN, particularly Seyferts
(e.g. \citealt{wrobel00,mundell09,giannios11,baldi15b,mattila18,williams19,nyland20}).

\subsection{Black hole -- host connection}
\label{bhhost}

The LeMMINGs survey detected nuclear radio activity from  high-  down to low-mass galaxies ($L_{\rm bulge}\sim 10^{9}-10^{10.5}$
$L_{\odot}$, \citealt{nagar05}), which harbour SMBHs with masses $\gtrsim$10$^{6.5}$ M$_{\odot}$. SF- and BH-driven radio emission are the two main mechanisms responsible for the $\mu$Jy-level radio core emission \citep{padovani15} with several pieces of evidence pointing to an increment of  AGN contribution to radio emission as $M_{\rm BH}$ increases. In fact,  the radio detection and jet fraction increase with BH mass for $\gtrsim$10$^{6.5}$ M$_{\odot}$ and the `bulgeness' of the host (i.e. favouring ETGs, see Paper~II).

One of the main results of our survey is the presence of a break of the empirical relations (radio, [\ion{O}{iii}], $M_{\rm BH}$, optical FPBHA)  between the star forming and active galaxies at $\sim$10$^{6.5}$ M$_{\odot}$ (conservatively, between $\sim$10$^{6}$ and $\sim$10$^{7}$ M$_{\odot}$). A BH mass of $\sim$10$^{6.5}$ M$_{\odot}$ seems to represent the turnover from a SF to an AGN regime as the BH mass increases. Such a break corresponds to the $M_{\rm BH}$ which roughly separates the BH mass function of local LTGs and ETGs \citep{davis14,greene20}, i.e. pseudo-bulges and bulges \citep{yesuf20} and SF- and AGN-dominated sources respectively (see \citealt{kelly12}, Appendix~\ref{app} and Fig.~\ref{BHmassfunction}). Both the fraction of galaxies which host a radio AGN and the radio core powers correlate with $M_{\rm BH}$ with different relations at the two sides of this BH mass turnover. These relationships possibly agree with galaxy evolution models which predict that high-mass galaxies evolve faster and host a radio-AGN which suppresses SF more effectively than low-mass galaxies \citep{fabian12}, which instead form stars in the current Universe (e.g. \citealt{lapi11,behroozi19}). However, it is important to point out that such a break could also be the consequence of an intrinsic $M_{\rm BH}$-$\sigma$ relation which flattens out at low BH masses \citep{mezcua17,shankar19}.

Substantial progress in understanding the process of
radio AGN phenomena in the nearby Universe has recently been made by
combining optical and radio surveys to determine the statistical
relationships between radio activity and galaxy/BH mass. It has been
found that the fraction of galaxies that host RL AGN (with
$L_{\rm radio}$ $>$ 10$^{21}$ W Hz$^{-1}$) is a strong function of
stellar mass $\propto M_{*}^{2.5}$ and BH mass $\propto M_{\rm BH}^{1.6}$
\citep{best05b,mauch07,sabater19,hardcastle19} with the most
massive ETGs `always' switched on at low radio powers ($\gtrsim 10^{21}$ W Hz$^{-1}$). In our statistically-complete heterogeneous sample of local galaxies, mostly consisting of
LTGs, we find a flatter dependence between the radio-AGN fraction $f_{d}$ and BH mass for the
active galaxies, ($\propto M_{\rm BH}^{0.15}$), independent of the radio
properties. The different slopes between our work and previous studies could be ascribed to three aspects: the higher angular resolution and higher sensitivity of our survey, crucial to identifying jets,  the identification of  genuine `jetted' RQ AGN with respect to RL AGN, and a possible presence of a break in the radio luminosity function at 10$^{19.5}$--10$^{20}$ W Hz$^{-1}$ \citep{nagar05}. This results in the inclusion of 
SMBHs with masses lower than those of standard RL AGN ($<$10$^{8}$ M$_{\odot}$), which can often emanate low-power jets on sub-kpc scale, despite their nominal radio-quietness definition, differently from the rarer powerful RGs. This selection eventually induces a much shallower dependence with BH mass. \citet{janssen12} also noted 
a  shallower $f_{d}$-$M_{\rm BH}$  relation  than previous studies, scaling as $\propto M_{*}^{1.5}$, for Seyfert-like radio sources, that remains still steeper than what we found for RQ AGN. Moreover, in our survey, we do not find a clear separation (within the large scatter)
between LINERs and Seyferts in the radio--$M_{\rm BH}$ plane. This likely indicates that accretion rate, which largely differs between the two classes,  plays a more complicate role in the radio production,  together with BH mass/spin (see Sect.~\ref{disc-jet}).

The observed correlation between $L_{\rm core}$ and BH mass or
host type can also be an indirect consequence of a more fundamental
relationship between radio luminosity and optical bulge luminosity (or
mass) \citep{falcke01,ho02,nagar02}: ETGs  show larger radio luminosities than LTGs \citep{laor00,decarli07,mauch07,gallo08}. The more massive elliptical
galaxies or bulge-dominated systems are more efficient at producing
radio emission as they have larger SMBHs, and hence, are more likely
to be highly RL than spiral galaxies for $M_{\rm BH} >$10$^{8}$
M$_{\odot}$ \citep{chiaberge11}. The high bulge-to-disc ratio is
achieved in more evolved galaxies, where spheroids and pseudo-bulges
dominate the central gravitational well. At their centres the extended jets
expelled from accreting SMBHs can more easily plough through the more rarefied ISM than in discy (spiral) galaxies. Furthermore, according to the spin and gap
paradigm (e.g. \citealt{moderski08,garofalo10,meier12}), a higher BH
spin and a larger gap region between the inner edge of the accretion
disc, typically achieved in ETGs, can cause a larger magnetic flux accumulation on the SMBH, and
consequently a maximisation of the BF jet mechanism. Such
a condition is reached with retrograde accretion. Typically in LTG, a
low-spinning SMBH and prograde disc have been suggested to account for
their limited jet production compared to the different RL
AGN classes, found commonly in ETGs (RQ AGN against FR classes, see, e.g. \citealt{tchekhovskoy12,garofalo14,garofalo20}).

\subsection{Star formation}
\label{anysfg}

Non-jetted H{\sc ii} galaxies clearly depart from the
AGN classes in the optical--radio plane and FPBHA at low $L_{\rm core}$ and $M_{\rm BH}<$10$^{6.5}$ M$_{\odot}$, showing a clear break
likely caused by the increased SF contributions to the pc-scale radio
emission. Such a sub-sample represents a much larger population of low-luminosity star-forming galaxies, which dominate the local Universe at $M_{\rm BH}<$10$^{6.5}$ M$_{\odot}$ (Appendix~\ref{app}, \citealt{greene07,greene20}), where stellar processes are generally the main source of energy.

Assuming that non-jetted H{\sc ii} galaxies are powered by SF, the relation between radio luminosities in such galaxies and $M_{\rm BH}$ can be deciphered as a link between thermal/non-thermal SF luminosity and BH mass, in the form $L_{\rm SF} \propto M_{\rm BH}^{0.61}$, which has the largest scatter of the observed linear regressions, due to the large uncertainties of $M_{\rm BH}$ estimates at low values and numerous radio non-detections. This unprecedented relation is likely the result of the observed link found between SF indicators such as radio, far-infrared, optical or line luminosities, and galaxy mass (e.g. \citealt{mauch07,gurkan18}). Moreover, the SF contribution in the radio band with respect to the AGN activity is expected to increase with galaxy mass \citep{aird19}. In fact the observed correlation suggests that the radio production due to SF broadly increases with BH mass, as the latter quantity is approximately a constant proportion of the galaxy mass (from a third down to a fifth in the lowest $M_{\rm BH}\sim$10$^{5}$ M$_{\odot}$, \citealt{haring04,reines15,martin18}), i.e. the stellar content: more stars likely produce more radio emission. This is consistent with the idea that SF, although mostly included in the disc \citep{bluck20} and known to scale with galaxy mass \citep{speagle14}, would possibly also scale with $M_{\rm BH}$. More precisely, the nuclear starburst is plausibly set by the available gas mass present in the NSC ($M_{\rm NSC}$, \citealt{fernandez09,neumayer20}), which scales with the galaxy dynamical mass ($M_{\rm NSC} \propto M _{dyn}^{0.55}$, \citealt{scott13}), which in turn respond to the SMBH gravitational well \citep{cen15,pitchford16}. Such a sequence of links eventually enacts the radio-$M_{\rm BH}$ and radio-optical relations observed for star-forming galaxies, e.g setting the fraction of radio emission with respect to the photoionising energy from young stellar populations and SN products.

Using the [\ion{O}{iii}] as a SF indicator \citep{suzuki16}, the observed optical luminosities of non-jetted H{\sc ii} galaxies can be interpreted by a SFR up to
10$^{1.1 \pm 0.5}$ M$_{\odot}$ yr$^{-1}$ in a nuclear starburst.
The [\ion{O}{iii}]-radio correlation  valid for non-jetted H{\sc ii} galaxies also represents an empirical method to predict the nuclear radio emission in star forming galaxies based on
the ubiquitous [\ion{O}{iii}] line data. The best fit approximately agrees with the [\ion{O}{iii}]--radio relation estimated from {\it Herschel}, {\it LOFAR} and {\it VLA} 
data ($L_{\rm 1.4~GHz}$ = [0.83$\pm$0.10] $\log$ $L_{\rm [\ion{O}{iii}]}$ + [7.9$\pm$2.0], \citealt{best05a,gurkan15,gurkan18}), although our
observations are at a higher resolution than any previous study on large samples of star forming galaxies.

\section{Reviewing the radio properties of local galaxies} 
\label{sect5}

\subsection{The LeMMINGs view}

The 1.5-GHz LeMMINGs survey has revealed pc-scale cores and jet-like structures in  more than a third of local optically active and inactive Palomar galaxies (Papers~I and II). Such nuclei are part of a large population of radio LLAGN  and faint nuclear starbursts ($\gtrsim$10$^{17.6}$\,W \,Hz$^{-1}$).  This result adds observational evidence that a large fraction of LLAGN (even RQ AGN) feature low-power galactic-scale jets \citep{webster21}, which may have a tremendous impact on their hosts by continuously injecting energy, a fundamental ingredient for galaxy feedback \citep{morganti17b,jarvis19,hardcastle20,smith20b,venturi21,jarvis21}, as supported by the state-of-the-art jet simulations
\citep{massaglia16,mukherjee18,mukerjee20}.

In this work we have investigated the nature of the pc-scale nuclear radio components of LINERs, Seyferts, ALG and H{\sc ii} galaxies. The emission-line [\ion{O}{iii}] luminosity, available for the entire sample, gives the optical nuclear counterpart of the 1.5-GHz cores we detected. The optical--radio connection and the optical FPBHA \citep{saikia15} probe the different physical processes involved at the centre of galaxies. In these diagnostic diagrams we have found that  optical classes follow distinct tracks: this is indicative of different physical processes responsible for the radio emission and broadly agrees with previous multi-band studies of LLAGN and star-forming galaxies (e.g. \citealt{koerding06,vitale15,tadhunter16,mancuso17,smith20}). Nonetheless, we have set more robust constraints on these trends with respect to previous studies and Paper~I, especially in the thus far poorly explored regime of low luminosities and BH masses, by leveraging the unprecedented completeness and depth of the LeMMINGs survey.

.

{\bf LINERs} are divided into two classes: RQ and RL. The RL LINERs ($\sim$19 per cent of LINERs) appear to represent the scaled-down version of FR~Is at lower radio and [\ion{O}{iii}] luminosities and BH masses down to $\sim$10$^{7.5}$ M$_{\odot}$.  Our results indicate that the class of RL LINERs belong to a large homogeneous population of RL LLAGN ($<$ 10$^{42}$ erg s$^{-1}$), characterized by a broad and continuous distribution of parameters of relativistic jets  (e.g. sizes, luminosities, $\Gamma_{\rm jet}$), powered by a sub-Eddington, RIAF-like disc, and hosted in ETGs \citep{heckman14}. This picture, in turn, suggests that a common jet-launching mechanism operates in all RL LINERs which smoothly connects low-power (sub)kpc-scale jetted LINERs to more powerful compact RGs (named FR~0s, \citealt{baldi15}) and full-fledged FR~Is and FR~IIs \citep{kharb14,mingo19,grandi21}.  Conversely, RQ LINERs consist of a more heterogeneous population of LLAGN, which can reside in either LTGs or ETGs. An intermediate disc between low-$\dot{m}$ ADAF and high-$\dot{m}$ SAD may be consistent with the observed characteristics of RQ LINERs, where radio emission is due to a low-power non-relativistic uncollimated jet \citep{gallimore06}.

{\bf Seyfert} nuclei radiate at a larger fraction of Eddington luminosity than LINERs. The common picture is that Seyferts are powered by a radiatively efficient SAD with higher accretion rates (Eddington ratios $\gtrsim$ 10$^{-3}$) than seen in LINERs. Nevertheless, Seyferts are not radio-silent at all, and can show compact or edge-brightened jetted morphologies. In order to reconcile their radio and optical properties, different disc--jet models,  intermediate between jetted LINERs and the `radio-silent' near-Eddington QSOs \citep{balick82,liu20} have been proposed. Low-$\dot{m}$ Seyferts are possibly the scaled-down version of the so-called red QSOs, which have intermediate radio loudness and have been found to have a higher incidence of compact and extended radio morphologies than blue QSOs (e.g. \citealt{klindt19,rosario20}), which instead are more similar to high-$\dot{m}$ Seyferts. We conclude that the change of disc--jet coupling from low-power Seyferts to powerful QSOs and the predominance of a specific radio-emitting physical process may depend on different factors \citep{garofalo20}: the jet efficiency with respect to the disc- or corona-wind radio production decreases with Eddington ratio and increases with BH mass, `bulgeness' of the host (favouring ETGs), and SMBH/disc spin.

Apart from a few irregular galaxies, the vast majority of {\bf ALGs} are hosted by massive ETGs with $M_{\rm BH}$ $>$ 10$^{7}$ M$_{\odot}$, which lack of  significant optical emission from the nucleus. They  have been interpreted as part of a large population 
of evolved massive galaxies \citep{salucci99}, which occasionally or recurrently go through periods of LINER-like low-$\dot{m}$ activity, during which they launch pc-scale jets. However, during this low-activity phenomenon, the optical line nucleus remains quiescent due to the combination of a low ionising radiation field produced by the disc (and jet) and the rarefaction of cold gas to photoionise \citep{capetti:cccriga}.

A dual population of {\bf H{\sc ii} galaxies} has been found in our sample. On one hand, the majority of the H{\sc ii} galaxies are hosted in LTGs with $M_{\rm BH}\lesssim$10$^{6.5}$ M$_{\odot}$ and their optical--radio properties reconcile with a SF origin. A SF rate of $\sim$10 M$_{\odot}$ yr$^{-1}$ in the central region can account for their compact cores, with SN-driven winds \citep{heckman15} or extended H{\sc ii} regions possibly causing their complex radio structures. On the other hand, $\sim$5 per cent of H{\sc ii} galaxies exhibit jet-like radio structures. These sources follow the diagnostic optical--radio sequence of AGN-classed sources with $M_{\rm BH}$ $\gtrsim$10$^{6.5}$ M$_{\odot}$. The co-existence of a LINER-like sub-Eddington accreting SMBH and a nuclear starburst, which are responsible for the radio/jet and the optical properties, respectively, is the simplest way to solve their puzzling nature \citep{padovani16}.

\subsection{Conclusions}

The LeMMINGs survey characterises radio activity (either from SF or AGN) from a complete sample of local galaxies, by
reaching a 100 per cent detection rate for galaxies with $M_{\rm BH}$ $>$ 10$^{7}$ M$_{\odot}$ down to luminosities $L_{\rm core}>$ 10$^{19.8}$ W Hz$^{-1}$. This limit represents the lowest luminosity for which galaxies appear to host a radio-active SMBH
in the local Universe.  The fraction of radio-active galaxies scales as $\propto$$M_{\rm BH}^{0.15}$, a relation that is flatter than what has been found in previous radio surveys (e.g. \citealt{best05b}), probably because of higher angular resolution and the inclusion of RQ AGN.

One of the main results of our survey is the presence of a break
at  $M_{\rm BH} \sim 10^{6.5}$ M$_{\odot}$ moving from the SF to AGN regime with increasing BH mass. Such a  $M_{\rm BH}$ value roughly separates the star-forming galaxy and (LL)AGN populations in the local Universe. At larger BH masses, mostly in ETGs, a link between SMBH activity and jet production is well established (e.g \citealt{cen12,heckman14}). In contrast, in  less-massive  (late-type)  galaxies,  SF makes a proportionally larger contribution to radio luminosity causing a flattening of the relations valid for AGN  at lower $M_{\rm BH}$. 

For galaxies with small BH masses, $\lesssim$10$^{6.5}$ M$_{\odot}$, thermal/non-thermal emission  from stellar processes in star forming galaxies is the most plausible scenario of the origin of their radio emission. An unprecedented empirical relation between their nuclear radio emission and BH masses, $L_{\rm core} \propto M_{\rm BH}^{0.61}$, has been inferred in our sample (although with large scatter).  This relation between the nuclear SF and the SMBH was found as a result of the {\it e-}MERLIN array, able to resolve out smooth, extended emission, and reveal the radio emission at the nuclear scale, where the central starburst is more likely to respond to the influence of the central SMBH.

The steepening of the empirical correlations found for galaxies with BH masses $\gtrsim$10$^{6.5}$ M$_{\odot}$ ($L_{\rm core}\propto$$M_{\rm BH}^{1.5-1.65}$ and  radio-AGN fraction $\propto$$M_{\rm BH}^{0.15}$)  with respect to those at lower BH masses (valid for star-forming galaxies) strongly invoke  an active SMBH as the main driver of their radio emission. The link between radio activity, SMBH properties, and local conditions (e.g. host type, gas availability) appears more complicated than the standard BH--disc bimodality deduced from galaxy--SMBH models \citep{heckman14}, or direct analogies with XRBs states \citep{remillard06,fender09}: low-$\dot{m}$/jet vs high-$\dot{m}$/no-jet. As a first approximation, the Eddington-scaled accretion rates ($\dot{m}$) of the sample correspond to two extreme radio and host categories. The most massive and bulged galaxies, which are the radio loudest with the (optically) faintest nuclei (LINERs, ALGs, and jetted H{\sc ii} galaxies), are powered by sub-Eddington ($\dot{m} \lesssim 10^{-3}$) RIAF-like discs and launch the most core-brightened jets. In contrast, LTGs can host high-$\dot{m}$ ($\dot{m} \gtrsim 10^{-3}$) standard disc and occasionally support sub-relativistic edge-brightened jets.  As a second-order approximation,  a closer inspection of the data reveal that  optical classifications do not simply map the accretion states and this can further complicate the comparison with XRBs. 
Therefore, to break the degeneracy between radio-loudness, accretion modes and host types, distinct disc--jet couplings (accretion/ejection states) that are observed among the various optical--radio classes (RL/RQ, LINER/Seyfert, SF/AGN-dominated) are required. The connection between disc and jet is probably determined by a combination of several factors (some not directly observable at present), including BH spin, magnetic field strength, accretion rotation, gas supply, small and large-scale environment, and host location within the environment (e.g. distance from cluster centre). These parameters play a significant role in setting the generic disc configuration (e.g. accretion mode, rate and radiative efficiency) and the radio production mechanism (e.g. jet, outflow, wind) for each optical/radio class.  Variation of these factors could favour the increase of the jet production efficiency and/or jet bulk speed \citep{lagos09}. This picture would account for the  continuous, rather than dichotomous, radio properties observed in LLAGN which are characterised by a broad distribution of sizes and luminosities of their extended radio emission, from weak RQ LINER (even ALG) to RGs and from jetted low-power Seyferts to  extremely-RQ high-$\dot{m}$ QSOs \citep{brown11,kharb14,garofalo14,baldi19rev,garofalo19}. A time-dependent evolution among different radio/optical states would add a 
further level of complexity to this picture, probably in analogy to XRB intermediate states \citep{ontiveros21}. An exhaustive comparison between LLAGN and XRBs will be addressed in a future work by using X-day data.

A complete census of the local SMBH population, offered by the forthcoming deep follow-up 5-GHz {\it e-}MERLIN observations along with complete {\it Chandra} X-ray and {\it HST} optical data, pushing to the rather unexplored regime of very low luminosities, $\sim$10$^{16}$ W Hz$^{-1}$ (comparable to Sgr~A*), will allow an unbiased view of the multi-band SMBH activity of the local Universe and provide even more robust  constraints on disc--jet coupling in LLAGN.

\vspace{-0.25cm}
\section*{Acknowledgements}

We thank the referee for her/his comments, which helped to improve the paper. This research was supported by a Newton Fund project, DARA (Development in Africa with Radio Astronomy), and awarded by the UK’s Science and Technology Facilities Council (STFC) - grant reference ST/R001103/1. This research was supported by European Commission Horizon 2020 Research and Innovation Programme under grant agreement No. 730884 (JUMPING JIVE). FS acknowledges support from a Leverhulme Trust Research Fellowship.
J.H.K. acknowledges support from the European Union's Horizon 2020 research and innovation programme under Marie Sk\l odowska-Curie grant agreement No 721463 to the SUNDIAL ITN network, from the State Research Agency (AEI) of the Spanish Ministry of Science and Innovation and the European Regional Development Fund (FEDER) under the grant with reference PID2019-105602GB-I00, and from IAC project P/300724, financed by the Ministry of Science and Innovation, through the State Budget and by the Canary Islands Department of Economy, Knowledge and Employment, through the Regional Budget of the Autonomous Community. IMV thanks the Generalitat Valenciana (funding from the GenT Project
CIDEGENT/2018/021) and the MICINN (funding from the Research Project
PID2019-108995GB-C22). AA and MAPT acknowledge support from the Spanish MCIU through grant PGC2018-098915-B-C21  and  from  the  State  Agency  for Research  of 
the  Spanish  MCIU  through  the  “Center  of Excellence  Severo  Ochoa” award  for  the  Instituto  de  Astrof\`isica de Andaluc\`ia
(SEV-2017-0709). B.T.D. acknowledges financial support from grant “Ayudas para la realizaci\'on de proyectos de I+D para j\'ovenes doctores 2019.” funded by Comunidad de Madrid and Universidad Complutense de Madrid under grant No. PR65/19-22417.'

\vspace{-0.25cm}
\section*{DATA AVAILABILITY}

The data on which this paper is based are publicly available from the {\it e}-MERLIN archive. Calibrated image products are available upon reasonable request to the corresponding author. These, along with other LeMMINGs advanced survey products, will be publicly hosted on upcoming project web-page which is being developed (\url{http://lemmingslegacy.pbworks.com}).




\bibliographystyle{mnras}
\bibliography{my.bib} 

\begin{thebibliography}{}
\makeatletter
\relax
\def\mn@urlcharsother{\let\do\@makeother \do\$\do\&\do\#\do\^\do\_\do\%\do\~}
\def\mn@doi{\begingroup\mn@urlcharsother \@ifnextchar [ {\mn@doi@}
  {\mn@doi@[]}}
\def\mn@doi@[#1]#2{\def\@tempa{#1}\ifx\@tempa\@empty \href
  {http://dx.doi.org/#2} {doi:#2}\else \href {http://dx.doi.org/#2} {#1}\fi
  \endgroup}
\def\mn@eprint#1#2{\mn@eprint@#1:#2::\@nil}
\def\mn@eprint@arXiv#1{\href {http://arxiv.org/abs/#1} {{\tt arXiv:#1}}}
\def\mn@eprint@dblp#1{\href {http://dblp.uni-trier.de/rec/bibtex/#1.xml}
  {dblp:#1}}
\def\mn@eprint@#1:#2:#3:#4\@nil{\def\@tempa {#1}\def\@tempb {#2}\def\@tempc
  {#3}\ifx \@tempc \@empty \let \@tempc \@tempb \let \@tempb \@tempa \fi \ifx
  \@tempb \@empty \def\@tempb {arXiv}\fi \@ifundefined
  {mn@eprint@\@tempb}{\@tempb:\@tempc}{\expandafter \expandafter \csname
  mn@eprint@\@tempb\endcsname \expandafter{\@tempc}}}

\bibitem[\protect\citeauthoryear{{Abazajian} et~al.,}{{Abazajian}
  et~al.}{2009}]{abazajian09}
{Abazajian} K.~N.,  et~al., 2009, \mn@doi [\apjs]
  {10.1088/0067-0049/182/2/543}, \href
  {http://adsabs.harvard.edu/abs/2009ApJS..182..543A} {182, 543}

\bibitem[\protect\citeauthoryear{{Aird}, {Coil}  \& {Georgakakis}}{{Aird}
  et~al.}{2019}]{aird19}
{Aird} J.,  {Coil} A.~L.,   {Georgakakis} A.,  2019, \mn@doi [\mnras]
  {10.1093/mnras/stz125}, \href
  {https://ui.adsabs.harvard.edu/abs/2019MNRAS.484.4360A} {484, 4360}

\bibitem[\protect\citeauthoryear{{Akritas}}{{Akritas}}{1989}]{akritas89}
{Akritas} M.,  1989, {Aligned Rank Tests for Regression With Censored Data}.
Penn State Dept. of Statistics Technical Report, 1989

\bibitem[\protect\citeauthoryear{{Akritas} \& {Siebert}}{{Akritas} \&
  {Siebert}}{1996}]{akritas96}
{Akritas} M.~G.,  {Siebert} J.,  1996, \mn@doi [\mnras]
  {10.1093/mnras/278.4.919}, \href
  {https://ui.adsabs.harvard.edu/abs/1996MNRAS.278..919A} {278, 919}

\bibitem[\protect\citeauthoryear{{Alexander} et~al.,}{{Alexander}
  et~al.}{2003}]{alexander03}
{Alexander} D.~M.,  et~al., 2003, \mn@doi [\aj] {10.1086/376473}, \href
  {https://ui.adsabs.harvard.edu/abs/2003AJ....126..539A} {126, 539}

\bibitem[\protect\citeauthoryear{{Allen}, {Dunn}, {Fabian}, {Taylor}  \&
  {Reynolds}}{{Allen} et~al.}{2006}]{allen06}
{Allen} S.~W.,  {Dunn} R.~J.~H.,  {Fabian} A.~C.,  {Taylor} G.~B.,   {Reynolds}
  C.~S.,  2006, \mn@doi [MNRAS] {10.1111/j.1365-2966.2006.10778.x}, \href
  {http://adsabs.harvard.edu/abs/2006MNRAS.372...21A} {372, 21}

\bibitem[\protect\citeauthoryear{{Allen}, {Groves}, {Dopita}, {Sutherland}  \&
  {Kewley}}{{Allen} et~al.}{2008}]{allen08}
{Allen} M.~G.,  {Groves} B.~A.,  {Dopita} M.~A.,  {Sutherland} R.~S.,
  {Kewley} L.~J.,  2008, \mn@doi [\apjs] {10.1086/589652}, \href
  {http://adsabs.harvard.edu/abs/2008ApJS..178...20A} {178, 20}

\bibitem[\protect\citeauthoryear{{Aller} \& {Richstone}}{{Aller} \&
  {Richstone}}{2002}]{aller02}
{Aller} M.~C.,  {Richstone} D.,  2002, \mn@doi [\aj] {10.1086/344484}, \href
  {https://ui.adsabs.harvard.edu/abs/2002AJ....124.3035A} {124, 3035}

\bibitem[\protect\citeauthoryear{{Baldi} \& {Capetti}}{{Baldi} \&
  {Capetti}}{2010}]{baldi10b}
{Baldi} R.~D.,  {Capetti} A.,  2010, \mn@doi [\aap]
  {10.1051/0004-6361/201014446}, \href
  {http://adsabs.harvard.edu/abs/2010A%26A...519A..48B} {519, A48}

\bibitem[\protect\citeauthoryear{{Baldi}, {Behar}, {Laor}  \& {Horesh}}{{Baldi}
  et~al.}{2015a}]{baldi15b}
{Baldi} R.~D.,  {Behar} E.,  {Laor} A.,   {Horesh} A.,  2015a, \mn@doi [\mnras]
  {10.1093/mnras/stv2284}, \href
  {https://ui.adsabs.harvard.edu/abs/2015MNRAS.454.4277B} {454, 4277}

\bibitem[\protect\citeauthoryear{{Baldi}, {Capetti}  \& {Giovannini}}{{Baldi}
  et~al.}{2015b}]{baldi15}
{Baldi} R.~D.,  {Capetti} A.,   {Giovannini} G.,  2015b, \mn@doi [\aap]
  {10.1051/0004-6361/201425426}, \href
  {http://adsabs.harvard.edu/abs/2015A%26A...576A..38B} {576, A38}

\bibitem[\protect\citeauthoryear{{Baldi} et~al.,}{{Baldi}
  et~al.}{2018}]{baldi18lem}
{Baldi} R.~D.,  et~al., 2018, \mn@doi [\mnras] {10.1093/mnras/sty342}, \href
  {https://ui.adsabs.harvard.edu/abs/2018MNRAS.476.3478B} {476, 3478}

\bibitem[\protect\citeauthoryear{{Baldi}, {Torresi}, {Migliori}  \&
  {Balmaverde}}{{Baldi} et~al.}{2019a}]{baldi19rev}
{Baldi} R.~D.,  {Torresi} E.,  {Migliori} G.,   {Balmaverde} B.,  2019a,
  \mn@doi [Galaxies] {10.3390/galaxies7030076}, \href
  {https://ui.adsabs.harvard.edu/abs/2019Galax...7...76B} {7, 76}

\bibitem[\protect\citeauthoryear{{Baldi}, {Rodr{\'\i}guez Zaur{\'\i}n},
  {Chiaberge}, {Capetti}, {Sparks}  \& {McHardy}}{{Baldi}
  et~al.}{2019b}]{baldi19b}
{Baldi} R.~D.,  {Rodr{\'\i}guez Zaur{\'\i}n} J.,  {Chiaberge} M.,  {Capetti}
  A.,  {Sparks} W.~B.,   {McHardy} I.~M.,  2019b, \mn@doi [\apj]
  {10.3847/1538-4357/aaf002}, \href
  {https://ui.adsabs.harvard.edu/abs/2019ApJ...870...53B} {870, 53}

\bibitem[\protect\citeauthoryear{{Baldi} et~al.,}{{Baldi}
  et~al.}{2021}]{baldi21}
{Baldi} R.~D.,  et~al., 2021, \mn@doi [\mnras] {10.1093/mnras/staa3519}, \href
  {https://ui.adsabs.harvard.edu/abs/2021MNRAS.500.4749B} {500, 4749}

\bibitem[\protect\citeauthoryear{{Baldwin}, {Phillips}  \&
  {Terlevich}}{{Baldwin} et~al.}{1981}]{baldwin81}
{Baldwin} J.~A.,  {Phillips} M.~M.,   {Terlevich} R.,  1981, \pasp, \href
  {http://adsabs.harvard.edu/abs/1981PASP...93....5B} {93, 5}

\bibitem[\protect\citeauthoryear{{Balick} \& {Heckman}}{{Balick} \&
  {Heckman}}{1982}]{balick82}
{Balick} B.,  {Heckman} T.~M.,  1982, \mn@doi [\araa]
  {10.1146/annurev.aa.20.090182.002243}, \href
  {http://adsabs.harvard.edu/abs/1982ARA%26A..20..431B} {20, 431}

\bibitem[\protect\citeauthoryear{{Balmaverde} \& {Capetti}}{{Balmaverde} \&
  {Capetti}}{2006}]{balmaverde06b}
{Balmaverde} B.,  {Capetti} A.,  2006, \mn@doi [A\&A]
  {10.1051/0004-6361:20054031}, \href
  {http://adsabs.harvard.edu/cgi-bin/nph-bib_query?bibcode=2006A%26A...447...97B&db_key=AST}
  {447, 97}

\bibitem[\protect\citeauthoryear{{Balmaverde}, {Capetti}  \&
  {Grandi}}{{Balmaverde} et~al.}{2006}]{balmaverde06a}
{Balmaverde} B.,  {Capetti} A.,   {Grandi} P.,  2006, \mn@doi [\aap]
  {10.1051/0004-6361:20053799}, \href
  {http://adsabs.harvard.edu/cgi-bin/nph-bib_query?bibcode=2006A%26A...451...35B&db_key=AST}
  {451, 35}

\bibitem[\protect\citeauthoryear{{Balmaverde}, {Baldi}  \&
  {Capetti}}{{Balmaverde} et~al.}{2008}]{balmaverde08}
{Balmaverde} B.,  {Baldi} R.~D.,   {Capetti} A.,  2008, \mn@doi [A\&A]
  {10.1051/0004-6361:200809810}, \href
  {http://adsabs.harvard.edu/abs/2008A%26A...486..119B} {486, 119}

\bibitem[\protect\citeauthoryear{{Barausse}, {Shankar}, {Bernardi}, {Dubois}
  \& {Sheth}}{{Barausse} et~al.}{2017}]{barausse17}
{Barausse} E.,  {Shankar} F.,  {Bernardi} M.,  {Dubois} Y.,   {Sheth} R.~K.,
  2017, \mn@doi [\mnras] {10.1093/mnras/stx799}, \href
  {http://adsabs.harvard.edu/abs/2017MNRAS.468.4782B} {468, 4782}

\bibitem[\protect\citeauthoryear{{Begelman}}{{Begelman}}{2012}]{begelman12}
{Begelman} M.~C.,  2012, \mn@doi [\mnras] {10.1111/j.1365-2966.2011.20071.x},
  \href {http://adsabs.harvard.edu/abs/2012MNRAS.420.2912B} {420, 2912}

\bibitem[\protect\citeauthoryear{{Behar} et~al.,}{{Behar}
  et~al.}{2020}]{behar20}
{Behar} E.,  et~al., 2020, \mn@doi [\mnras] {10.1093/mnras/stz3273}, \href
  {https://ui.adsabs.harvard.edu/abs/2020MNRAS.491.3523B} {491, 3523}

\bibitem[\protect\citeauthoryear{{Behroozi}, {Wechsler}, {Hearin}  \&
  {Conroy}}{{Behroozi} et~al.}{2019}]{behroozi19}
{Behroozi} P.,  {Wechsler} R.~H.,  {Hearin} A.~P.,   {Conroy} C.,  2019,
  \mn@doi [\mnras] {10.1093/mnras/stz1182}, \href
  {https://ui.adsabs.harvard.edu/abs/2019MNRAS.488.3143B} {488, 3143}

\bibitem[\protect\citeauthoryear{{Bell} et~al.,}{{Bell} et~al.}{2011}]{bell11}
{Bell} M.~E.,  et~al., 2011, \mn@doi [\mnras]
  {10.1111/j.1365-2966.2010.17692.x}, \href
  {https://ui.adsabs.harvard.edu/abs/2011MNRAS.411..402B} {411, 402}

\bibitem[\protect\citeauthoryear{{Bennett}}{{Bennett}}{1962}]{bennett62a}
{Bennett} A.~S.,  1962, \memras, \href
  {http://adsabs.harvard.edu/cgi-bin/nph-bib_query?bibcode=1962MmRAS..68..163B&db_key=AST}
  {68, 163}

\bibitem[\protect\citeauthoryear{{Bernardi} et~al.,}{{Bernardi}
  et~al.}{2003}]{bernardi03}
{Bernardi} M.,  et~al., 2003, \mn@doi [\aj] {10.1086/374256}, \href
  {http://adsabs.harvard.edu/abs/2003AJ....125.1849B} {125, 1849}

\bibitem[\protect\citeauthoryear{{Best} \& {Heckman}}{{Best} \&
  {Heckman}}{2012}]{best12}
{Best} P.~N.,  {Heckman} T.~M.,  2012, \mn@doi [\mnras]
  {10.1111/j.1365-2966.2012.20414.x}, \href
  {http://adsabs.harvard.edu/abs/2012MNRAS.421.1569B} {421, 1569}

\bibitem[\protect\citeauthoryear{{Best}, {Kauffmann}, {Heckman}  \&
  {Ivezi{\'c}}}{{Best} et~al.}{2005a}]{best05a}
{Best} P.~N.,  {Kauffmann} G.,  {Heckman} T.~M.,   {Ivezi{\'c}} {\v Z}.,
  2005a, \mn@doi [MNRAS] {10.1111/j.1365-2966.2005.09283.x}, \href
  {http://adsabs.harvard.edu/abs/2005MNRAS.362....9B} {362, 9}

\bibitem[\protect\citeauthoryear{{Best}, {Kauffmann}, {Heckman}, {Brinchmann},
  {Charlot}, {Ivezi{\'c}}  \& {White}}{{Best} et~al.}{2005b}]{best05b}
{Best} P.~N.,  {Kauffmann} G.,  {Heckman} T.~M.,  {Brinchmann} J.,  {Charlot}
  S.,  {Ivezi{\'c}} {\v Z}.,   {White} S.~D.~M.,  2005b, \mn@doi [MNRAS]
  {10.1111/j.1365-2966.2005.09192.x}, \href
  {http://adsabs.harvard.edu/abs/2005MNRAS.362...25B} {362, 25}

\bibitem[\protect\citeauthoryear{{Beswick}, {Argo}, {Evans}, {McHardy},
  {Williams}  \& {Westcott}}{{Beswick} et~al.}{2014}]{beswick14}
{Beswick} R.,  {Argo} M.~K.,  {Evans} R.,  {McHardy} I.,  {Williams} D.~R.~A.,
   {Westcott} J.,  2014, in Proceedings of the 12th European VLBI Network
  Symposium and Users Meeting (EVN 2014). 7-10 October 2014. Cagliari, Italy.
  p.~10

\bibitem[\protect\citeauthoryear{{Binette}, {Magris}, {Stasi{\'n}ska}  \&
  {Bruzual}}{{Binette} et~al.}{1994}]{binette94}
{Binette} L.,  {Magris} C.~G.,  {Stasi{\'n}ska} G.,   {Bruzual} A.~G.,  1994,
  A\&A, \href {http://adsabs.harvard.edu/abs/1994A%26A...292...13B} {292, 13}

\bibitem[\protect\citeauthoryear{{Bisogni}, {Marconi}  \& {Risaliti}}{{Bisogni}
  et~al.}{2017}]{bisogni17}
{Bisogni} S.,  {Marconi} A.,   {Risaliti} G.,  2017, \mn@doi [\mnras]
  {10.1093/mnras/stw2324}, \href
  {http://adsabs.harvard.edu/abs/2017MNRAS.464..385B} {464, 385}

\bibitem[\protect\citeauthoryear{{Blandford} \& {Znajek}}{{Blandford} \&
  {Znajek}}{1977}]{blandford77}
{Blandford} R.~D.,  {Znajek} R.~L.,  1977, \mnras, \href
  {http://adsabs.harvard.edu/abs/1977MNRAS.179..433B} {179, 433}

\bibitem[\protect\citeauthoryear{{Blandford}, {Meier}  \&
  {Readhead}}{{Blandford} et~al.}{2019}]{blandford19}
{Blandford} R.,  {Meier} D.,   {Readhead} A.,  2019, \mn@doi [\araa]
  {10.1146/annurev-astro-081817-051948}, \href
  {https://ui.adsabs.harvard.edu/abs/2019ARA&A..57..467B} {57, 467}

\bibitem[\protect\citeauthoryear{{Bluck} et~al.,}{{Bluck}
  et~al.}{2020}]{bluck20}
{Bluck} A. F.~L.,  et~al., 2020, \mn@doi [\mnras] {10.1093/mnras/staa2806},
  \href {https://ui.adsabs.harvard.edu/abs/2020MNRAS.499..230B} {499, 230}

\bibitem[\protect\citeauthoryear{{Bonchi}, {La Franca}, {Melini}, {Bongiorno}
  \& {Fiore}}{{Bonchi} et~al.}{2013}]{bonchi13}
{Bonchi} A.,  {La Franca} F.,  {Melini} G.,  {Bongiorno} A.,   {Fiore} F.,
  2013, \mn@doi [\mnras] {10.1093/mnras/sts456}, \href
  {http://adsabs.harvard.edu/abs/2013MNRAS.429.1970B} {429, 1970}

\bibitem[\protect\citeauthoryear{{Bonzini}, {Padovani}, {Mainieri},
  {Kellermann}, {Miller}, {Rosati}, {Tozzi}  \& {Vattakunnel}}{{Bonzini}
  et~al.}{2013}]{bonzini13}
{Bonzini} M.,  {Padovani} P.,  {Mainieri} V.,  {Kellermann} K.~I.,  {Miller}
  N.,  {Rosati} P.,  {Tozzi} P.,   {Vattakunnel} S.,  2013, \mn@doi [\mnras]
  {10.1093/mnras/stt1879}, \href
  {http://adsabs.harvard.edu/abs/2013MNRAS.436.3759B} {436, 3759}

\bibitem[\protect\citeauthoryear{{Brinchmann}, {Charlot}, {White}, {Tremonti},
  {Kauffmann}, {Heckman}  \& {Brinkmann}}{{Brinchmann}
  et~al.}{2004}]{brinchmann04}
{Brinchmann} J.,  {Charlot} S.,  {White} S.~D.~M.,  {Tremonti} C.,  {Kauffmann}
  G.,  {Heckman} T.,   {Brinkmann} J.,  2004, \mn@doi [\mnras]
  {10.1111/j.1365-2966.2004.07881.x}, \href
  {https://ui.adsabs.harvard.edu/abs/2004MNRAS.351.1151B} {351, 1151}

\bibitem[\protect\citeauthoryear{{Brown}, {Jannuzi}, {Floyd}  \&
  {Mould}}{{Brown} et~al.}{2011}]{brown11}
{Brown} M. J.~I.,  {Jannuzi} B.~T.,  {Floyd} D. J.~E.,   {Mould} J.~R.,  2011,
  \mn@doi [\apjl] {10.1088/2041-8205/731/2/L41}, \href
  {https://ui.adsabs.harvard.edu/abs/2011ApJ...731L..41B} {731, L41}

\bibitem[\protect\citeauthoryear{{Bundy} et~al.,}{{Bundy}
  et~al.}{2015}]{bundy15}
{Bundy} K.,  et~al., 2015, \mn@doi [\apj] {10.1088/0004-637X/798/1/7}, \href
  {https://ui.adsabs.harvard.edu/abs/2015ApJ...798....7B} {798, 7}

\bibitem[\protect\citeauthoryear{{Buttiglione}, {Capetti}, {Celotti}, {Axon},
  {Chiaberge}, {Macchetto}  \& {Sparks}}{{Buttiglione}
  et~al.}{2009}]{buttiglione09}
{Buttiglione} S.,  {Capetti} A.,  {Celotti} A.,  {Axon} D.~J.,  {Chiaberge} M.,
   {Macchetto} F.~D.,   {Sparks} W.~B.,  2009, \mn@doi [\aap]
  {10.1051/0004-6361:200811102}, \href
  {http://adsabs.harvard.edu/abs/2009A%26A...495.1033B} {495, 1033}

\bibitem[\protect\citeauthoryear{{Buttiglione}, {Capetti}, {Celotti}, {Axon},
  {Chiaberge}, {Macchetto}  \& {Sparks}}{{Buttiglione}
  et~al.}{2010}]{buttiglione10}
{Buttiglione} S.,  {Capetti} A.,  {Celotti} A.,  {Axon} D.~J.,  {Chiaberge} M.,
   {Macchetto} F.~D.,   {Sparks} W.~B.,  2010, \mn@doi [\aap]
  {10.1051/0004-6361/200913290}, \href
  {http://adsabs.harvard.edu/abs/2010A%26A...509A...6B} {509, A6}

\bibitem[\protect\citeauthoryear{{Cao}}{{Cao}}{2007}]{cao07}
{Cao} X.,  2007, \mn@doi [\apj] {10.1086/512116}, \href
  {https://ui.adsabs.harvard.edu/abs/2007ApJ...659..950C} {659, 950}

\bibitem[\protect\citeauthoryear{{Capetti}}{{Capetti}}{2011}]{capetti11c}
{Capetti} A.,  2011, \mn@doi [\aap] {10.1051/0004-6361/201117937}, \href
  {http://adsabs.harvard.edu/abs/2011A%26A...535A..28C} {535, A28}

\bibitem[\protect\citeauthoryear{{Capetti} \& {Baldi}}{{Capetti} \&
  {Baldi}}{2011}]{capetti11b}
{Capetti} A.,  {Baldi} R.~D.,  2011, \mn@doi [A\&A]
  {10.1051/0004-6361/201016388}, \href
  {http://adsabs.harvard.edu/abs/2011A%26A...529A.126C} {529, A126}

\bibitem[\protect\citeauthoryear{{Capetti} \& {Balmaverde}}{{Capetti} \&
  {Balmaverde}}{2006}]{capetti06}
{Capetti} A.,  {Balmaverde} B.,  2006, \mn@doi [A\&A]
  {10.1051/0004-6361:20054490}, \href
  {http://adsabs.harvard.edu/cgi-bin/nph-bib_query?bibcode=2006A%26A...453...27C&db_key=AST}
  {453, 27 }

\bibitem[\protect\citeauthoryear{{Capetti} \& {Balmaverde}}{{Capetti} \&
  {Balmaverde}}{2007}]{capetti07}
{Capetti} A.,  {Balmaverde} B.,  2007, \mn@doi [\aap]
  {10.1051/0004-6361:20066684}, \href
  {https://ui.adsabs.harvard.edu/abs/2007A&A...469...75C} {469, 75}

\bibitem[\protect\citeauthoryear{{Capetti}, {Verdoes Kleijn}  \&
  {Chiaberge}}{{Capetti} et~al.}{2005}]{capetti:cccriga}
{Capetti} A.,  {Verdoes Kleijn} G.~A.,   {Chiaberge} M.,  2005, \mn@doi [\aap]
  {10.1051/0004-6361:20041609}, \href
  {http://adsabs.harvard.edu/cgi-bin/nph-bib_query?bibcode=2005A%26A...439..935C&db_key=AST}
  {439, 935}

\bibitem[\protect\citeauthoryear{{Capetti}, {Robinson}, {Baldi}, {Buttiglione},
  {Axon}, {Celotti}  \& {Chiaberge}}{{Capetti} et~al.}{2013}]{capetti13}
{Capetti} A.,  {Robinson} A.,  {Baldi} R.~D.,  {Buttiglione} S.,  {Axon} D.~J.,
   {Celotti} A.,   {Chiaberge} M.,  2013, \mn@doi [\aap]
  {10.1051/0004-6361/201220662}, \href
  {https://ui.adsabs.harvard.edu/abs/2013A&A...551A..55C} {551, A55}

\bibitem[\protect\citeauthoryear{{Capetti}, {Massaro}  \& {Baldi}}{{Capetti}
  et~al.}{2017}]{capetti17a}
{Capetti} A.,  {Massaro} F.,   {Baldi} R.~D.,  2017, \mn@doi [\aap]
  {10.1051/0004-6361/201629287}, \href
  {https://ui.adsabs.harvard.edu/abs/2017A&A...598A..49C} {598, A49}

\bibitem[\protect\citeauthoryear{{Cappi} et~al.,}{{Cappi}
  et~al.}{2006}]{cappi06}
{Cappi} M.,  et~al., 2006, \mn@doi [\aap] {10.1051/0004-6361:20053893}, \href
  {https://ui.adsabs.harvard.edu/abs/2006A&A...446..459C} {446, 459}

\bibitem[\protect\citeauthoryear{{Cattaneo} \& {Best}}{{Cattaneo} \&
  {Best}}{2009}]{cattaneo09}
{Cattaneo} A.,  {Best} P.~N.,  2009, \mn@doi [\mnras]
  {10.1111/j.1365-2966.2009.14557.x}, \href
  {http://adsabs.harvard.edu/abs/2009MNRAS.395..518C} {395, 518}

\bibitem[\protect\citeauthoryear{{Cen}}{{Cen}}{2012}]{cen12}
{Cen} R.,  2012, \mn@doi [\apj] {10.1088/0004-637X/755/1/28}, \href
  {https://ui.adsabs.harvard.edu/abs/2012ApJ...755...28C} {755, 28}

\bibitem[\protect\citeauthoryear{{Cen}}{{Cen}}{2015}]{cen15}
{Cen} R.,  2015, \mn@doi [\apjl] {10.1088/2041-8205/805/1/L9}, \href
  {https://ui.adsabs.harvard.edu/abs/2015ApJ...805L...9C} {805, L9}

\bibitem[\protect\citeauthoryear{{Chang}, {Xie}, {Liu}, {Ho}, {Dong}, {Han}  \&
  {Wang}}{{Chang} et~al.}{2021}]{chang21}
{Chang} N.,  {Xie} F.~G.,  {Liu} X.,  {Ho} L.~C.,  {Dong} A.~J.,  {Han} Z.~H.,
   {Wang} X.,  2021, \mn@doi [\mnras] {10.1093/mnras/stab521}, \href
  {https://ui.adsabs.harvard.edu/abs/2021MNRAS.503.1987C} {503, 1987}

\bibitem[\protect\citeauthoryear{{Chen}, {Halpern}  \& {Filippenko}}{{Chen}
  et~al.}{1989}]{chen89}
{Chen} K.,  {Halpern} J.~P.,   {Filippenko} A.~V.,  1989, \mn@doi [\apj]
  {10.1086/167332}, \href {http://adsabs.harvard.edu/abs/1989ApJ...339..742C}
  {339, 742}

\bibitem[\protect\citeauthoryear{{Chen} et~al.,}{{Chen} et~al.}{2021}]{chen21}
{Chen} Y.,  et~al., 2021, \mn@doi [\apj] {10.3847/1538-4357/abf4ff}, \href
  {https://ui.adsabs.harvard.edu/abs/2021ApJ...913...93C} {913, 93}

\bibitem[\protect\citeauthoryear{{Chiaberge} \& {Marconi}}{{Chiaberge} \&
  {Marconi}}{2011}]{chiaberge11}
{Chiaberge} M.,  {Marconi} A.,  2011, \mn@doi [\mnras]
  {10.1111/j.1365-2966.2011.19079.x}, \href
  {http://adsabs.harvard.edu/abs/2011MNRAS.416..917C} {416, 917}

\bibitem[\protect\citeauthoryear{{Chiaberge}, {Capetti}  \&
  {Celotti}}{{Chiaberge} et~al.}{1999}]{chiaberge:ccc}
{Chiaberge} M.,  {Capetti} A.,   {Celotti} A.,  1999, \aap, \href
  {http://adsabs.harvard.edu/cgi-bin/nph-bib_query?bibcode=1999A%26A...349...77C&amp;db_key=AST}
  {349, 77}

\bibitem[\protect\citeauthoryear{{Chiaberge}, {Capetti}  \&
  {Macchetto}}{{Chiaberge} et~al.}{2005}]{chiaberge05}
{Chiaberge} M.,  {Capetti} A.,   {Macchetto} F.~D.,  2005, \mn@doi [\apj]
  {10.1086/429612}, \href
  {http://adsabs.harvard.edu/cgi-bin/nph-bib_query?bibcode=2005\apj...625..716C&db_key=AST}
  {625, 716}

\bibitem[\protect\citeauthoryear{{Cirasuolo}, {Magliocchetti}, {Celotti}  \&
  {Danese}}{{Cirasuolo} et~al.}{2003}]{cirasuolo03}
{Cirasuolo} M.,  {Magliocchetti} M.,  {Celotti} A.,   {Danese} L.,  2003,
  \mn@doi [\mnras] {10.1046/j.1365-8711.2003.06485.x}, \href
  {https://ui.adsabs.harvard.edu/abs/2003MNRAS.341..993C} {341, 993}

\bibitem[\protect\citeauthoryear{{Condon}}{{Condon}}{1992}]{condon92}
{Condon} J.~J.,  1992, \mn@doi [\araa] {10.1146/annurev.aa.30.090192.003043},
  \href {http://adsabs.harvard.edu/abs/1992ARA%26A..30..575C} {30, 575}

\bibitem[\protect\citeauthoryear{{Coriat} et~al.,}{{Coriat}
  et~al.}{2011}]{coriat11}
{Coriat} M.,  et~al., 2011, \mn@doi [\mnras]
  {10.1111/j.1365-2966.2011.18433.x}, \href
  {https://ui.adsabs.harvard.edu/abs/2011MNRAS.414..677C} {414, 677}

\bibitem[\protect\citeauthoryear{{Czerny}, {Siemiginowska}, {Janiuk},
  {Nikiel-Wroczy{\'n}ski}  \& {Stawarz}}{{Czerny} et~al.}{2009}]{czerny09}
{Czerny} B.,  {Siemiginowska} A.,  {Janiuk} A.,  {Nikiel-Wroczy{\'n}ski} B.,
  {Stawarz} {\L}.,  2009, \mn@doi [\apj] {10.1088/0004-637X/698/1/840}, \href
  {https://ui.adsabs.harvard.edu/abs/2009ApJ...698..840C} {698, 840}

\bibitem[\protect\citeauthoryear{{Davis} \& {Tchekhovskoy}}{{Davis} \&
  {Tchekhovskoy}}{2020}]{davis20}
{Davis} S.~W.,  {Tchekhovskoy} A.,  2020, \mn@doi [\araa]
  {10.1146/annurev-astro-081817-051905}, \href
  {https://ui.adsabs.harvard.edu/abs/2020ARA&A..58..407D} {58, 407}

\bibitem[\protect\citeauthoryear{{Davis} et~al.,}{{Davis}
  et~al.}{2014}]{davis14}
{Davis} B.~L.,  et~al., 2014, \mn@doi [\apj] {10.1088/0004-637X/789/2/124},
  \href {https://ui.adsabs.harvard.edu/abs/2014ApJ...789..124D} {789, 124}

\bibitem[\protect\citeauthoryear{{Decarli}, {Gavazzi}, {Arosio}, {Cortese},
  {Boselli}, {Bonfanti}  \& {Colpi}}{{Decarli} et~al.}{2007}]{decarli07}
{Decarli} R.,  {Gavazzi} G.,  {Arosio} I.,  {Cortese} L.,  {Boselli} A.,
  {Bonfanti} C.,   {Colpi} M.,  2007, \mn@doi [\mnras]
  {10.1111/j.1365-2966.2007.12208.x}, \href
  {http://adsabs.harvard.edu/abs/2007MNRAS.381..136D} {381, 136}

\bibitem[\protect\citeauthoryear{{Dickson}, {Tadhunter}, {Shaw}, {Clark}  \&
  {Morganti}}{{Dickson} et~al.}{1995}]{dickson95}
{Dickson} R.,  {Tadhunter} C.,  {Shaw} M.,  {Clark} N.,   {Morganti} R.,  1995,
  \mn@doi [\mnras] {10.1093/mnras/273.1.L29}, \href
  {https://ui.adsabs.harvard.edu/abs/1995MNRAS.273L..29D} {273, L29}

\bibitem[\protect\citeauthoryear{{Donea} \& {Biermann}}{{Donea} \&
  {Biermann}}{2002}]{donea02}
{Donea} A.-C.,  {Biermann} P.~L.,  2002, \mn@doi [\pasa] {10.1071/AS01078},
  \href {http://adsabs.harvard.edu/abs/2002PASA...19..125D} {19, 125}

\bibitem[\protect\citeauthoryear{{Dopita}, {Koratkar}, {Allen}, {Tsvetanov},
  {Ford}, {Bicknell}  \& {Sutherland}}{{Dopita} et~al.}{1997}]{dopita:m87}
{Dopita} M.~A.,  {Koratkar} A.~P.,  {Allen} M.~G.,  {Tsvetanov} Z.~I.,  {Ford}
  H.~C.,  {Bicknell} G.~V.,   {Sutherland} R.~S.,  1997, \apj, \href
  {http://adsabs.harvard.edu/cgi-bin/nph-bib_query?bibcode=1997\apj...490..202D&amp;db_key=AST}
  {490, 202}

\bibitem[\protect\citeauthoryear{{Drappeau} et~al.,}{{Drappeau}
  et~al.}{2017}]{drappeau17}
{Drappeau} S.,  et~al., 2017, \mn@doi [\mnras] {10.1093/mnras/stw3277}, \href
  {https://ui.adsabs.harvard.edu/abs/2017MNRAS.466.4272D} {466, 4272}

\bibitem[\protect\citeauthoryear{{Dugan}, {Bryan}, {Gaibler}, {Silk}  \&
  {Haas}}{{Dugan} et~al.}{2014}]{dugan14}
{Dugan} Z.,  {Bryan} S.,  {Gaibler} V.,  {Silk} J.,   {Haas} M.,  2014, \mn@doi
  [\apj] {10.1088/0004-637X/796/2/113}, \href
  {https://ui.adsabs.harvard.edu/abs/2014ApJ...796..113D} {796, 113}

\bibitem[\protect\citeauthoryear{{Dullo}, {Bouquin}, {Gil de Paz}, {Knapen}  \&
  {Gorgas}}{{Dullo} et~al.}{2020}]{dullo20}
{Dullo} B.~T.,  {Bouquin} A. Y.~K.,  {Gil de Paz} A.,  {Knapen} J.~H.,
  {Gorgas} J.,  2020, \mn@doi [\apj] {10.3847/1538-4357/ab9dff}, \href
  {https://ui.adsabs.harvard.edu/abs/2020ApJ...898...83D} {898, 83}

\bibitem[\protect\citeauthoryear{{Dunn} \& {Fabian}}{{Dunn} \&
  {Fabian}}{2006}]{dunn06}
{Dunn} R.~J.~H.,  {Fabian} A.~C.,  2006, \mn@doi [\mnras]
  {10.1111/j.1365-2966.2006.11080.x}, \href
  {https://ui.adsabs.harvard.edu/abs/2006MNRAS.373..959D} {373, 959}

\bibitem[\protect\citeauthoryear{{Esin}, {McClintock}  \& {Narayan}}{{Esin}
  et~al.}{1997}]{esin97}
{Esin} A.~A.,  {McClintock} J.~E.,   {Narayan} R.,  1997, \mn@doi [\apj]
  {10.1086/304829}, \href
  {https://ui.adsabs.harvard.edu/abs/1997ApJ...489..865E} {489, 865}

\bibitem[\protect\citeauthoryear{{Esparza-Arredondo}, {Osorio-Clavijo},
  {Gonz{\'a}lez-Mart{\'\i}n}, {Victoria-Ceballos}, {Haro-Corzo},
  {Reyes-Amador}, {L{\'o}pez-S{\'a}nchez}  \& {Pasetto}}{{Esparza-Arredondo}
  et~al.}{2020}]{esparza20}
{Esparza-Arredondo} D.,  {Osorio-Clavijo} N.,  {Gonz{\'a}lez-Mart{\'\i}n} O.,
  {Victoria-Ceballos} C.,  {Haro-Corzo} S. A.~R.,  {Reyes-Amador} O.~U.,
  {L{\'o}pez-S{\'a}nchez} J.,   {Pasetto} A.,  2020, \mn@doi [\apj]
  {10.3847/1538-4357/abc425}, \href
  {https://ui.adsabs.harvard.edu/abs/2020ApJ...905...29E} {905, 29}

\bibitem[\protect\citeauthoryear{{Fabian}}{{Fabian}}{2012}]{fabian12}
{Fabian} A.~C.,  2012, \mn@doi [\araa] {10.1146/annurev-astro-081811-125521},
  \href {https://ui.adsabs.harvard.edu/abs/2012ARA&A..50..455F} {50, 455}

\bibitem[\protect\citeauthoryear{{Fabian} \& {Canizares}}{{Fabian} \&
  {Canizares}}{1988}]{fabian88}
{Fabian} A.~C.,  {Canizares} C.~R.,  1988, \mn@doi [\nat] {10.1038/333829a0},
  \href {http://adsabs.harvard.edu/abs/1988Natur.333..829F} {333, 829}

\bibitem[\protect\citeauthoryear{{Falcke}, {Nagar}, {Wilson}  \&
  {Ulvestad}}{{Falcke} et~al.}{2000}]{falcke00}
{Falcke} H.,  {Nagar} N.~M.,  {Wilson} A.~S.,   {Ulvestad} J.~S.,  2000, \apj,
  \href
  {http://adsabs.harvard.edu/cgi-bin/nph-bib_query?bibcode=2000\apj...542..197F&db_key=AST}
  {542, 197}

\bibitem[\protect\citeauthoryear{{Falcke}, {Nagar}, {Wilson}, {Ho}  \&
  {Ulvestad}}{{Falcke} et~al.}{2001}]{falcke01}
{Falcke} H.,  {Nagar} N.~M.,  {Wilson} A.~S.,  {Ho} L.~C.,   {Ulvestad} J.~S.,
  2001, in {Kaper} L.,  {Heuvel} E. P.~J. V.~D.,   {Woudt} P.~A.,  eds, Black
  Holes in Binaries and Galactic Nuclei. Springer-Verlag, p.~218,
  \mn@doi{10.1007/10720995_45}

\bibitem[\protect\citeauthoryear{{Falcke}, {K{\"o}rding}  \&
  {Markoff}}{{Falcke} et~al.}{2004}]{falcke04}
{Falcke} H.,  {K{\"o}rding} E.,   {Markoff} S.,  2004, \mn@doi [\aap]
  {10.1051/0004-6361:20031683}, \href
  {http://adsabs.harvard.edu/abs/2004A%26A...414..895F} {414, 895}

\bibitem[\protect\citeauthoryear{{Fanaroff} \& {Riley}}{{Fanaroff} \&
  {Riley}}{1974}]{fanaroff74}
{Fanaroff} B.~L.,  {Riley} J.~M.,  1974, \mn@doi [\mnras]
  {10.1093/mnras/167.1.31P}, \href
  {https://ui.adsabs.harvard.edu/abs/1974MNRAS.167P..31F} {167, 31P}

\bibitem[\protect\citeauthoryear{{Fanti}, {Fanti}, {de Ruiter}  \&
  {Parma}}{{Fanti} et~al.}{1987}]{fanti87}
{Fanti} C.,  {Fanti} R.,  {de Ruiter} H.~R.,   {Parma} P.,  1987, \aaps, \href
  {http://adsabs.harvard.edu/cgi-bin/nph-bib_query?bibcode=1987A%26AS...69...57F&amp;db_key=AST}
  {69, 57}

\bibitem[\protect\citeauthoryear{{Faucher-Gigu{\`e}re} \&
  {Quataert}}{{Faucher-Gigu{\`e}re} \& {Quataert}}{2012}]{faucher12}
{Faucher-Gigu{\`e}re} C.-A.,  {Quataert} E.,  2012, \mn@doi [\mnras]
  {10.1111/j.1365-2966.2012.21512.x}, \href
  {https://ui.adsabs.harvard.edu/abs/2012MNRAS.425..605F} {425, 605}

\bibitem[\protect\citeauthoryear{{Feigelson} \& {Nelson}}{{Feigelson} \&
  {Nelson}}{1985}]{feigelson85}
{Feigelson} E.~D.,  {Nelson} P.~I.,  1985, \apj, \href
  {http://adsabs.harvard.edu/cgi-bin/nph-bib_query?bibcode=1985\apj...293..192F&amp;db_key=AST}
  {293, 192}

\bibitem[\protect\citeauthoryear{{Fender}}{{Fender}}{2001}]{fender01}
{Fender} R.~P.,  2001, \mn@doi [\mnras] {10.1046/j.1365-8711.2001.04080.x},
  \href {https://ui.adsabs.harvard.edu/abs/2001MNRAS.322...31F} {322, 31}

\bibitem[\protect\citeauthoryear{{Fender}, {Gallo}  \& {Jonker}}{{Fender}
  et~al.}{2003}]{fender03}
{Fender} R.~P.,  {Gallo} E.,   {Jonker} P.~G.,  2003, \mn@doi [\mnras]
  {10.1046/j.1365-8711.2003.06950.x}, \href
  {https://ui.adsabs.harvard.edu/abs/2003MNRAS.343L..99F} {343, L99}

\bibitem[\protect\citeauthoryear{{Fender}, {Belloni}  \& {Gallo}}{{Fender}
  et~al.}{2004}]{fender04}
{Fender} R.~P.,  {Belloni} T.~M.,   {Gallo} E.,  2004, \mn@doi [\mnras]
  {10.1111/j.1365-2966.2004.08384.x}, \href
  {https://ui.adsabs.harvard.edu/abs/2004MNRAS.355.1105F} {355, 1105}

\bibitem[\protect\citeauthoryear{{Fender}, {Homan}  \& {Belloni}}{{Fender}
  et~al.}{2009}]{fender09}
{Fender} R.~P.,  {Homan} J.,   {Belloni} T.~M.,  2009, \mn@doi [\mnras]
  {10.1111/j.1365-2966.2009.14841.x}, \href
  {https://ui.adsabs.harvard.edu/abs/2009MNRAS.396.1370F} {396, 1370}

\bibitem[\protect\citeauthoryear{{Fern{\'a}ndez-Ontiveros} \&
  {Mu{\~n}oz-Darias}}{{Fern{\'a}ndez-Ontiveros} \&
  {Mu{\~n}oz-Darias}}{2021}]{ontiveros21}
{Fern{\'a}ndez-Ontiveros} J.~A.,  {Mu{\~n}oz-Darias} T.,  2021, \mn@doi
  [\mnras] {10.1093/mnras/stab1108}, \href
  {https://ui.adsabs.harvard.edu/abs/2021MNRAS.504.5726F} {504, 5726}

\bibitem[\protect\citeauthoryear{{Fern{\'a}ndez-Ontiveros}, {Prieto}  \&
  {Acosta-Pulido}}{{Fern{\'a}ndez-Ontiveros} et~al.}{2009}]{fernandez09}
{Fern{\'a}ndez-Ontiveros} J.~A.,  {Prieto} M.~A.,   {Acosta-Pulido} J.~A.,
  2009, \mn@doi [\mnras] {10.1111/j.1745-3933.2008.00575.x}, \href
  {https://ui.adsabs.harvard.edu/abs/2009MNRAS.392L..16F} {392, L16}

\bibitem[\protect\citeauthoryear{{Ferreira}}{{Ferreira}}{1997}]{ferreira97}
{Ferreira} J.,  1997, \aap, \href
  {https://ui.adsabs.harvard.edu/abs/1997A&A...319..340F} {319, 340}

\bibitem[\protect\citeauthoryear{{Ferreira}, {Petrucci}, {Murphy}, {Zanni}  \&
  {Henri}}{{Ferreira} et~al.}{2010}]{ferreira10}
{Ferreira} J.,  {Petrucci} P.~O.,  {Murphy} G.,  {Zanni} C.,   {Henri} G.,
  2010, {Jet Emitting Discs: a New Accretion Flow Solution}.
p.~49

\bibitem[\protect\citeauthoryear{{Filho}, {Barthel}  \& {Ho}}{{Filho}
  et~al.}{2006}]{filho06}
{Filho} M.~E.,  {Barthel} P.~D.,   {Ho} L.~C.,  2006, \mn@doi [\aap]
  {10.1051/0004-6361:20054510}, \href
  {http://adsabs.harvard.edu/abs/2006A%26A...451...71F} {451, 71}

\bibitem[\protect\citeauthoryear{{Fischer} et~al.,}{{Fischer}
  et~al.}{2021}]{fischer21}
{Fischer} T.~C.,  et~al., 2021, \mn@doi [\apj] {10.3847/1538-4357/abca3c},
  \href {https://ui.adsabs.harvard.edu/abs/2021ApJ...906...88F} {906, 88}

\bibitem[\protect\citeauthoryear{{Flohic}, {Eracleous}, {Chartas}, {Shields}
  \& {Moran}}{{Flohic} et~al.}{2006}]{flohic06}
{Flohic} H. M.~L.~G.,  {Eracleous} M.,  {Chartas} G.,  {Shields} J.~C.,
  {Moran} E.~C.,  2006, \mn@doi [\apj] {10.1086/505296}, \href
  {https://ui.adsabs.harvard.edu/abs/2006ApJ...647..140F} {647, 140}

\bibitem[\protect\citeauthoryear{{Forman}, {Jones}  \& {Tucker}}{{Forman}
  et~al.}{1985}]{forman85}
{Forman} W.,  {Jones} C.,   {Tucker} W.,  1985, \mn@doi [\apj]
  {10.1086/163218}, \href
  {https://ui.adsabs.harvard.edu/abs/1985ApJ...293..102F} {293, 102}

\bibitem[\protect\citeauthoryear{{Gaibler}, {Khochfar}, {Krause}  \&
  {Silk}}{{Gaibler} et~al.}{2012}]{gaibler12}
{Gaibler} V.,  {Khochfar} S.,  {Krause} M.,   {Silk} J.,  2012, \mn@doi
  [\mnras] {10.1111/j.1365-2966.2012.21479.x}, \href
  {https://ui.adsabs.harvard.edu/abs/2012MNRAS.425..438G} {425, 438}

\bibitem[\protect\citeauthoryear{{Gallimore}, {Axon}, {O'Dea}, {Baum}  \&
  {Pedlar}}{{Gallimore} et~al.}{2006}]{gallimore06}
{Gallimore} J.~F.,  {Axon} D.~J.,  {O'Dea} C.~P.,  {Baum} S.~A.,   {Pedlar} A.,
   2006, \mn@doi [\aj] {10.1086/504593}, \href
  {http://adsabs.harvard.edu/abs/2006AJ....132..546G} {132, 546}

\bibitem[\protect\citeauthoryear{{Gallo} \& {Sesana}}{{Gallo} \&
  {Sesana}}{2019}]{gallo19}
{Gallo} E.,  {Sesana} A.,  2019, \mn@doi [\apjl] {10.3847/2041-8213/ab40c6},
  \href {https://ui.adsabs.harvard.edu/abs/2019ApJ...883L..18G} {883, L18}

\bibitem[\protect\citeauthoryear{{Gallo}, {Treu}, {Jacob}, {Woo}, {Marshall}
  \& {Antonucci}}{{Gallo} et~al.}{2008}]{gallo08}
{Gallo} E.,  {Treu} T.,  {Jacob} J.,  {Woo} J.-H.,  {Marshall} P.~J.,
  {Antonucci} R.,  2008, \mn@doi [\apj] {10.1086/588012}, \href
  {http://adsabs.harvard.edu/abs/2008ApJ...680..154G} {680, 154}

\bibitem[\protect\citeauthoryear{{Garofalo} \& {Bishop}}{{Garofalo} \&
  {Bishop}}{2020}]{garofalo20}
{Garofalo} D.,  {Bishop} K.,  2020, \mn@doi [\pasp] {10.1088/1538-3873/abb999},
  \href {https://ui.adsabs.harvard.edu/abs/2020PASP..132k4103G} {132, 114103}

\bibitem[\protect\citeauthoryear{{Garofalo} \& {Singh}}{{Garofalo} \&
  {Singh}}{2019}]{garofalo19}
{Garofalo} D.,  {Singh} C.~B.,  2019, \mn@doi [\apj]
  {10.3847/1538-4357/aaf056}, \href
  {https://ui.adsabs.harvard.edu/abs/2019ApJ...871..259G} {871, 259}

\bibitem[\protect\citeauthoryear{{Garofalo}, {Evans}  \& {Sambruna}}{{Garofalo}
  et~al.}{2010}]{garofalo10}
{Garofalo} D.,  {Evans} D.~A.,   {Sambruna} R.~M.,  2010, \mn@doi [\mnras]
  {10.1111/j.1365-2966.2010.16797.x}, \href
  {http://adsabs.harvard.edu/abs/2010MNRAS.406..975G} {406, 975}

\bibitem[\protect\citeauthoryear{{Garofalo}, {Kim}  \& {Christian}}{{Garofalo}
  et~al.}{2014}]{garofalo14}
{Garofalo} D.,  {Kim} M.~I.,   {Christian} D.~J.,  2014, \mn@doi [\mnras]
  {10.1093/mnras/stu1086}, \href
  {https://ui.adsabs.harvard.edu/abs/2014MNRAS.442.3097G} {442, 3097}

\bibitem[\protect\citeauthoryear{{Gebhardt} et~al.,}{{Gebhardt}
  et~al.}{2000}]{gebhardt00}
{Gebhardt} K.,  et~al., 2000, ApJL, \href
  {http://adsabs.harvard.edu/cgi-bin/nph-bib_query?bibcode=2000\apj...539L..13G&amp;db_key=AST}
  {539, L13}

\bibitem[\protect\citeauthoryear{{Giannios} \& {Metzger}}{{Giannios} \&
  {Metzger}}{2011}]{giannios11}
{Giannios} D.,  {Metzger} B.~D.,  2011, \mn@doi [\mnras]
  {10.1111/j.1365-2966.2011.19188.x}, \href
  {https://ui.adsabs.harvard.edu/abs/2011MNRAS.416.2102G} {416, 2102}

\bibitem[\protect\citeauthoryear{{Giovannini}}{{Giovannini}}{2003}]{giovannini03}
{Giovannini} G.,  2003, \mn@doi [\nar] {10.1016/S1387-6473(03)00091-5}, \href
  {https://ui.adsabs.harvard.edu/abs/2003NewAR..47..551G} {47, 551}

\bibitem[\protect\citeauthoryear{{Giovannini}}{{Giovannini}}{2004}]{giovannini04}
{Giovannini} G.,  2004, \mn@doi [\apss] {10.1023/B:ASTR.0000044647.04235.d4},
  \href {https://ui.adsabs.harvard.edu/abs/2004Ap&SS.293....1G} {293, 1}

\bibitem[\protect\citeauthoryear{{Giovannini}, {Taylor}, {Feretti}, {Cotton},
  {Lara}  \& {Venturi}}{{Giovannini} et~al.}{2005}]{giovannini05}
{Giovannini} G.,  {Taylor} G.~B.,  {Feretti} L.,  {Cotton} W.~D.,  {Lara} L.,
  {Venturi} T.,  2005, \mn@doi [\apj] {10.1086/426106}, \href
  {http://adsabs.harvard.edu/abs/2005ApJ...618..635G} {618, 635}

\bibitem[\protect\citeauthoryear{{Giroletti} \& {Panessa}}{{Giroletti} \&
  {Panessa}}{2009}]{giroletti09}
{Giroletti} M.,  {Panessa} F.,  2009, \mn@doi [\apjl]
  {10.1088/0004-637X/706/2/L260}, \href
  {http://adsabs.harvard.edu/abs/2009ApJ...706L.260G} {706, L260}

\bibitem[\protect\citeauthoryear{{Gonz{\'a}lez-Mart{\'\i}n}, {Masegosa},
  {M{\'a}rquez}, {Guerrero}  \& {Dultzin-Hacyan}}{{Gonz{\'a}lez-Mart{\'\i}n}
  et~al.}{2006}]{gonzalez06}
{Gonz{\'a}lez-Mart{\'\i}n} O.,  {Masegosa} J.,  {M{\'a}rquez} I.,  {Guerrero}
  M.~A.,   {Dultzin-Hacyan} D.,  2006, \mn@doi [\aap]
  {10.1051/0004-6361:20054756}, \href
  {https://ui.adsabs.harvard.edu/abs/2006A&A...460...45G} {460, 45}

\bibitem[\protect\citeauthoryear{{Grandi}, {Torresi}, {Macconi}, {Boccardi}  \&
  {Capetti}}{{Grandi} et~al.}{2021}]{grandi21}
{Grandi} P.,  {Torresi} E.,  {Macconi} D.,  {Boccardi} B.,   {Capetti} A.,
  2021, \mn@doi [\apj] {10.3847/1538-4357/abe776}, \href
  {https://ui.adsabs.harvard.edu/abs/2021ApJ...911...17G} {911, 17}

\bibitem[\protect\citeauthoryear{{Greene}}{{Greene}}{2012}]{greene12}
{Greene} J.~E.,  2012, \mn@doi [Nature Communications] {10.1038/ncomms2314},
  \href {https://ui.adsabs.harvard.edu/abs/2012NatCo...3.1304G} {3, 1304}

\bibitem[\protect\citeauthoryear{{Greene} \& {Ho}}{{Greene} \&
  {Ho}}{2007}]{greene07}
{Greene} J.~E.,  {Ho} L.~C.,  2007, \mn@doi [\apj] {10.1086/520497}, \href
  {https://ui.adsabs.harvard.edu/abs/2007ApJ...667..131G} {667, 131}

\bibitem[\protect\citeauthoryear{{Greene}, {Strader}  \& {Ho}}{{Greene}
  et~al.}{2020}]{greene20}
{Greene} J.~E.,  {Strader} J.,   {Ho} L.~C.,  2020, \mn@doi [\araa]
  {10.1146/annurev-astro-032620-021835}, \href
  {https://ui.adsabs.harvard.edu/abs/2020ARA&A..58..257G} {58, 257}

\bibitem[\protect\citeauthoryear{{G{\"u}ltekin}, {King}, {Cackett}, {Nyland},
  {Miller}, {Di Matteo}, {Markoff}  \& {Rupen}}{{G{\"u}ltekin}
  et~al.}{2019}]{gultekin19}
{G{\"u}ltekin} K.,  {King} A.~L.,  {Cackett} E.~M.,  {Nyland} K.,  {Miller}
  J.~M.,  {Di Matteo} T.,  {Markoff} S.,   {Rupen} M.~P.,  2019, \mn@doi [\apj]
  {10.3847/1538-4357/aaf6b9}, \href
  {https://ui.adsabs.harvard.edu/abs/2019ApJ...871...80G} {871, 80}

\bibitem[\protect\citeauthoryear{{G{\"u}rkan} et~al.,}{{G{\"u}rkan}
  et~al.}{2015}]{gurkan15}
{G{\"u}rkan} G.,  et~al., 2015, \mn@doi [\mnras] {10.1093/mnras/stv1502}, \href
  {https://ui.adsabs.harvard.edu/abs/2015MNRAS.452.3776G} {452, 3776}

\bibitem[\protect\citeauthoryear{{G{\"u}rkan} et~al.,}{{G{\"u}rkan}
  et~al.}{2018}]{gurkan18}
{G{\"u}rkan} G.,  et~al., 2018, \mn@doi [\mnras] {10.1093/mnras/sty016}, \href
  {https://ui.adsabs.harvard.edu/abs/2018MNRAS.475.3010G} {475, 3010}

\bibitem[\protect\citeauthoryear{{Hardcastle}}{{Hardcastle}}{2018}]{hardcastle18}
{Hardcastle} M.,  2018, \mn@doi [Nature Astronomy] {10.1038/s41550-018-0424-1},
  \href {https://ui.adsabs.harvard.edu/abs/2018NatAs...2..273H} {2, 273}

\bibitem[\protect\citeauthoryear{Hardcastle \& Croston}{Hardcastle \&
  Croston}{2020}]{hardcastle20}
Hardcastle M.,  Croston J.,  2020, \mn@doi [New Astronomy Reviews]
  {https://doi.org/10.1016/j.newar.2020.101539}, 88, 101539

\bibitem[\protect\citeauthoryear{{Hardcastle} et~al.,}{{Hardcastle}
  et~al.}{2019}]{hardcastle19}
{Hardcastle} M.~J.,  et~al., 2019, \mn@doi [\aap]
  {10.1051/0004-6361/201833893}, \href
  {https://ui.adsabs.harvard.edu/abs/2019A&A...622A..12H} {622, A12}

\bibitem[\protect\citeauthoryear{{H{\"a}ring} \& {Rix}}{{H{\"a}ring} \&
  {Rix}}{2004}]{haring04}
{H{\"a}ring} N.,  {Rix} H.-W.,  2004, \mn@doi [\apjl] {10.1086/383567}, \href
  {http://adsabs.harvard.edu/cgi-bin/nph-bib_query?bibcode=2004\apj...604L..89H&db_key=AST}
  {604, L89}

\bibitem[\protect\citeauthoryear{{Heckman}}{{Heckman}}{1980}]{heckman80}
{Heckman} T.~M.,  1980, \aap, \href
  {https://ui.adsabs.harvard.edu/abs/1980A&A....87..152H} {500, 187}

\bibitem[\protect\citeauthoryear{{Heckman} \& {Best}}{{Heckman} \&
  {Best}}{2014}]{heckman14}
{Heckman} T.~M.,  {Best} P.~N.,  2014, \mn@doi [\araa]
  {10.1146/annurev-astro-081913-035722}, \href
  {http://adsabs.harvard.edu/abs/2014ARA%26A..52..589H} {52, 589}

\bibitem[\protect\citeauthoryear{{Heckman}, {Lehnert}  \& {Armus}}{{Heckman}
  et~al.}{1993}]{heckman93}
{Heckman} T.~M.,  {Lehnert} M.~D.,   {Armus} L.,  1993, in {Shull} J.~M.,
  {Thronson} H.~A.,  eds,  Astrophysics and Space Science Library Vol. 188, The
  Environment and Evolution of Galaxies. p.~455,
  \mn@doi{10.1007/978-94-011-1882-8_25}

\bibitem[\protect\citeauthoryear{{Heckman}, {Kauffmann}, {Brinchmann},
  {Charlot}, {Tremonti}  \& {White}}{{Heckman} et~al.}{2004}]{heckman04}
{Heckman} T.~M.,  {Kauffmann} G.,  {Brinchmann} J.,  {Charlot} S.,  {Tremonti}
  C.,   {White} S.~D.~M.,  2004, \apj, \href
  {http://adsabs.harvard.edu/cgi-bin/nph-bib_query?bibcode=2004\apj...613..109H&db_key=AST}
  {613, 109}

\bibitem[\protect\citeauthoryear{{Heckman}, {Alexandroff}, {Borthakur},
  {Overzier}  \& {Leitherer}}{{Heckman} et~al.}{2015}]{heckman15}
{Heckman} T.~M.,  {Alexandroff} R.~M.,  {Borthakur} S.,  {Overzier} R.,
  {Leitherer} C.,  2015, \mn@doi [\apj] {10.1088/0004-637X/809/2/147}, \href
  {https://ui.adsabs.harvard.edu/abs/2015ApJ...809..147H} {809, 147}

\bibitem[\protect\citeauthoryear{{Heinz} \& {Sunyaev}}{{Heinz} \&
  {Sunyaev}}{2003}]{heinz03}
{Heinz} S.,  {Sunyaev} R.~A.,  2003, \mn@doi [\mnras]
  {10.1046/j.1365-8711.2003.06918.x}, \href
  {http://adsabs.harvard.edu/abs/2003MNRAS.343L..59H} {343, L59}

\bibitem[\protect\citeauthoryear{{Heinz}, {Merloni}  \& {Schwab}}{{Heinz}
  et~al.}{2007}]{heinz07}
{Heinz} S.,  {Merloni} A.,   {Schwab} J.,  2007, \mn@doi [\apjl]
  {10.1086/513507}, \href
  {https://ui.adsabs.harvard.edu/abs/2007ApJ...658L...9H} {658, L9}

\bibitem[\protect\citeauthoryear{{Herpich}, {Stasi{\'n}ska}, {Mateus}, {Vale
  Asari}  \& {Cid Fernandes}}{{Herpich} et~al.}{2018}]{herpich18}
{Herpich} F.,  {Stasi{\'n}ska} G.,  {Mateus} A.,  {Vale Asari} N.,   {Cid
  Fernandes} R.,  2018, \mn@doi [\mnras] {10.1093/mnras/sty2391}, \href
  {https://ui.adsabs.harvard.edu/abs/2018MNRAS.481.1774H} {481, 1774}

\bibitem[\protect\citeauthoryear{{Heywood} et~al.,}{{Heywood}
  et~al.}{2019}]{heywood19}
{Heywood} I.,  et~al., 2019, \mn@doi [\nat] {10.1038/s41586-019-1532-5}, \href
  {https://ui.adsabs.harvard.edu/abs/2019Natur.573..235H} {573, 235}

\bibitem[\protect\citeauthoryear{{Ho}}{{Ho}}{1999}]{ho99}
{Ho} L.~C.,  1999, \mn@doi [\apj] {10.1086/307137}, \href
  {http://adsabs.harvard.edu/abs/1999ApJ...516..672H} {516, 672}

\bibitem[\protect\citeauthoryear{{Ho}}{{Ho}}{2002}]{ho02}
{Ho} L.~C.,  2002, \apj, \href
  {http://adsabs.harvard.edu/cgi-bin/nph-bib_query?bibcode=2002\apj...564..120H&db_key=AST}
  {564, 120}

\bibitem[\protect\citeauthoryear{{Ho}}{{Ho}}{2008}]{ho08}
{Ho} L.~C.,  2008, \mn@doi [\araa] {10.1146/annurev.astro.45.051806.110546},
  \href {http://adsabs.harvard.edu/abs/2008ARA%26A..46..475H} {46, 475}

\bibitem[\protect\citeauthoryear{{Ho} \& {Ulvestad}}{{Ho} \&
  {Ulvestad}}{2001}]{ho01a}
{Ho} L.~C.,  {Ulvestad} J.~S.,  2001, \mn@doi [\apjs] {10.1086/319185}, \href
  {http://adsabs.harvard.edu/abs/2001ApJS..133...77H} {133, 77}

\bibitem[\protect\citeauthoryear{{Ho}, {Filippenko}  \& {Sargent}}{{Ho}
  et~al.}{1995}]{ho95}
{Ho} L.~C.,  {Filippenko} A.~V.,   {Sargent} W.~L.,  1995, \mn@doi [ApJS]
  {10.1086/192170}, \href {http://adsabs.harvard.edu/abs/1995ApJS...98..477H}
  {98, 477}

\bibitem[\protect\citeauthoryear{{Ho}, {Filippenko}  \& {Sargent}}{{Ho}
  et~al.}{1997a}]{ho97a}
{Ho} L.~C.,  {Filippenko} A.~V.,   {Sargent} W.~L.~W.,  1997a, \mn@doi [\apjs]
  {10.1086/313041}, \href {http://adsabs.harvard.edu/abs/1997ApJS..112..315H}
  {112, 315}

\bibitem[\protect\citeauthoryear{{Ho}, {Filippenko}  \& {Sargent}}{{Ho}
  et~al.}{1997b}]{ho97b}
{Ho} L.~C.,  {Filippenko} A.~V.,   {Sargent} W.~L.~W.,  1997b, \apj, \href
  {http://adsabs.harvard.edu/cgi-bin/nph-bib_query?bibcode=1997\apj...487..568H&amp;db_key=AST}
  {487, 568}

\bibitem[\protect\citeauthoryear{{Ho}, {Greene}, {Filippenko}  \&
  {Sargent}}{{Ho} et~al.}{2009}]{ho09}
{Ho} L.~C.,  {Greene} J.~E.,  {Filippenko} A.~V.,   {Sargent} W.~L.~W.,  2009,
  \mn@doi [\apjs] {10.1088/0067-0049/183/1/1}, \href
  {http://adsabs.harvard.edu/abs/2009ApJS..183....1H} {183, 1}

\bibitem[\protect\citeauthoryear{{Huchra} \& {Burg}}{{Huchra} \&
  {Burg}}{1992}]{huchra92}
{Huchra} J.,  {Burg} R.,  1992, \apj, \href
  {http://adsabs.harvard.edu/cgi-bin/nph-bib_query?bibcode=1992\apj...393...90H&db_key=AST}
  {393, 90}

\bibitem[\protect\citeauthoryear{{Inayoshi}, {Ichikawa}  \& {Ho}}{{Inayoshi}
  et~al.}{2020}]{inayoshi20}
{Inayoshi} K.,  {Ichikawa} K.,   {Ho} L.~C.,  2020, \mn@doi [\apj]
  {10.3847/1538-4357/ab8569}, \href
  {https://ui.adsabs.harvard.edu/abs/2020ApJ...894..141I} {894, 141}

\bibitem[\protect\citeauthoryear{{Ishibashi}, {Auger}, {Zhang}  \&
  {Fabian}}{{Ishibashi} et~al.}{2014}]{ishibashi14}
{Ishibashi} W.,  {Auger} M.~W.,  {Zhang} D.,   {Fabian} A.~C.,  2014, \mn@doi
  [\mnras] {10.1093/mnras/stu1236}, \href
  {https://ui.adsabs.harvard.edu/abs/2014MNRAS.443.1339I} {443, 1339}

\bibitem[\protect\citeauthoryear{{Janssen}, {R{\"o}ttgering}, {Best}  \&
  {Brinchmann}}{{Janssen} et~al.}{2012}]{janssen12}
{Janssen} R.~M.~J.,  {R{\"o}ttgering} H.~J.~A.,  {Best} P.~N.,   {Brinchmann}
  J.,  2012, \mn@doi [\aap] {10.1051/0004-6361/201219052}, \href
  {https://ui.adsabs.harvard.edu/abs/2012A&A...541A..62J} {541, A62}

\bibitem[\protect\citeauthoryear{{Jarvis} et~al.,}{{Jarvis}
  et~al.}{2019}]{jarvis19}
{Jarvis} M.~E.,  et~al., 2019, \mn@doi [\mnras] {10.1093/mnras/stz556}, \href
  {https://ui.adsabs.harvard.edu/abs/2019MNRAS.485.2710J} {485, 2710}

\bibitem[\protect\citeauthoryear{{Jarvis} et~al.,}{{Jarvis}
  et~al.}{2021}]{jarvis21}
{Jarvis} M.~E.,  et~al., 2021, \mn@doi [\mnras] {10.1093/mnras/stab549}, \href
  {https://ui.adsabs.harvard.edu/abs/2021MNRAS.503.1780J} {503, 1780}

\bibitem[\protect\citeauthoryear{{Jiang}, {Fan}, {Ivezi{\'c}}, {Richards},
  {Schneider}, {Strauss}  \& {Kelly}}{{Jiang} et~al.}{2007}]{jiang07}
{Jiang} L.,  {Fan} X.,  {Ivezi{\'c}} {\v{Z}}.,  {Richards} G.~T.,  {Schneider}
  D.~P.,  {Strauss} M.~A.,   {Kelly} B.~C.,  2007, \mn@doi [\apj]
  {10.1086/510831}, \href
  {https://ui.adsabs.harvard.edu/abs/2007ApJ...656..680J} {656, 680}

\bibitem[\protect\citeauthoryear{{Kauffmann} et~al.,}{{Kauffmann}
  et~al.}{2003}]{kauffmann03c}
{Kauffmann} G.,  et~al., 2003, \mn@doi [\mnras]
  {10.1111/j.1365-2966.2003.07154.x}, \href
  {http://adsabs.harvard.edu/abs/2003MNRAS.346.1055K} {346, 1055}

\bibitem[\protect\citeauthoryear{{Kauffmann}, {Heckman}  \& {Best}}{{Kauffmann}
  et~al.}{2008}]{kauffmann08}
{Kauffmann} G.,  {Heckman} T.~M.,   {Best} P.~N.,  2008, \mn@doi [\mnras]
  {10.1111/j.1365-2966.2007.12752.x}, \href
  {http://adsabs.harvard.edu/abs/2008MNRAS.384..953K} {384, 953}

\bibitem[\protect\citeauthoryear{{Kellermann}, {Sramek}, {Schmidt}, {Shaffer}
  \& {Green}}{{Kellermann} et~al.}{1989}]{kellerman89}
{Kellermann} K.~I.,  {Sramek} R.,  {Schmidt} M.,  {Shaffer} D.~B.,   {Green}
  R.,  1989, \mn@doi [\aj] {10.1086/115207}, \href
  {http://adsabs.harvard.edu/abs/1989AJ.....98.1195K} {98, 1195}

\bibitem[\protect\citeauthoryear{{Kelly} \& {Merloni}}{{Kelly} \&
  {Merloni}}{2012}]{kelly12}
{Kelly} B.~C.,  {Merloni} A.,  2012, \mn@doi [Advances in Astronomy]
  {10.1155/2012/970858}, \href
  {https://ui.adsabs.harvard.edu/abs/2012AdAst2012E...7K} {2012, 970858}

\bibitem[\protect\citeauthoryear{{Kendall}}{{Kendall}}{1938}]{kendall38}
{Kendall} M.,  1938, Biometrika, 30, 81

\bibitem[\protect\citeauthoryear{{Kennicutt}}{{Kennicutt}}{1998}]{kennicutt98}
{Kennicutt} Jr. R.~C.,  1998, \mn@doi [\araa] {10.1146/annurev.astro.36.1.189},
  \href {http://adsabs.harvard.edu/abs/1998ARA%26A..36..189K} {36, 189}

\bibitem[\protect\citeauthoryear{{Kewley}, {Groves}, {Kauffmann}  \&
  {Heckman}}{{Kewley} et~al.}{2006}]{kewley06}
{Kewley} L.~J.,  {Groves} B.,  {Kauffmann} G.,   {Heckman} T.,  2006, \mn@doi
  [MNRAS] {10.1111/j.1365-2966.2006.10859.x}, \href
  {http://adsabs.harvard.edu/abs/2006MNRAS.372..961K} {372, 961}

\bibitem[\protect\citeauthoryear{{Kewley}, {Nicholls}  \&
  {Sutherland}}{{Kewley} et~al.}{2019}]{kewley19}
{Kewley} L.~J.,  {Nicholls} D.~C.,   {Sutherland} R.~S.,  2019, \mn@doi [\araa]
  {10.1146/annurev-astro-081817-051832}, \href
  {https://ui.adsabs.harvard.edu/abs/2019ARA&A..57..511K} {57, 511}

\bibitem[\protect\citeauthoryear{{Kharb}, {O'Dea}, {Baum}, {Colbert}  \&
  {Xu}}{{Kharb} et~al.}{2006}]{kharb06}
{Kharb} P.,  {O'Dea} C.~P.,  {Baum} S.~A.,  {Colbert} E.~J.~M.,   {Xu} C.,
  2006, \mn@doi [\apj] {10.1086/507945}, \href
  {http://adsabs.harvard.edu/abs/2006ApJ...652..177K} {652, 177}

\bibitem[\protect\citeauthoryear{{Kharb} et~al.,}{{Kharb}
  et~al.}{2012}]{kharb12}
{Kharb} P.,  et~al., 2012, \mn@doi [\aj] {10.1088/0004-6256/143/4/78}, \href
  {http://adsabs.harvard.edu/abs/2012AJ....143...78K} {143, 78}

\bibitem[\protect\citeauthoryear{{Kharb}, {O'Dea}, {Baum}, {Hardcastle},
  {Dicken}, {Croston}, {Mingo}  \& {Noel-Storr}}{{Kharb}
  et~al.}{2014}]{kharb14}
{Kharb} P.,  {O'Dea} C.~P.,  {Baum} S.~A.,  {Hardcastle} M.~J.,  {Dicken} D.,
  {Croston} J.~H.,  {Mingo} B.,   {Noel-Storr} J.,  2014, \mn@doi [\mnras]
  {10.1093/mnras/stu421}, \href
  {http://adsabs.harvard.edu/abs/2014MNRAS.440.2976K} {440, 2976}

\bibitem[\protect\citeauthoryear{{Kharb}, {Subramanian}, {Vaddi}, {Das}  \&
  {Paragi}}{{Kharb} et~al.}{2017}]{kharb17}
{Kharb} P.,  {Subramanian} S.,  {Vaddi} S.,  {Das} M.,   {Paragi} Z.,  2017,
  \mn@doi [\apj] {10.3847/1538-4357/aa8321}, \href
  {http://adsabs.harvard.edu/abs/2017ApJ...846...12K} {846, 12}

\bibitem[\protect\citeauthoryear{{King} et~al.,}{{King} et~al.}{2011}]{king11}
{King} A.~L.,  et~al., 2011, \mn@doi [\apj] {10.1088/0004-637X/729/1/19}, \href
  {http://adsabs.harvard.edu/abs/2011ApJ...729...19K} {729, 19}

\bibitem[\protect\citeauthoryear{{Klindt}, {Alexander}, {Rosario}, {Lusso}  \&
  {Fotopoulou}}{{Klindt} et~al.}{2019}]{klindt19}
{Klindt} L.,  {Alexander} D.~M.,  {Rosario} D.~J.,  {Lusso} E.,   {Fotopoulou}
  S.,  2019, \mn@doi [\mnras] {10.1093/mnras/stz1771}, \href
  {https://ui.adsabs.harvard.edu/abs/2019MNRAS.488.3109K} {488, 3109}

\bibitem[\protect\citeauthoryear{{K{\"o}rding}, {Fender}  \&
  {Migliari}}{{K{\"o}rding} et~al.}{2006}]{koerding06}
{K{\"o}rding} E.~G.,  {Fender} R.~P.,   {Migliari} S.,  2006, \mn@doi [\mnras]
  {10.1111/j.1365-2966.2006.10383.x}, \href
  {https://ui.adsabs.harvard.edu/abs/2006MNRAS.369.1451K} {369, 1451}

\bibitem[\protect\citeauthoryear{{K{\"o}rding}, {Jester}  \&
  {Fender}}{{K{\"o}rding} et~al.}{2008}]{koerding08}
{K{\"o}rding} E.~G.,  {Jester} S.,   {Fender} R.,  2008, \mn@doi [\mnras]
  {10.1111/j.1365-2966.2007.12529.x}, \href
  {http://adsabs.harvard.edu/abs/2008MNRAS.383..277K} {383, 277}

\bibitem[\protect\citeauthoryear{{Kormendy} \& {Ho}}{{Kormendy} \&
  {Ho}}{2001}]{kormendy01}
{Kormendy} J.,  {Ho} L.,  2001, {Supermassive Black Holes in Inactive
  Galaxies}, \mn@doi{10.1888/0333750888/2635.
}

\bibitem[\protect\citeauthoryear{{Kozie{\l}-Wierzbowska}, {Stasi{\'n}ska},
  {Vale Asari}, {Sikora}, {Goettems}  \&
  {W{\'o}jtowicz}}{{Kozie{\l}-Wierzbowska} et~al.}{2017}]{koziel17}
{Kozie{\l}-Wierzbowska} D.,  {Stasi{\'n}ska} G.,  {Vale Asari} N.,  {Sikora}
  M.,  {Goettems} E.,   {W{\'o}jtowicz} A.,  2017, \mn@doi [Frontiers in
  Astronomy and Space Sciences] {10.3389/fspas.2017.00039}, \href
  {https://ui.adsabs.harvard.edu/abs/2017FrASS...4...39K} {4, 39}

\bibitem[\protect\citeauthoryear{{Kukula}, {Ghosh}, {Pedlar}, {Schilizzi},
  {Miley}, {de Bruyn}  \& {Saikia}}{{Kukula} et~al.}{1993}]{kukula93}
{Kukula} M.~J.,  {Ghosh} T.,  {Pedlar} A.,  {Schilizzi} R.~T.,  {Miley} G.~K.,
  {de Bruyn} A.~G.,   {Saikia} D.~J.,  1993, \mn@doi [\mnras]
  {10.1093/mnras/264.4.893}, \href
  {http://adsabs.harvard.edu/abs/1993MNRAS.264..893K} {264, 893}

\bibitem[\protect\citeauthoryear{{Kukula}, {Pedlar}, {Baum}  \&
  {O'Dea}}{{Kukula} et~al.}{1995}]{kukula95}
{Kukula} M.~J.,  {Pedlar} A.,  {Baum} S.~A.,   {O'Dea} C.~P.,  1995, \mn@doi
  [\mnras] {10.1093/mnras/276.4.1262}, \href
  {http://adsabs.harvard.edu/abs/1995MNRAS.276.1262K} {276, 1262}

\bibitem[\protect\citeauthoryear{{Kukula}, {Ghosh}, {Pedlar}  \&
  {Schilizzi}}{{Kukula} et~al.}{1999}]{kukula99}
{Kukula} M.~J.,  {Ghosh} T.,  {Pedlar} A.,   {Schilizzi} R.~T.,  1999, \mn@doi
  [\apj] {10.1086/307254}, \href
  {http://adsabs.harvard.edu/abs/1999ApJ...518..117K} {518, 117}

\bibitem[\protect\citeauthoryear{{Lagos}, {Padilla}  \& {Cora}}{{Lagos}
  et~al.}{2009}]{lagos09}
{Lagos} C. D.~P.,  {Padilla} N.~D.,   {Cora} S.~A.,  2009, \mn@doi [\mnras]
  {10.1111/j.1365-2966.2009.14451.x}, \href
  {https://ui.adsabs.harvard.edu/abs/2009MNRAS.395..625L} {395, 625}

\bibitem[\protect\citeauthoryear{{Lamastra}, {Bianchi}, {Matt}, {Perola},
  {Barcons}  \& {Carrera}}{{Lamastra} et~al.}{2009}]{lamastra09}
{Lamastra} A.,  {Bianchi} S.,  {Matt} G.,  {Perola} G.~C.,  {Barcons} X.,
  {Carrera} F.~J.,  2009, \mn@doi [\aap] {10.1051/0004-6361/200912023}, \href
  {https://ui.adsabs.harvard.edu/abs/2009A&A...504...73L} {504, 73}

\bibitem[\protect\citeauthoryear{{Laor}}{{Laor}}{2000}]{laor00}
{Laor} A.,  2000, \mn@doi [\apjl] {10.1086/317280}, \href
  {http://adsabs.harvard.edu/abs/2000ApJ...543L.111L} {543, L111}

\bibitem[\protect\citeauthoryear{{Laor} \& {Behar}}{{Laor} \&
  {Behar}}{2008}]{laor08}
{Laor} A.,  {Behar} E.,  2008, \mn@doi [MNRAS]
  {10.1111/j.1365-2966.2008.13806.x}, \href
  {http://adsabs.harvard.edu/abs/2008MNRAS.390..847L} {390, 847}

\bibitem[\protect\citeauthoryear{{Laor}, {Baldi}  \& {Behar}}{{Laor}
  et~al.}{2019}]{laor19}
{Laor} A.,  {Baldi} R.~D.,   {Behar} E.,  2019, \mn@doi [\mnras]
  {10.1093/mnras/sty3098}, \href
  {https://ui.adsabs.harvard.edu/abs/2019MNRAS.482.5513L} {482, 5513}

\bibitem[\protect\citeauthoryear{{Lapi} et~al.,}{{Lapi} et~al.}{2011}]{lapi11}
{Lapi} A.,  et~al., 2011, \mn@doi [\apj] {10.1088/0004-637X/742/1/24}, \href
  {https://ui.adsabs.harvard.edu/abs/2011ApJ...742...24L} {742, 24}

\bibitem[\protect\citeauthoryear{{Lavalley}, {Isobe}  \&
  {Feigelson}}{{Lavalley} et~al.}{1992}]{lavalley92}
{Lavalley} M.,  {Isobe} T.,   {Feigelson} E.,  1992, {ASURV: Astronomy Survival
  Analysis Package}.
Astronomical Society of the Pacific Conference Series, p.~245

\bibitem[\protect\citeauthoryear{{Liu}, {Chang}, {Han}  \& {Wang}}{{Liu}
  et~al.}{2020}]{liu20}
{Liu} X.,  {Chang} N.,  {Han} Z.,   {Wang} X.,  2020, \mn@doi [Universe]
  {10.3390/universe6050068}, \href
  {https://ui.adsabs.harvard.edu/abs/2020Univ....6...68L} {6, 68}

\bibitem[\protect\citeauthoryear{{Maccarone}}{{Maccarone}}{2003}]{Maccarone_state_trans}
{Maccarone} T.~J.,  2003, \mn@doi [\aap] {10.1051/0004-6361:20031146}, \href
  {https://ui.adsabs.harvard.edu/abs/2003A&A...409..697M} {409, 697}

\bibitem[\protect\citeauthoryear{{Macfarlane} et~al.,}{{Macfarlane}
  et~al.}{2021}]{macfarlane21}
{Macfarlane} C.,  et~al., 2021, \mn@doi [\mnras] {10.1093/mnras/stab1998},
  \href {https://ui.adsabs.harvard.edu/abs/2021MNRAS.506.5888M} {506, 5888}

\bibitem[\protect\citeauthoryear{{Magorrian} et~al.,}{{Magorrian}
  et~al.}{1998}]{magorrian98}
{Magorrian} J.,  et~al., 1998, \mn@doi [\aj] {10.1086/300353}, \href
  {https://ui.adsabs.harvard.edu/abs/1998AJ....115.2285M} {115, 2285}

\bibitem[\protect\citeauthoryear{{Mancuso} et~al.,}{{Mancuso}
  et~al.}{2017}]{mancuso17}
{Mancuso} C.,  et~al., 2017, \mn@doi [\apj] {10.3847/1538-4357/aa745d}, \href
  {https://ui.adsabs.harvard.edu/abs/2017ApJ...842...95M} {842, 95}

\bibitem[\protect\citeauthoryear{{Maoz}}{{Maoz}}{2007}]{maoz07}
{Maoz} D.,  2007, \mn@doi [\mnras] {10.1111/j.1365-2966.2007.11735.x}, \href
  {https://ui.adsabs.harvard.edu/abs/2007MNRAS.377.1696M} {377, 1696}

\bibitem[\protect\citeauthoryear{{Marconi}, {Risaliti}, {Gilli}, {Hunt},
  {Maiolino}  \& {Salvati}}{{Marconi} et~al.}{2004}]{marconi04}
{Marconi} A.,  {Risaliti} G.,  {Gilli} R.,  {Hunt} L.~K.,  {Maiolino} R.,
  {Salvati} M.,  2004, \mn@doi [\mnras] {10.1111/j.1365-2966.2004.07765.x},
  \href {https://ui.adsabs.harvard.edu/abs/2004MNRAS.351..169M} {351, 169}

\bibitem[\protect\citeauthoryear{{Markoff}, {Nowak}, {Corbel}, {Fender}  \&
  {Falcke}}{{Markoff} et~al.}{2003}]{markoff03}
{Markoff} S.,  {Nowak} M.,  {Corbel} S.,  {Fender} R.,   {Falcke} H.,  2003,
  \mn@doi [\nar] {10.1016/S1387-6473(03)00078-2}, \href
  {https://ui.adsabs.harvard.edu/abs/2003NewAR..47..491M} {47, 491}

\bibitem[\protect\citeauthoryear{{Markoff}, {Nowak}  \& {Wilms}}{{Markoff}
  et~al.}{2005}]{markoff05}
{Markoff} S.,  {Nowak} M.~A.,   {Wilms} J.,  2005, \mn@doi [\apj]
  {10.1086/497628}, \href {http://adsabs.harvard.edu/abs/2005ApJ...635.1203M}
  {635, 1203}

\bibitem[\protect\citeauthoryear{{Markowitz} et~al.,}{{Markowitz}
  et~al.}{2003}]{Markowitz2003}
{Markowitz} A.,  et~al., 2003, \mn@doi [\apj] {10.1086/375330}, \href
  {https://ui.adsabs.harvard.edu/abs/2003ApJ...593...96M} {593, 96}

\bibitem[\protect\citeauthoryear{{Mart{\'\i}n-Navarro} \&
  {Mezcua}}{{Mart{\'\i}n-Navarro} \& {Mezcua}}{2018}]{martin18}
{Mart{\'\i}n-Navarro} I.,  {Mezcua} M.,  2018, \mn@doi [\apjl]
  {10.3847/2041-8213/aab103}, \href
  {https://ui.adsabs.harvard.edu/abs/2018ApJ...855L..20M} {855, L20}

\bibitem[\protect\citeauthoryear{{Mascoop}, {Anderson}, {Wenger}, {Makai},
  {Armentrout}, {Balser}  \& {Bania}}{{Mascoop} et~al.}{2021}]{mascoop21}
{Mascoop} J.~L.,  {Anderson} L.~D.,  {Wenger} T.~V.,  {Makai} Z.,  {Armentrout}
  W.~P.,  {Balser} D.~S.,   {Bania} T.~M.,  2021, \mn@doi [\apj]
  {10.3847/1538-4357/abe532}, \href
  {https://ui.adsabs.harvard.edu/abs/2021ApJ...910..159M} {910, 159}

\bibitem[\protect\citeauthoryear{{Massaglia}, {Bodo}, {Rossi}, {Capetti}  \&
  {Mignone}}{{Massaglia} et~al.}{2016}]{massaglia16}
{Massaglia} S.,  {Bodo} G.,  {Rossi} P.,  {Capetti} S.,   {Mignone} A.,  2016,
  \mn@doi [\aap] {10.1051/0004-6361/201629375}, \href
  {http://adsabs.harvard.edu/abs/2016A%26A...596A..12M} {596, A12}

\bibitem[\protect\citeauthoryear{{Mattila} et~al.,}{{Mattila}
  et~al.}{2018}]{mattila18}
{Mattila} S.,  et~al., 2018, \mn@doi [Science] {10.1126/science.aao4669}, \href
  {https://ui.adsabs.harvard.edu/abs/2018Sci...361..482M} {361, 482}

\bibitem[\protect\citeauthoryear{{Mauch} \& {Sadler}}{{Mauch} \&
  {Sadler}}{2007}]{mauch07}
{Mauch} T.,  {Sadler} E.~M.,  2007, \mn@doi [MNRAS]
  {10.1111/j.1365-2966.2006.11353.x}, \href
  {http://adsabs.harvard.edu/abs/2007MNRAS.375..931M} {375, 931}

\bibitem[\protect\citeauthoryear{{McKinney}, {Tchekhovskoy}  \&
  {Blandford}}{{McKinney} et~al.}{2012}]{mckinney12}
{McKinney} J.~C.,  {Tchekhovskoy} A.,   {Blandford} R.~D.,  2012, \mn@doi
  [\mnras] {10.1111/j.1365-2966.2012.21074.x}, \href
  {http://adsabs.harvard.edu/abs/2012MNRAS.423.3083M} {423, 3083}

\bibitem[\protect\citeauthoryear{{Meert}, {Vikram}  \& {Bernardi}}{{Meert}
  et~al.}{2015}]{meert15}
{Meert} A.,  {Vikram} V.,   {Bernardi} M.,  2015, \mn@doi [\mnras]
  {10.1093/mnras/stu2333}, \href
  {https://ui.adsabs.harvard.edu/abs/2015MNRAS.446.3943M} {446, 3943}

\bibitem[\protect\citeauthoryear{{Meier}}{{Meier}}{2001}]{meier01}
{Meier} D.~L.,  2001, \mn@doi [\apjl] {10.1086/318921}, \href
  {http://adsabs.harvard.edu/abs/2001ApJ...548L...9M} {548, L9}

\bibitem[\protect\citeauthoryear{{Meier}}{{Meier}}{2012}]{meier12}
{Meier} D.~L.,  2012, {Black Hole Astrophysics: The Engine Paradigm}.
Springer-Verlag Berlin Heidelberg

\bibitem[\protect\citeauthoryear{{Mendel}, {Simard}, {Palmer}, {Ellison}  \&
  {Patton}}{{Mendel} et~al.}{2014}]{mendel14}
{Mendel} J.~T.,  {Simard} L.,  {Palmer} M.,  {Ellison} S.~L.,   {Patton} D.~R.,
   2014, \mn@doi [\apjs] {10.1088/0067-0049/210/1/3}, \href
  {https://ui.adsabs.harvard.edu/abs/2014ApJS..210....3M} {210, 3}

\bibitem[\protect\citeauthoryear{{Merloni} \& {Heinz}}{{Merloni} \&
  {Heinz}}{2007}]{merloni07}
{Merloni} A.,  {Heinz} S.,  2007, \mn@doi [\mnras]
  {10.1111/j.1365-2966.2007.12253.x}, \href
  {https://ui.adsabs.harvard.edu/abs/2007MNRAS.381..589M} {381, 589}

\bibitem[\protect\citeauthoryear{{Merloni}, {Heinz}  \& {di Matteo}}{{Merloni}
  et~al.}{2003}]{merloni03}
{Merloni} A.,  {Heinz} S.,   {di Matteo} T.,  2003, \mn@doi [\mnras]
  {10.1046/j.1365-2966.2003.07017.x}, \href
  {https://ui.adsabs.harvard.edu/abs/2003MNRAS.345.1057M} {345, 1057}

\bibitem[\protect\citeauthoryear{{Mezcua}}{{Mezcua}}{2017}]{mezcua17}
{Mezcua} M.,  2017, \mn@doi [International Journal of Modern Physics D]
  {10.1142/S021827181730021X}, \href
  {https://ui.adsabs.harvard.edu/abs/2017IJMPD..2630021M} {26, 1730021}

\bibitem[\protect\citeauthoryear{{Mezcua} \& {Prieto}}{{Mezcua} \&
  {Prieto}}{2014}]{mezcua14}
{Mezcua} M.,  {Prieto} M.~A.,  2014, \mn@doi [\apj]
  {10.1088/0004-637X/787/1/62}, \href
  {https://ui.adsabs.harvard.edu/abs/2014ApJ...787...62M} {787, 62}

\bibitem[\protect\citeauthoryear{{Middelberg} et~al.,}{{Middelberg}
  et~al.}{2004}]{middleberg04}
{Middelberg} E.,  et~al., 2004, \mn@doi [\aap] {10.1051/0004-6361:20040019},
  \href {http://adsabs.harvard.edu/abs/2004A%26A...417..925M} {417, 925}

\bibitem[\protect\citeauthoryear{{Mingo} et~al.,}{{Mingo}
  et~al.}{2019}]{mingo19}
{Mingo} B.,  et~al., 2019, \mn@doi [\mnras] {10.1093/mnras/stz1901}, \href
  {https://ui.adsabs.harvard.edu/abs/2019MNRAS.488.2701M} {488, 2701}

\bibitem[\protect\citeauthoryear{{Moderski}, {Sikora}  \& {Lasota}}{{Moderski}
  et~al.}{1998}]{moderski08}
{Moderski} R.,  {Sikora} M.,   {Lasota} J.~P.,  1998, \mn@doi [\mnras]
  {10.1046/j.1365-8711.1998.02009.x}, \href
  {https://ui.adsabs.harvard.edu/abs/1998MNRAS.301..142M} {301, 142}

\bibitem[\protect\citeauthoryear{{Morganti}}{{Morganti}}{2017}]{morganti17b}
{Morganti} R.,  2017, \mn@doi [Frontiers in Astronomy and Space Sciences]
  {10.3389/fspas.2017.00042}, \href
  {https://ui.adsabs.harvard.edu/abs/2017FrASS...4...42M} {4, 42}

\bibitem[\protect\citeauthoryear{{Morganti}, {Fanti}, {Fanti}, {Parma}  \& {de
  Ruiter}}{{Morganti} et~al.}{1987}]{morganti87}
{Morganti} R.,  {Fanti} C.,  {Fanti} R.,  {Parma} P.,   {de Ruiter} H.~R.,
  1987, \aap, \href {http://adsabs.harvard.edu/abs/1987A%26A...183..203M} {183,
  203}

\bibitem[\protect\citeauthoryear{{Morganti}, {Tsvetanov}, {Gallimore}  \&
  {Allen}}{{Morganti} et~al.}{1999}]{morganti99}
{Morganti} R.,  {Tsvetanov} Z.~I.,  {Gallimore} J.,   {Allen} M.~G.,  1999,
  \mn@doi [\aaps] {10.1051/aas:1999258}, \href
  {http://adsabs.harvard.edu/abs/1999A%26AS..137..457M} {137, 457}

\bibitem[\protect\citeauthoryear{{Moustakas}, {Kennicutt}  \&
  {Tremonti}}{{Moustakas} et~al.}{2006}]{moustakas06}
{Moustakas} J.,  {Kennicutt} R.~C. J.,   {Tremonti} C.~A.,  2006, \mn@doi
  [\apj] {10.1086/500964}, \href
  {https://ui.adsabs.harvard.edu/abs/2006ApJ...642..775M} {642, 775}

\bibitem[\protect\citeauthoryear{{Mukherjee}, {Bicknell}, {Wagner},
  {Sutherland}  \& {Silk}}{{Mukherjee} et~al.}{2018}]{mukherjee18}
{Mukherjee} D.,  {Bicknell} G.~V.,  {Wagner} A.~Y.,  {Sutherland} R.~S.,
  {Silk} J.,  2018, \mn@doi [\mnras] {10.1093/mnras/sty1776}, \href
  {https://ui.adsabs.harvard.edu/abs/2018MNRAS.479.5544M} {479, 5544}

\bibitem[\protect\citeauthoryear{{Mukherjee}, {Bodo}, {Mignone}, {Rossi}  \&
  {Vaidya}}{{Mukherjee} et~al.}{2020}]{mukerjee20}
{Mukherjee} D.,  {Bodo} G.,  {Mignone} A.,  {Rossi} P.,   {Vaidya} B.,  2020,
  \mn@doi [\mnras] {10.1093/mnras/staa2934}, \href
  {https://ui.adsabs.harvard.edu/abs/2020MNRAS.499..681M} {499, 681}

\bibitem[\protect\citeauthoryear{{Mundell}, {Wilson}, {Ulvestad}  \&
  {Roy}}{{Mundell} et~al.}{2000}]{mundell00}
{Mundell} C.~G.,  {Wilson} A.~S.,  {Ulvestad} J.~S.,   {Roy} A.~L.,  2000,
  \mn@doi [\apj] {10.1086/308318}, \href
  {https://ui.adsabs.harvard.edu/abs/2000ApJ...529..816M} {529, 816}

\bibitem[\protect\citeauthoryear{{Mundell}, {Ferruit}, {Nagar}  \&
  {Wilson}}{{Mundell} et~al.}{2009}]{mundell09}
{Mundell} C.~G.,  {Ferruit} P.,  {Nagar} N.,   {Wilson} A.~S.,  2009, \mn@doi
  [\apj] {10.1088/0004-637X/703/1/802}, \href
  {https://ui.adsabs.harvard.edu/abs/2009ApJ...703..802M} {703, 802}

\bibitem[\protect\citeauthoryear{{Muxlow} et~al.,}{{Muxlow}
  et~al.}{2020}]{muxlow20}
{Muxlow} T.~W.~B.,  et~al., 2020, \mn@doi [\mnras] {10.1093/mnras/staa1279},
  \href {https://ui.adsabs.harvard.edu/abs/2020MNRAS.495.1188M} {495, 1188}

\bibitem[\protect\citeauthoryear{{Nagar}, {Falcke}, {Wilson}  \& {Ho}}{{Nagar}
  et~al.}{2000}]{nagar00}
{Nagar} N.~M.,  {Falcke} H.,  {Wilson} A.~S.,   {Ho} L.~C.,  2000, \mn@doi
  [\apj] {10.1086/309524}, \href
  {http://adsabs.harvard.edu/cgi-bin/nph-bib_query?bibcode=2000\apj...542..186N&db_key=AST}
  {542, 186}

\bibitem[\protect\citeauthoryear{{Nagar}, {Falcke}, {Wilson}  \&
  {Ulvestad}}{{Nagar} et~al.}{2002}]{nagar02}
{Nagar} N.~M.,  {Falcke} H.,  {Wilson} A.~S.,   {Ulvestad} J.~S.,  2002,
  \mn@doi [\aap] {10.1051/0004-6361:20020874}, \href
  {https://ui.adsabs.harvard.edu/abs/2002A&A...392...53N} {392, 53}

\bibitem[\protect\citeauthoryear{{Nagar}, {Falcke}  \& {Wilson}}{{Nagar}
  et~al.}{2005}]{nagar05}
{Nagar} N.~M.,  {Falcke} H.,   {Wilson} A.~S.,  2005, \mn@doi [\aap]
  {10.1051/0004-6361:20042277}, \href
  {http://adsabs.harvard.edu/abs/2005A%26A...435..521N} {435, 521}

\bibitem[\protect\citeauthoryear{{Narayan}}{{Narayan}}{2005}]{narayan05}
{Narayan} R.,  2005, \mn@doi [\apss] {10.1007/s10509-005-1178-7}, \href
  {https://ui.adsabs.harvard.edu/abs/2005Ap&SS.300..177N} {300, 177}

\bibitem[\protect\citeauthoryear{{Narayan} \& {McClintock}}{{Narayan} \&
  {McClintock}}{2008}]{narayan08}
{Narayan} R.,  {McClintock} J.~E.,  2008, \mn@doi [\nar]
  {10.1016/j.newar.2008.03.002}, \href
  {http://adsabs.harvard.edu/abs/2008NewAR..51..733N} {51, 733}

\bibitem[\protect\citeauthoryear{{Narayan} \& {Yi}}{{Narayan} \&
  {Yi}}{1994}]{narayan94}
{Narayan} R.,  {Yi} I.,  1994, \mn@doi [\apjl] {10.1086/187381}, \href
  {https://ui.adsabs.harvard.edu/abs/1994ApJ...428L..13N} {428, L13}

\bibitem[\protect\citeauthoryear{{Narayan} \& {Yi}}{{Narayan} \&
  {Yi}}{1995}]{narayan95}
{Narayan} R.,  {Yi} I.,  1995, \apj, \href
  {http://adsabs.harvard.edu/cgi-bin/nph-bib_query?bibcode=1995\apj...444..231N&amp;db_key=AST}
  {444, 231}

\bibitem[\protect\citeauthoryear{{Nelson}}{{Nelson}}{2000}]{nelson00}
{Nelson} C.~H.,  2000, \mn@doi [\apjl] {10.1086/317314}, \href
  {https://ui.adsabs.harvard.edu/abs/2000ApJ...544L..91N} {544, L91}

\bibitem[\protect\citeauthoryear{{Nemmen}, {Bower}, {Babul}  \&
  {Storchi-Bergmann}}{{Nemmen} et~al.}{2007}]{nemmen07}
{Nemmen} R.~S.,  {Bower} R.~G.,  {Babul} A.,   {Storchi-Bergmann} T.,  2007,
  \mn@doi [\mnras] {10.1111/j.1365-2966.2007.11726.x}, \href
  {http://adsabs.harvard.edu/abs/2007MNRAS.377.1652N} {377, 1652}

\bibitem[\protect\citeauthoryear{{Nemmen}, {Storchi-Bergmann}  \&
  {Eracleous}}{{Nemmen} et~al.}{2014}]{nemmen14}
{Nemmen} R.~S.,  {Storchi-Bergmann} T.,   {Eracleous} M.,  2014, \mn@doi
  [\mnras] {10.1093/mnras/stt2388}, \href
  {http://adsabs.harvard.edu/abs/2014MNRAS.438.2804N} {438, 2804}

\bibitem[\protect\citeauthoryear{{Netzer}}{{Netzer}}{2009}]{netzer09}
{Netzer} H.,  2009, \mn@doi [\mnras] {10.1111/j.1365-2966.2009.15434.x}, \href
  {https://ui.adsabs.harvard.edu/abs/2009MNRAS.399.1907N} {399, 1907}

\bibitem[\protect\citeauthoryear{{Neumayer}, {Seth}  \& {B{\"o}ker}}{{Neumayer}
  et~al.}{2020}]{neumayer20}
{Neumayer} N.,  {Seth} A.,   {B{\"o}ker} T.,  2020, \mn@doi [\aapr]
  {10.1007/s00159-020-00125-0}, \href
  {https://ui.adsabs.harvard.edu/abs/2020A&ARv..28....4N} {28, 4}

\bibitem[\protect\citeauthoryear{{Nyland} et~al.,}{{Nyland}
  et~al.}{2020}]{nyland20}
{Nyland} K.,  et~al., 2020, \mn@doi [\apj] {10.3847/1538-4357/abc341}, \href
  {https://ui.adsabs.harvard.edu/abs/2020ApJ...905...74N} {905, 74}

\bibitem[\protect\citeauthoryear{{Padovani}}{{Padovani}}{2016}]{padovani16}
{Padovani} P.,  2016, \mn@doi [\aapr] {10.1007/s00159-016-0098-6}, \href
  {https://ui.adsabs.harvard.edu/abs/2016A&ARv..24...13P} {24, 13}

\bibitem[\protect\citeauthoryear{{Padovani}, {Bonzini}, {Kellermann}, {Miller},
  {Mainieri}  \& {Tozzi}}{{Padovani} et~al.}{2015}]{padovani15}
{Padovani} P.,  {Bonzini} M.,  {Kellermann} K.~I.,  {Miller} N.,  {Mainieri}
  V.,   {Tozzi} P.,  2015, \mn@doi [\mnras] {10.1093/mnras/stv1375}, \href
  {https://ui.adsabs.harvard.edu/abs/2015MNRAS.452.1263P} {452, 1263}

\bibitem[\protect\citeauthoryear{{Panessa}, {Bassani}, {Cappi}, {Dadina},
  {Barcons}, {Carrera}, {Ho}  \& {Iwasawa}}{{Panessa} et~al.}{2006}]{panessa06}
{Panessa} F.,  {Bassani} L.,  {Cappi} M.,  {Dadina} M.,  {Barcons} X.,
  {Carrera} F.~J.,  {Ho} L.~C.,   {Iwasawa} K.,  2006, \mn@doi [\aap]
  {10.1051/0004-6361:20064894}, \href
  {http://adsabs.harvard.edu/abs/2006A%26A...455..173P} {455, 173}

\bibitem[\protect\citeauthoryear{{Panessa}, {Barcons}, {Bassani}, {Cappi},
  {Carrera}, {Ho}  \& {Pellegrini}}{{Panessa} et~al.}{2007}]{panessa07}
{Panessa} F.,  {Barcons} X.,  {Bassani} L.,  {Cappi} M.,  {Carrera} F.~J.,
  {Ho} L.~C.,   {Pellegrini} S.,  2007, \mn@doi [\aap]
  {10.1051/0004-6361:20066943}, \href
  {https://ui.adsabs.harvard.edu/abs/2007A&A...467..519P} {467, 519}

\bibitem[\protect\citeauthoryear{{Panessa}, {Baldi}, {Laor}, {Padovani},
  {Behar}  \& {McHardy}}{{Panessa} et~al.}{2019}]{panessa19}
{Panessa} F.,  {Baldi} R.~D.,  {Laor} A.,  {Padovani} P.,  {Behar} E.,
  {McHardy} I.,  2019, \mn@doi [Nature Astronomy] {10.1038/s41550-019-0765-4},
  \href {https://ui.adsabs.harvard.edu/abs/2019NatAs...3..387P} {3, 387}

\bibitem[\protect\citeauthoryear{{Parma}, {de Ruiter}, {Fanti}  \&
  {Fanti}}{{Parma} et~al.}{1986}]{parma86}
{Parma} P.,  {de Ruiter} H.~R.,  {Fanti} C.,   {Fanti} R.,  1986, \aaps, \href
  {http://adsabs.harvard.edu/abs/1986A%26AS...64..135P} {64, 135}

\bibitem[\protect\citeauthoryear{{Paudel} \& {Yoon}}{{Paudel} \&
  {Yoon}}{2020}]{paudel20}
{Paudel} S.,  {Yoon} S.-J.,  2020, \mn@doi [\apjl] {10.3847/2041-8213/aba6ed},
  \href {https://ui.adsabs.harvard.edu/abs/2020ApJ...898L..47P} {898, L47}

\bibitem[\protect\citeauthoryear{{Pearson}}{{Pearson}}{1895}]{pearson95}
{Pearson} K.,  1895, Proceedings of the Royal Society of London Series I, \href
  {https://ui.adsabs.harvard.edu/abs/1895RSPS...58..240P} {58, 240}

\bibitem[\protect\citeauthoryear{{Pedlar}, {Unger}  \& {Dyson}}{{Pedlar}
  et~al.}{1985}]{pedlar85}
{Pedlar} A.,  {Unger} S.~W.,   {Dyson} J.~E.,  1985, \mn@doi [\mnras]
  {10.1093/mnras/214.4.463}, \href
  {http://adsabs.harvard.edu/abs/1985MNRAS.214..463P} {214, 463}

\bibitem[\protect\citeauthoryear{{Perlman}, {Biretta}, {Zhou}, {Sparks}  \&
  {Macchetto}}{{Perlman} et~al.}{1999}]{perlman99}
{Perlman} E.~S.,  {Biretta} J.~A.,  {Zhou} F.,  {Sparks} W.~B.,   {Macchetto}
  F.~D.,  1999, \mn@doi [\aj] {10.1086/300844}, \href
  {https://ui.adsabs.harvard.edu/abs/1999AJ....117.2185P} {117, 2185}

\bibitem[\protect\citeauthoryear{{Pitchford} et~al.,}{{Pitchford}
  et~al.}{2016}]{pitchford16}
{Pitchford} L.~K.,  et~al., 2016, \mn@doi [\mnras] {10.1093/mnras/stw1840},
  \href {https://ui.adsabs.harvard.edu/abs/2016MNRAS.462.4067P} {462, 4067}

\bibitem[\protect\citeauthoryear{{Plotkin}, {Markoff}, {Kelly}, {K{\"o}rding}
  \& {Anderson}}{{Plotkin} et~al.}{2012}]{plotkin12}
{Plotkin} R.~M.,  {Markoff} S.,  {Kelly} B.~C.,  {K{\"o}rding} E.,   {Anderson}
  S.~F.,  2012, \mn@doi [\mnras] {10.1111/j.1365-2966.2011.19689.x}, \href
  {http://adsabs.harvard.edu/abs/2012MNRAS.419..267P} {419, 267}

\bibitem[\protect\citeauthoryear{{Radcliffe} et~al.,}{{Radcliffe}
  et~al.}{2018}]{radcliffe18}
{Radcliffe} J.~F.,  et~al., 2018, \mn@doi [\aap] {10.1051/0004-6361/201833399},
  \href {https://ui.adsabs.harvard.edu/abs/2018A&A...619A..48R} {619, A48}

\bibitem[\protect\citeauthoryear{{Raginski} \& {Laor}}{{Raginski} \&
  {Laor}}{2016}]{raginski16}
{Raginski} I.,  {Laor} A.,  2016, \mn@doi [\mnras] {10.1093/mnras/stw772},
  \href {https://ui.adsabs.harvard.edu/abs/2016MNRAS.459.2082R} {459, 2082}

\bibitem[\protect\citeauthoryear{{Reines} \& {Volonteri}}{{Reines} \&
  {Volonteri}}{2015}]{reines15}
{Reines} A.~E.,  {Volonteri} M.,  2015, \mn@doi [\apj]
  {10.1088/0004-637X/813/2/82}, \href
  {https://ui.adsabs.harvard.edu/abs/2015ApJ...813...82R} {813, 82}

\bibitem[\protect\citeauthoryear{{Remillard} \& {McClintock}}{{Remillard} \&
  {McClintock}}{2006}]{remillard06}
{Remillard} R.~A.,  {McClintock} J.~E.,  2006, \mn@doi [\araa]
  {10.1146/annurev.astro.44.051905.092532}, \href
  {https://ui.adsabs.harvard.edu/abs/2006ARA&A..44...49R} {44, 49}

\bibitem[\protect\citeauthoryear{{Reynolds} \& {Begelman}}{{Reynolds} \&
  {Begelman}}{1997}]{reynolds97a}
{Reynolds} C.~S.,  {Begelman} M.~C.,  1997, \mn@doi [\apjl] {10.1086/310894},
  \href {https://ui.adsabs.harvard.edu/abs/1997ApJ...487L.135R} {487, L135}

\bibitem[\protect\citeauthoryear{{Risaliti}, {Salvati}  \&
  {Marconi}}{{Risaliti} et~al.}{2011}]{risaliti11}
{Risaliti} G.,  {Salvati} M.,   {Marconi} A.,  2011, \mn@doi [\mnras]
  {10.1111/j.1365-2966.2010.17843.x}, \href
  {http://adsabs.harvard.edu/abs/2011MNRAS.411.2223R} {411, 2223}

\bibitem[\protect\citeauthoryear{{Rosario}, {Fawcett}, {Klindt}, {Alexander},
  {Morabito}, {Fotopoulou}, {Lusso}  \& {Calistro Rivera}}{{Rosario}
  et~al.}{2020}]{rosario20}
{Rosario} D.~J.,  {Fawcett} V.~A.,  {Klindt} L.,  {Alexander} D.~M.,
  {Morabito} L.~K.,  {Fotopoulou} S.,  {Lusso} E.,   {Calistro Rivera} G.,
  2020, \mn@doi [\mnras] {10.1093/mnras/staa866}, \href
  {https://ui.adsabs.harvard.edu/abs/2020MNRAS.494.3061R} {494, 3061}

\bibitem[\protect\citeauthoryear{{Rossi}, {Bodo}, {Massaglia}  \&
  {Capetti}}{{Rossi} et~al.}{2020}]{rossi20}
{Rossi} P.,  {Bodo} G.,  {Massaglia} S.,   {Capetti} A.,  2020, \mn@doi [\aap]
  {10.1051/0004-6361/202038725}, \href
  {https://ui.adsabs.harvard.edu/abs/2020A&A...642A..69R} {642, A69}

\bibitem[\protect\citeauthoryear{{Roy} et~al.,}{{Roy} et~al.}{2021}]{roy21}
{Roy} N.,  et~al., 2021, arXiv e-prints, \href
  {https://ui.adsabs.harvard.edu/abs/2021arXiv210902609R} {p. arXiv:2109.02609}

\bibitem[\protect\citeauthoryear{{Rushton} et~al.,}{{Rushton}
  et~al.}{2012}]{rushton12}
{Rushton} A.,  et~al., 2012, \mn@doi [\mnras]
  {10.1111/j.1365-2966.2011.19959.x}, \href
  {https://ui.adsabs.harvard.edu/abs/2012MNRAS.419.3194R} {419, 3194}

\bibitem[\protect\citeauthoryear{{Rusinek}, {Sikora}, {Kozie{\l}-Wierzbowska}
  \& {Gupta}}{{Rusinek} et~al.}{2020}]{rusinek20}
{Rusinek} K.,  {Sikora} M.,  {Kozie{\l}-Wierzbowska} D.,   {Gupta} M.,  2020,
  \mn@doi [\apj] {10.3847/1538-4357/aba75f}, \href
  {https://ui.adsabs.harvard.edu/abs/2020ApJ...900..125R} {900, 125}

\bibitem[\protect\citeauthoryear{{Sabater} et~al.,}{{Sabater}
  et~al.}{2019}]{sabater19}
{Sabater} J.,  et~al., 2019, \mn@doi [\aap] {10.1051/0004-6361/201833883},
  \href {https://ui.adsabs.harvard.edu/abs/2019A&A...622A..17S} {622, A17}

\bibitem[\protect\citeauthoryear{{Saikia}, {K{\"o}rding}  \& {Falcke}}{{Saikia}
  et~al.}{2015}]{saikia15}
{Saikia} P.,  {K{\"o}rding} E.,   {Falcke} H.,  2015, \mn@doi [\mnras]
  {10.1093/mnras/stv731}, \href
  {http://adsabs.harvard.edu/abs/2015MNRAS.450.2317S} {450, 2317}

\bibitem[\protect\citeauthoryear{{Saikia}, {K{\"o}rding}  \& {Dibi}}{{Saikia}
  et~al.}{2018a}]{saikia18a}
{Saikia} P.,  {K{\"o}rding} E.,   {Dibi} S.,  2018a, \mn@doi [\mnras]
  {10.1093/mnras/sty754}, \href
  {https://ui.adsabs.harvard.edu/abs/2018MNRAS.477.2119S} {477, 2119}

\bibitem[\protect\citeauthoryear{{Saikia}, {K{\"o}rding}, {Coppejans},
  {Falcke}, {Williams}, {Baldi}, {McHardy}  \& {Beswick}}{{Saikia}
  et~al.}{2018b}]{saikia18}
{Saikia} P.,  {K{\"o}rding} E.,  {Coppejans} D.~L.,  {Falcke} H.,  {Williams}
  D.,  {Baldi} R.~D.,  {McHardy} I.,   {Beswick} R.,  2018b, \mn@doi [\aap]
  {10.1051/0004-6361/201833233}, \href
  {https://ui.adsabs.harvard.edu/abs/2018A&A...616A.152S} {616, A152}

\bibitem[\protect\citeauthoryear{{Salucci}, {Szuszkiewicz}, {Monaco}  \&
  {Danese}}{{Salucci} et~al.}{1999}]{salucci99}
{Salucci} P.,  {Szuszkiewicz} E.,  {Monaco} P.,   {Danese} L.,  1999, \mn@doi
  [\mnras] {10.1046/j.1365-8711.1999.02659.x}, \href
  {https://ui.adsabs.harvard.edu/abs/1999MNRAS.307..637S} {307, 637}

\bibitem[\protect\citeauthoryear{{Sandage} \& {Tammann}}{{Sandage} \&
  {Tammann}}{1981}]{sandage81}
{Sandage} A.,  {Tammann} G.~A.,  1981, in Carnegie Inst. of Washington, Publ.
  635.

\bibitem[\protect\citeauthoryear{{Santini} et~al.,}{{Santini}
  et~al.}{2012}]{santini12}
{Santini} P.,  et~al., 2012, \mn@doi [\aap] {10.1051/0004-6361/201118266},
  \href {https://ui.adsabs.harvard.edu/abs/2012A&A...540A.109S} {540, A109}

\bibitem[\protect\citeauthoryear{{Sarzi} et~al.,}{{Sarzi}
  et~al.}{2010}]{sarzi10}
{Sarzi} M.,  et~al., 2010, \mn@doi [\mnras] {10.1111/j.1365-2966.2009.16039.x},
  \href {http://adsabs.harvard.edu/abs/2010MNRAS.402.2187S} {402, 2187}

\bibitem[\protect\citeauthoryear{{Schmitt}}{{Schmitt}}{1985}]{schmitt85}
{Schmitt} J.~H.~M.~M.,  1985, \mn@doi [\apj] {10.1086/163224}, \href
  {http://cdsads.u-strasbg.fr/cgi-bin/nph-bib_query?bibcode=1985\apj...293..178S&db_key=AST}
  {293, 178}

\bibitem[\protect\citeauthoryear{{Schulze} \& {Wisotzki}}{{Schulze} \&
  {Wisotzki}}{2010}]{schulze10}
{Schulze} A.,  {Wisotzki} L.,  2010, \mn@doi [\aap]
  {10.1051/0004-6361/201014193}, \href
  {https://ui.adsabs.harvard.edu/abs/2010A&A...516A..87S} {516, A87}

\bibitem[\protect\citeauthoryear{{Scott} \& {Graham}}{{Scott} \&
  {Graham}}{2013}]{scott13}
{Scott} N.,  {Graham} A.~W.,  2013, \mn@doi [\apj]
  {10.1088/0004-637X/763/2/76}, \href
  {https://ui.adsabs.harvard.edu/abs/2013ApJ...763...76S} {763, 76}

\bibitem[\protect\citeauthoryear{{Shakura} \& {Sunyaev}}{{Shakura} \&
  {Sunyaev}}{1973}]{shakura73}
{Shakura} N.~I.,  {Sunyaev} R.~A.,  1973, \aap, \href
  {http://adsabs.harvard.edu/abs/1973A%26A....24..337S} {24, 337}

\bibitem[\protect\citeauthoryear{{Shankar}}{{Shankar}}{2009}]{shankar09}
{Shankar} F.,  2009, \mn@doi [\nar] {10.1016/j.newar.2009.07.006}, \href
  {https://ui.adsabs.harvard.edu/abs/2009NewAR..53...57S} {53, 57}

\bibitem[\protect\citeauthoryear{{Shankar}}{{Shankar}}{2013}]{shankar13}
{Shankar} F.,  2013, \mn@doi [Classical and Quantum Gravity]
  {10.1088/0264-9381/30/24/244001}, \href
  {https://ui.adsabs.harvard.edu/abs/2013CQGra..30x4001S} {30, 244001}

\bibitem[\protect\citeauthoryear{{Shankar}, {Salucci}, {Granato}, {De Zotti}
  \& {Danese}}{{Shankar} et~al.}{2004}]{shankar04}
{Shankar} F.,  {Salucci} P.,  {Granato} G.~L.,  {De Zotti} G.,   {Danese} L.,
  2004, \mn@doi [\mnras] {10.1111/j.1365-2966.2004.08261.x}, \href
  {https://ui.adsabs.harvard.edu/abs/2004MNRAS.354.1020S} {354, 1020}

\bibitem[\protect\citeauthoryear{{Shankar} et~al.,}{{Shankar}
  et~al.}{2019}]{shankar19}
{Shankar} F.,  et~al., 2019, \mn@doi [\mnras] {10.1093/mnras/stz376}, \href
  {https://ui.adsabs.harvard.edu/abs/2019MNRAS.485.1278S} {485, 1278}

\bibitem[\protect\citeauthoryear{{Shankar} et~al.,}{{Shankar}
  et~al.}{2020}]{shankar20}
{Shankar} F.,  et~al., 2020, \mn@doi [Nature Astronomy]
  {10.1038/s41550-019-0949-y}, \href
  {https://ui.adsabs.harvard.edu/abs/2020NatAs...4..282S} {4, 282}

\bibitem[\protect\citeauthoryear{{Silpa}, {Kharb}, {Ho}, {Ishwara-Chandra},
  {Jarvis}  \& {Harrison}}{{Silpa} et~al.}{2020}]{silpa20}
{Silpa} S.,  {Kharb} P.,  {Ho} L.~C.,  {Ishwara-Chandra} C.~H.,  {Jarvis}
  M.~E.,   {Harrison} C.,  2020, \mn@doi [\mnras] {10.1093/mnras/staa2970},
  \href {https://ui.adsabs.harvard.edu/abs/2020MNRAS.tmp.2786S} {}

\bibitem[\protect\citeauthoryear{{Singh}, {Shastri}  \& {Risaliti}}{{Singh}
  et~al.}{2011}]{singh11}
{Singh} V.,  {Shastri} P.,   {Risaliti} G.,  2011, \mn@doi [\aap]
  {10.1051/0004-6361/201016387}, \href
  {https://ui.adsabs.harvard.edu/abs/2011A&A...532A..84S} {532, A84}

\bibitem[\protect\citeauthoryear{{Singh} et~al.,}{{Singh}
  et~al.}{2013}]{singh13}
{Singh} R.,  et~al., 2013, \mn@doi [\aap] {10.1051/0004-6361/201322062}, \href
  {http://adsabs.harvard.edu/abs/2013A%26A...558A..43S} {558, A43}

\bibitem[\protect\citeauthoryear{{Smith} et~al.,}{{Smith}
  et~al.}{2020a}]{smith20}
{Smith} K.~L.,  et~al., 2020a, \mn@doi [\mnras] {10.1093/mnras/stz3608}, \href
  {https://ui.adsabs.harvard.edu/abs/2020MNRAS.492.4216S} {492, 4216}

\bibitem[\protect\citeauthoryear{{Smith}, {Koss}, {Mushotzky}, {Wong},
  {Shimizu}, {Ricci}  \& {Ricci}}{{Smith} et~al.}{2020b}]{smith20b}
{Smith} K.~L.,  {Koss} M.,  {Mushotzky} R.,  {Wong} O.~I.,  {Shimizu} T.~T.,
  {Ricci} C.,   {Ricci} F.,  2020b, \mn@doi [\apj] {10.3847/1538-4357/abc3c4},
  \href {https://ui.adsabs.harvard.edu/abs/2020ApJ...904...83S} {904, 83}

\bibitem[\protect\citeauthoryear{{Smol{\v{c}}i{\'c}}
  et~al.,}{{Smol{\v{c}}i{\'c}} et~al.}{2008}]{smolcic08}
{Smol{\v{c}}i{\'c}} V.,  et~al., 2008, \mn@doi [\apjs] {10.1086/588028}, \href
  {https://ui.adsabs.harvard.edu/abs/2008ApJS..177...14S} {177, 14}

\bibitem[\protect\citeauthoryear{{Soltan}}{{Soltan}}{1982}]{soltan82}
{Soltan} A.,  1982, \mn@doi [\mnras] {10.1093/mnras/200.1.115}, \href
  {https://ui.adsabs.harvard.edu/abs/1982MNRAS.200..115S} {200, 115}

\bibitem[\protect\citeauthoryear{{Speagle}, {Steinhardt}, {Capak}  \&
  {Silverman}}{{Speagle} et~al.}{2014}]{speagle14}
{Speagle} J.~S.,  {Steinhardt} C.~L.,  {Capak} P.~L.,   {Silverman} J.~D.,
  2014, \mn@doi [\apjs] {10.1088/0067-0049/214/2/15}, \href
  {https://ui.adsabs.harvard.edu/abs/2014ApJS..214...15S} {214, 15}

\bibitem[\protect\citeauthoryear{{Suzuki} et~al.,}{{Suzuki}
  et~al.}{2016}]{suzuki16}
{Suzuki} T.~L.,  et~al., 2016, \mn@doi [\mnras] {10.1093/mnras/stw1655}, \href
  {https://ui.adsabs.harvard.edu/abs/2016MNRAS.462..181S} {462, 181}

\bibitem[\protect\citeauthoryear{{Tadhunter}}{{Tadhunter}}{2016}]{tadhunter16}
{Tadhunter} C.,  2016, \mn@doi [\aapr] {10.1007/s00159-016-0094-x}, \href
  {https://ui.adsabs.harvard.edu/abs/2016A&ARv..24...10T} {24, 10}

\bibitem[\protect\citeauthoryear{{Talbot}, {Bourne}  \& {Sijacki}}{{Talbot}
  et~al.}{2021}]{talbot21}
{Talbot} R.~Y.,  {Bourne} M.~A.,   {Sijacki} D.,  2021, \mn@doi [\mnras]
  {10.1093/mnras/stab804}, \href
  {https://ui.adsabs.harvard.edu/abs/2021MNRAS.504.3619T} {504, 3619}

\bibitem[\protect\citeauthoryear{{Tchekhovskoy} \& {McKinney}}{{Tchekhovskoy}
  \& {McKinney}}{2012}]{tchekhovskoy12}
{Tchekhovskoy} A.,  {McKinney} J.~C.,  2012, \mn@doi [\mnras]
  {10.1111/j.1745-3933.2012.01256.x}, \href
  {http://adsabs.harvard.edu/abs/2012MNRAS.423L..55T} {423, L55}

\bibitem[\protect\citeauthoryear{{Tchekhovskoy}, {Narayan}  \&
  {McKinney}}{{Tchekhovskoy} et~al.}{2011}]{tchekhovskoy11}
{Tchekhovskoy} A.,  {Narayan} R.,   {McKinney} J.~C.,  2011, \mn@doi [\mnras]
  {10.1111/j.1745-3933.2011.01147.x}, \href
  {http://adsabs.harvard.edu/abs/2011MNRAS.418L..79T} {418, L79}

\bibitem[\protect\citeauthoryear{{Terashima} \& {Wilson}}{{Terashima} \&
  {Wilson}}{2003}]{terashima03}
{Terashima} Y.,  {Wilson} A.~S.,  2003, ApJ, \href
  {http://adsabs.harvard.edu/cgi-bin/nph-bib_query?bibcode=2003\apj...583..145T&db_key=AST}
  {583, 145}

\bibitem[\protect\citeauthoryear{{Terashima}, {Iyomoto}, {Ho}  \&
  {Ptak}}{{Terashima} et~al.}{2002}]{terashima02}
{Terashima} Y.,  {Iyomoto} N.,  {Ho} L.~C.,   {Ptak} A.~F.,  2002, \mn@doi
  [\apjs] {10.1086/324373}, \href
  {https://ui.adsabs.harvard.edu/abs/2002ApJS..139....1T} {139, 1}

\bibitem[\protect\citeauthoryear{{Thean}, {Pedlar}, {Kukula}, {Baum}  \&
  {O'Dea}}{{Thean} et~al.}{2000}]{thean00}
{Thean} A.,  {Pedlar} A.,  {Kukula} M.~J.,  {Baum} S.~A.,   {O'Dea} C.~P.,
  2000, \mn@doi [\mnras] {10.1046/j.1365-8711.2000.03401.x}, \href
  {http://ukads.nottingham.ac.uk/abs/2000MNRAS.314..573T} {314, 573}

\bibitem[\protect\citeauthoryear{{Tremaine} et~al.,}{{Tremaine}
  et~al.}{2002}]{tremaine02}
{Tremaine} S.,  et~al., 2002, \apj, \href
  {http://adsabs.harvard.edu/cgi-bin/nph-bib_query?bibcode=2002\apj...574..740T&amp;db_key=AST}
  {574, 740}

\bibitem[\protect\citeauthoryear{{Trippe}}{{Trippe}}{2014}]{trippe14}
{Trippe} S.,  2014, \mn@doi [Journal of Korean Astronomical Society]
  {10.5303/JKAS.2014.47.4.159}, \href
  {http://adsabs.harvard.edu/abs/2014JKAS...47..159T} {47, 159}

\bibitem[\protect\citeauthoryear{{Ulvestad}, {Wilson}  \& {Sramek}}{{Ulvestad}
  et~al.}{1981}]{ulvestad81}
{Ulvestad} J.~S.,  {Wilson} A.~S.,   {Sramek} R.~A.,  1981, \mn@doi [\apj]
  {10.1086/159051}, \href {http://adsabs.harvard.edu/abs/1981ApJ...247..419U}
  {247, 419}

\bibitem[\protect\citeauthoryear{{Vahdat Motlagh}, {Kalemci}  \&
  {Maccarone}}{{Vahdat Motlagh} et~al.}{2019}]{Vahdat2019}
{Vahdat Motlagh} A.,  {Kalemci} E.,   {Maccarone} T.~J.,  2019, \mn@doi
  [\mnras] {10.1093/mnras/stz569}, \href
  {https://ui.adsabs.harvard.edu/abs/2019MNRAS.485.2744V} {485, 2744}

\bibitem[\protect\citeauthoryear{{Vaughan}, {Iwasawa}, {Fabian}  \&
  {Hayashida}}{{Vaughan} et~al.}{2005}]{Vaughn_4395}
{Vaughan} S.,  {Iwasawa} K.,  {Fabian} A.~C.,   {Hayashida} K.,  2005, \mn@doi
  [\mnras] {10.1111/j.1365-2966.2004.08463.x}, \href
  {https://ui.adsabs.harvard.edu/abs/2005MNRAS.356..524V} {356, 524}

\bibitem[\protect\citeauthoryear{{Venturi} et~al.,}{{Venturi}
  et~al.}{2021}]{venturi21}
{Venturi} G.,  et~al., 2021, \mn@doi [\aap] {10.1051/0004-6361/202039869},
  \href {https://ui.adsabs.harvard.edu/abs/2021A&A...648A..17V} {648, A17}

\bibitem[\protect\citeauthoryear{{Vitale}, {Fuhrmann},
  {Garc{\'\i}a-Mar{\'\i}n}, {Eckart}, {Zuther}  \& {Hopkins}}{{Vitale}
  et~al.}{2015}]{vitale15}
{Vitale} M.,  {Fuhrmann} L.,  {Garc{\'\i}a-Mar{\'\i}n} M.,  {Eckart} A.,
  {Zuther} J.,   {Hopkins} A.~M.,  2015, \mn@doi [\aap]
  {10.1051/0004-6361/201423993}, \href
  {https://ui.adsabs.harvard.edu/abs/2015A&A...573A..93V} {573, A93}

\bibitem[\protect\citeauthoryear{{Walker}, {Romney}  \& {Benson}}{{Walker}
  et~al.}{1994}]{walker94}
{Walker} R.~C.,  {Romney} J.~D.,   {Benson} J.~M.,  1994, \mn@doi [\apjl]
  {10.1086/187434}, \href
  {https://ui.adsabs.harvard.edu/abs/1994ApJ...430L..45W} {430, L45}

\bibitem[\protect\citeauthoryear{{Weaver}, {McCray}, {Castor}, {Shapiro}  \&
  {Moore}}{{Weaver} et~al.}{1977}]{weaver77}
{Weaver} R.,  {McCray} R.,  {Castor} J.,  {Shapiro} P.,   {Moore} R.,  1977,
  \mn@doi [\apj] {10.1086/155692}, \href
  {https://ui.adsabs.harvard.edu/abs/1977ApJ...218..377W} {218, 377}

\bibitem[\protect\citeauthoryear{{Webster} et~al.,}{{Webster}
  et~al.}{2021}]{webster21}
{Webster} B.,  et~al., 2021, \mn@doi [\mnras] {10.1093/mnras/staa3437}, \href
  {https://ui.adsabs.harvard.edu/abs/2021MNRAS.500.4921W} {500, 4921}

\bibitem[\protect\citeauthoryear{{Werner}, {McNamara}, {Churazov}  \&
  {Scannapieco}}{{Werner} et~al.}{2019}]{werner19}
{Werner} N.,  {McNamara} B.~R.,  {Churazov} E.,   {Scannapieco} E.,  2019,
  \mn@doi [\ssr] {10.1007/s11214-018-0571-9}, \href
  {https://ui.adsabs.harvard.edu/abs/2019SSRv..215....5W} {215, 5}

\bibitem[\protect\citeauthoryear{{White}, {Quataert}  \& {Gammie}}{{White}
  et~al.}{2020}]{white20}
{White} C.~J.,  {Quataert} E.,   {Gammie} C.~F.,  2020, \mn@doi [\apj]
  {10.3847/1538-4357/ab718e}, \href
  {https://ui.adsabs.harvard.edu/abs/2020ApJ...891...63W} {891, 63}

\bibitem[\protect\citeauthoryear{{Williams} et~al.,}{{Williams}
  et~al.}{2019}]{williams19}
{Williams} D.~R.~A.,  et~al., 2019, \mn@doi [\mnras] {10.1093/mnras/stz1135},
  \href {https://ui.adsabs.harvard.edu/abs/2019MNRAS.486.4962W} {486, 4962}

\bibitem[\protect\citeauthoryear{{W{\'o}jtowicz}, {Stawarz}, {Cheung},
  {Ostorero}, {Kosmaczewski}  \& {Siemiginowska}}{{W{\'o}jtowicz}
  et~al.}{2020}]{wojtowicz20}
{W{\'o}jtowicz} A.,  {Stawarz} {\L}.,  {Cheung} C.~C.,  {Ostorero} L.,
  {Kosmaczewski} E.,   {Siemiginowska} A.,  2020, \mn@doi [\apj]
  {10.3847/1538-4357/ab7930}, \href
  {https://ui.adsabs.harvard.edu/abs/2020ApJ...892..116W} {892, 116}

\bibitem[\protect\citeauthoryear{{Wrobel}}{{Wrobel}}{2000}]{wrobel00}
{Wrobel} J.~M.,  2000, \mn@doi [\apj] {10.1086/308519}, \href
  {http://adsabs.harvard.edu/abs/2000ApJ...531..716W} {531, 716}

\bibitem[\protect\citeauthoryear{{Yang} et~al.,}{{Yang} et~al.}{2020}]{yang20}
{Yang} X.,  et~al., 2020, \mn@doi [\apj] {10.3847/1538-4357/abb775}, \href
  {https://ui.adsabs.harvard.edu/abs/2020ApJ...904..200Y} {904, 200}

\bibitem[\protect\citeauthoryear{{Yang}, {Yuan}, {Yuan}  \& {White}}{{Yang}
  et~al.}{2021}]{yang21}
{Yang} H.,  {Yuan} F.,  {Yuan} Y.-F.,   {White} C.~J.,  2021, \mn@doi [\apj]
  {10.3847/1538-4357/abfe63}, \href
  {https://ui.adsabs.harvard.edu/abs/2021ApJ...914..131Y} {914, 131}

\bibitem[\protect\citeauthoryear{{Yesuf}, {Faber}, {Koo}, {Woo}, {Primack}  \&
  {Luo}}{{Yesuf} et~al.}{2020}]{yesuf20}
{Yesuf} H.~M.,  {Faber} S.~M.,  {Koo} D.~C.,  {Woo} J.,  {Primack} J.~R.,
  {Luo} Y.,  2020, \mn@doi [\apj] {10.3847/1538-4357/ab5fe1}, \href
  {https://ui.adsabs.harvard.edu/abs/2020ApJ...889...14Y} {889, 14}

\bibitem[\protect\citeauthoryear{{Yi} \& {Boughn}}{{Yi} \&
  {Boughn}}{1998}]{yi98}
{Yi} I.,  {Boughn} S.~P.,  1998, \mn@doi [\apj] {10.1086/305631}, \href
  {https://ui.adsabs.harvard.edu/abs/1998ApJ...499..198Y} {499, 198}

\bibitem[\protect\citeauthoryear{{Yi} \& {Boughn}}{{Yi} \&
  {Boughn}}{1999}]{yi99}
{Yi} I.,  {Boughn} S.~P.,  1999, \mn@doi [\apj] {10.1086/307041}, \href
  {https://ui.adsabs.harvard.edu/abs/1999ApJ...515..576Y} {515, 576}

\bibitem[\protect\citeauthoryear{{York} et~al.,}{{York} et~al.}{2000}]{york00}
{York} D.~G.,  et~al., 2000, \mn@doi [\aj] {10.1086/301513}, \href
  {http://adsabs.harvard.edu/abs/2000AJ....120.1579Y} {120, 1579}

\bibitem[\protect\citeauthoryear{{Yuan} \& {Narayan}}{{Yuan} \&
  {Narayan}}{2014}]{yuan14}
{Yuan} F.,  {Narayan} R.,  2014, \mn@doi [\araa]
  {10.1146/annurev-astro-082812-141003}, \href
  {http://adsabs.harvard.edu/abs/2014ARA%26A..52..529Y} {52, 529}

\bibitem[\protect\citeauthoryear{{Yuan}, {Bu}  \& {Wu}}{{Yuan}
  et~al.}{2012a}]{yuan12b}
{Yuan} F.,  {Bu} D.,   {Wu} M.,  2012a, \mn@doi [\apj]
  {10.1088/0004-637X/761/2/130}, \href
  {http://adsabs.harvard.edu/abs/2012ApJ...761..130Y} {761, 130}

\bibitem[\protect\citeauthoryear{{Yuan}, {Wu}  \& {Bu}}{{Yuan}
  et~al.}{2012b}]{yuan12a}
{Yuan} F.,  {Wu} M.,   {Bu} D.,  2012b, \mn@doi [\apj]
  {10.1088/0004-637X/761/2/129}, \href
  {https://ui.adsabs.harvard.edu/abs/2012ApJ...761..129Y} {761, 129}

\bibitem[\protect\citeauthoryear{{Zakamska} \& {Greene}}{{Zakamska} \&
  {Greene}}{2014}]{zakamska14}
{Zakamska} N.~L.,  {Greene} J.~E.,  2014, \mn@doi [\mnras]
  {10.1093/mnras/stu842}, \href
  {https://ui.adsabs.harvard.edu/abs/2014MNRAS.442..784Z} {442, 784}

\bibitem[\protect\citeauthoryear{{de Vaucouleurs}, {de Vaucouleurs}, {Corwin},
  {Buta}, {Paturel}  \& {Fouqu{\'e}}}{{de Vaucouleurs}
  et~al.}{1991}]{devaucouleurs91}
{de Vaucouleurs} G.,  {de Vaucouleurs} A.,  {Corwin} Jr. H.~G.,  {Buta} R.~J.,
  {Paturel} G.,   {Fouqu{\'e}} P.,  1991, {Third Reference Catalogue of Bright
  Galaxies. Volume I--III}.
Springer, New York

\bibitem[\protect\citeauthoryear{{van den Bosch}}{{van den
  Bosch}}{2016}]{vandenbosch16}
{van den Bosch} R. C.~E.,  2016, \mn@doi [\apj] {10.3847/0004-637X/831/2/134},
  \href {https://ui.adsabs.harvard.edu/abs/2016ApJ...831..134V} {831, 134}

\makeatother
\end{thebibliography}

\vspace{3mm}
\noindent
{\small 
$^{1}$ Istituto di Radioastronomia - INAF, Via P. Gobetti 101, I-40129 Bologna, Italy\\
$^{2}$ School of Physics and Astronomy, University of Southampton, Southampton, SO17 1BJ, UK\\
$^{3}$ Jodrell Bank Centre for Astrophysics, School of Physics and Astronomy, The University of Manchester, Manchester, M13 9PL, UK\\
$^{4}$ Departamento de F\'isica de la Tierra y Astrof\'isica, Instituto de F\'isica de Part\'iculas y del Cosmos IPARCOS, Universidad Complutense de Madrid, E-28040 Madrid, Spain \\
$^{5}$ Instituto de Astrof\'isica de Canarias, Via L\'actea S/N, E-38205, La Laguna, Tenerife, Spain\\
$^{6}$ Departamento de Astrof\'isica, Universidad de La Laguna, E-38206, La Laguna, Tenerife, Spain\\
$^{7}$ Jeremiah Horrocks Institute, University of Central Lancashire, Preston PR1 2HE, UK\\ 
$^{8}$ Department of Space, Earth and Environment
Chalmers University of Technology
SE-412 96  Gothenburg, Sweden\\
$^{9}$ Instituto de Astrof\`isica de Andaluc\'ia (IAA, CSIC), Glorieta de la Astronom\'ia s/n, 18008-Granada, Spain\\
$^{10}$ Netherlands Institute for Radio Astronomy, ASTRON, Dwingeloo, The Netherlands\\
$^{11}$ UK ALMA Regional Centre Node, Jodrell Bank Centre for Astrophysics\\
$^{12}$ Department of Physics \& Astronomy, University College London, Gower Street, London WC1E 6BT, UK\\
$^{13}$ Astrophysics Group, Cavendish Laboratory, 19 J.~J.~Thomson Avenue, Cambridge CB3 0HE, UK\\
$^{14}$ Max-Planck-Institut f\''{u}r Radioastronomie, Auf dem H\''{u}gel 69, 53121 Bonn, Germany \\
$^{15}$ Department of Astrophysics/IMAPP, Radboud University, P.O. Box 9010, 6500 GL Nijmegen, The Netherlands\\
$^{16}$ Department of Physics, Box 41051, Science Building, Texas Tech University, Lubbock, TX 79409-1051, US\\
$^{17}$ Real Academia de Ciencias, C/ Valverde 22, 28004 Madrid\\
$^{18}$ Technical University of Kenya, P.O. Box 52428 - 00200, Nairobi- Kenya\\
$^{19}$ INAF - Istituto di Astrofisica e Planetologia Spaziali, via Fosso del Cavaliere 100, I-00133 Roma, Italy\\ 
$^{20}$ Institute of Astronomy and Astrophysics, Academia Sinica, 11F of Astronomy-Mathematics Building,AS/NTU No. 1, Sec. 4, Roosevelt Rd, Taipei 10617, Taiwan, R.O.C\\
$^{21}$ Inter-University Centre for Astronomy and Astrophysics (IUCAA), Ganeshkhind P.O., Pune 411007, India\\
$^{22}$ Center for Astro, Particle and Planetary Physics, New York University Abu Dhabi, PO Box 129188, Abu Dhabi, UAE\\
$^{23}$ School of Physics and Astronomy, University of Birmingham, Edgbaston, Birmingham B15 2TT, UK\\
$^{24}$ Anton Pannekoek Institute for Astronomy, University of Amsterdam, Science Park 904, 1098 XH Amsterdam, the Netherlands\\
$^{25}$  Centre for Astrophysics Research, University of Hertfordshire, College Lane, Hatfield, AL10 9AB, UK\\
$^{26}$ Station de Radioastronomie de Nan\c{c}ay, Observatoire de Paris, PSL Research University, CNRS,Universit\'e Orl\'{e}ans, 18330 Nan\c{c}ay, France\\
$^{27}$ AIM, CEA, CNRS, Universit\'e  de Paris, Universit\'e Paris Saclay, F-91191 Gif-sur-Yvette, France\\
$^{28}$ Dpt. Astronomia i Astrof\'isica, Unversitat de Val\`encia, C/Dr. Moliner 50, 46100 Burjassot (Valencia, Spain)\\
$^{29}$ Department of Physics, University of Bath, Claverton Down, Bath, BA2 7AY, UK\\
$^{30}$ Department of Physics, Indian Institute of Technology, Hyderabad 502285, India\\
$^{31}$ Centre for Extragalactic Astronomy, Department of Physics, Durham University, Durham DH1 3LE.
}
%






\appendix
\section{Local black hole mass function}
\label{app}

One of the main results of our present work is the presence of a break in the radio and accretion properties of local galaxies at $M_{\rm BH}\sim 10^{6.5} M_{\odot}$, which correspond to a $\sigma \sim$ 70 km$^{-1}$ according to the scaling relation used in this work from \citet{tremaine02}. More precisely, we note an increasing AGN contribution to the radio core emission and a steepening of the radio-optical relations associated with AGN-dominated galaxies with respect to those relative to SF-dominated galaxies at larger BH masses.

If such a $M_{\rm BH}$ break is real, on the assumption of an underlying universal $M_{\rm BH}$--$\sigma$ relation, we would expect to witness a similar transition occurring in the local BH mass function (BHMF), with a predominance of active galaxies in more massive BHs progressively turning into more star-forming galaxies at less massive BHs.
Unfortunately, the lack of samples of local star-forming galaxies statistically complete at low $\sigma$ severely limits the computation of the BHMF at lower BH masses (see, e.g., \citealt{bernardi03} and also \citealt{shankar13,shankar20} on the low mass-end of the BHMF). Therefore, to overcome this problem, we have
focused on the SFR-$\sigma$ scaling relation rather than on the BHMF of star-forming galaxies, by using  the local galaxy population taken from  
Sloan Digital Sky survey (SDSS, \citealt{york00}) Data Release 7 \citep{abazajian09} at $z<$0.3, as presented in \citet{meert15}. The scaling relations of galaxies should still hold even at lower velocity dispersions irrespective of the degree of incompleteness in SDSS, as long as no strong biases are expected in the subsample of galaxies with measured velocity dispersions. Stellar masses are computed using the best-fitting \texttt{S\'ersic-Exponential} or \texttt{S\'ersic} photometry of $r$-band observations, and by adopting the mass-to-light ratios by \citet{mendel14}. We further adopt the SFR estimates from \citet{brinchmann04} and compute the specific star formation rate sSFR $\equiv$ SFR/$M_{\rm star}$. We convert the available measurements of the velocity dispersion $\sigma$ to BH mass, $M_{\rm BH}$, by using the mean $M_{BH}-\sigma$ relation from \citet{tremaine02}, as used throughout this work.

\begin{figure}
	\includegraphics[width=0.5\textwidth]{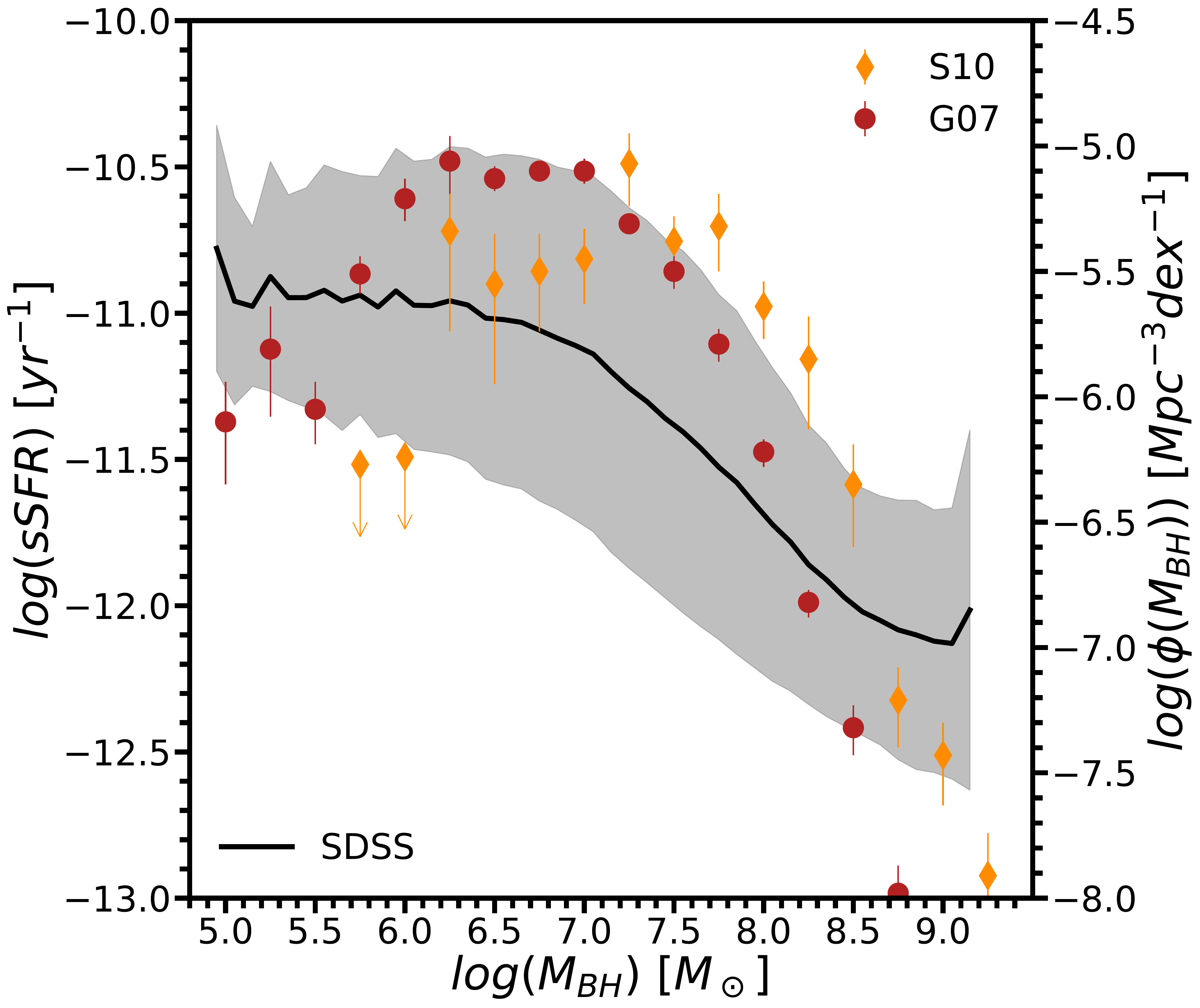}
        \caption[]{Specific star-formation rate (sSFR [yr$^{-1}$], left y-axis) versus $M_{\rm BH}$ (M$_{\odot}$) for star-forming  galaxies  and the local BHMF (Mpc$^{-3}$ dex$^{-1}$, right y-axis) of active galaxies. The black solid line and its shaded area represents the sSFR function computed by taking the local (z$<$0.3) star-forming galaxies from SDSS Data Release 7 \citep{abazajian09} and its 1-$\sigma$ scatter. The red and orange points are the data points of the BHMF of local active SMBHs taken respectively from \citet{schulze10} and \citet{greene07}. We note that sSFR steeply decreases at $M_{\rm BH} \gtrsim $ 10$^{6.5}$ M$_{\odot}$, where conversely the BHMF of active galaxies peaks. }
    \label{BHmassfunction}
\end{figure}

In Figure~\ref{BHmassfunction} we plot sSFR versus $M_{\rm BH}$ for star-forming  galaxies and the local BHMF of active galaxies from \citet{greene07} and \citet{schulze10}. We note that sSFR steeply decreases at larger BH masses, with a clear break at $\sim$10$^{6-7}$ M$_{\odot}$, where the BHMF of active galaxies peaks. In agreement, \citet{dullo20} reported that more massive BHs are hosted by galaxies with redder UV--[3.6 $\mu$m]) color (i.e., lower sSFR). They also revealed that the $M_{\rm BH}$--sSFR relations are morphology dependent where the SMBH masses for LTGs exhibit a steeper dependence on sSFR than those for ETGs. In conclusion, the observed break in the sSFR-$M_{\rm BH}$ relation (Fig.~\ref{BHmassfunction}), consistent with the one found in our optical-radio relations (Sect.~\ref{sect3}), provides independent support to our result in the present work: the action/energetics of the central SMBH becomes progressively more important in more massive BHs (and galaxies), $\gtrsim 10^{6.5} M_{\odot}$.
A detailed analysis of this study and the possible consequences of this result (e.g. hypothesis of a gradually more efficient role of radio AGN feedback at $M_{\rm BH} \gtrsim 10^{6.5} M_{\odot}$) will be subject of a forthcoming work.

\setcounter{table}{0}
\onecolumn
\begin{center}
\begin{longtable}{lC{2.5cm}ccccccccc}
\caption[Properties of the  sample.]{Radio and optical properties of the LeMMINGs (Palomar) sample.} 
\label{tab1} \\

\hline 
\hline
Name & Hubble    & class  & $\sigma$   & log $M_{\rm BH}$ & log $L_{\rm [O~III]}$ &  log Edd  &  det & morph &  log $L_{\rm core}$  & log $L_{\rm Total}$  \\
           &              &  BPT      &    km s$^{-1}$ &  M$_{\odot}$  & erg s$^{-1}$  & ratio &    &         &  erg s$^{-1}$ & erg s$^{-1}$\\     
(1)       & (2)         & (3)       & (4)             & (5)               & (6)                 & (7)       & (8)   &  (9)          &  (10)           &  (11)  \\
\hline	
\endfirsthead

\multicolumn{3}{c}{{\tablename} \thetable{} -- Continued} \\[0.5ex]
\hline
\hline
 Name & Hubble    & class  & $\sigma$   & log $M_{\rm BH}$ & log $L_{\rm [O~III]}$ &  log Edd  &  det & morph &  log $L_{\rm core}$  & log $L_{\rm total}$  \\
           &              &  BPT      &    km s$^{-1}$ &  M$_{\odot}$  & erg s$^{-1}$  & ratio &    &         &  erg s$^{-1}$ & erg s$^{-1}$\\ 
\hline 

\endhead

\hline
  \multicolumn{11}{c}{{Continued on Next Page}} \\
\endfoot

 \\[-1.8ex] 
\endlastfoot

NGC~7817 & SAbc    & H          &66.7  & 6.21  &  39.29 & $-$1.51  &  U & $-$  & $<$35.64 & $-$ \\
IC~10 	  & IBm?      & H        &35.5  & 5.11  &   37.13 & $-$2.57 & U  & $-$  & $<$32.91 & $-$  \\
NGC~147   &  dE5 pec & ALG   &  22.0    & 4.28  &  $-$    &  $-$       &  U  & $-$     &  $<$32.39  & $-$ \\
NGC~185   & dE3 pec  &  L       &  19.9  & 4.10  & 34.63 &  $-$4.06 & U & $-$  &  $<$32.33 & $-$   \\
NGC~205   &  dE5 pec  & ALG  & 23.3    & 4.34$^{*}$ &  $-$    &  $-$  &U & $-$  &  $<$32.36 & $-$  \\
NGC~221   & E2           &  ALG  &  72.1   & 6.36  & $-$    &  $-$    &  U & $-$   &  $<$32.33 & $-$    \\
NGC~224   & SAb        &  ALG   &  169.8 & 7.84  & $-$    &  $-$   &   U & $-$   &  $<$32.32 & $-$ \\
NGC~266   & SBab      & L         &  229.6 & 8.37  & 39.43  & $-$3.53 &  I   & A  & 36.94  & 37.02 \\
NGC~278   & SABb      & H        &  47.6   & 5.62  & 37.47  & $-$2.74 & I    & A  &  34.81 & 35.83 \\
NGC~315   &  E+        &  L$^{RL}$        &  303.7  & 8.92$^{*}$ & 39.43 & $-$4.02  &  I   & A  & 39.58 &  39.61 \\
NGC~404   & SA0       &  L        & 40.0        & 5.65$^{*}$ & 37.16  & $-$3.08  & U  & $-$   & $<$33.28  &   $-$ \\
NGC~410   &  E+        &  L        & 299.7 &  8.84   & $<$39.32 &   $<$-4.11  & I   & A  & 37.26 &  37.53 \\
NGC~507   & SA0       &  ALG    & 307.7  & 8.88   &  $-$    &  $-$   &  I    & A   & 36.86  & 37.07  \\
NGC~598   & SAcd     &   H       &  21.0       &  4.20  &  $<$34.63  &  $<-$4.16  & U & $-$  & $<$32.43  & $-$  \\
IC~1727     & SBM      &   L        &  136.8  & 7.47   &  37.34    &  $-$4.72  &  U & $-$   &  $<$34.55 &  $-$  \\
NGC~672   &  SBcd    &  H        &  $<$64.3  & $<$6.15 &  37.66  & $-$3.08  &  U & $-$  &  $<$34.55  & $-$ \\
NGC~697   &  SABc    &  H        &  75.0     &  6.42     &  37.86    &  $-$3.15   &   U &  $-$  &   $<$35.95  &  $-$ \\
NGC~777   & E1         &  L$^{RL}$    &  324.1  & 8.97     & 38.38   & $-$5.18      &  I    & A      &  36.77 & 36.98  \\
NGC~783   & SAc      & H        & 101.4 & 6.94        & 38.81    & $-$2.68    & U & $-$    & $<$36.12  & $-$     \\
NGC~784   & SBdm    &  H      &  35.5   & 5.11       & 37.68    & $-$2.02   &  U & $-$    &  $<$34.19  & $-$  \\
NGC~812   & SAB0/a pec &  H  &  120.9  &  7.25   & 38.68  & $-$3.16    &   U & $-$    &  $<$36.70  & $-$  \\
NGC~818   & SABc     &      H  &   151.3   & 7.64    & 38.46  & $-$3.77   &   U  & $-$     &  $<$36.14  & $-$ \\
NGC~841   & SABab   &  L      &  159.2     & 7.73    & 38.74  & $-$3.58   &    U & $-$     &  $<$36.19  & $-$ \\
NGC~890   & SAB0    &  ALG   &  210.9    &  8.22   &    $-$    &  $-$      &   U & $-$     &   $<$36.06   & $-$  \\
NGC~891   & SAb?     &  H      &  73.1      & 6.37     &  36.29  & $-$4.66  & unI &  $-$   & $<$34.65   &  $-$ \\
NGC~925   & SABd     &  H     &  71.9      &  6.34     &  37.21  &  $-$3.72  & U & $-$ &  $<$34.55  &  $-$   \\
NGC~959   & SAdm   &   H    &   43.6     &  5.47     &  37.40   & $-$2.66  &  U &  $-$ &  $<$34.65  &  $-$   \\
NGC~972   & SAab     &  j\,H    &   102.8   & 6.97     &  38.64    &  $-$2.92  & I   & B      &  35.31  &  36.06   \\
IC~239       & SAB(rs)cd    & H        & 92.3  & 6.78            & $<$36.93    & $<-$4.40 & U &$-$      & $<$34.98 &  $-$         \\
NGC~1003 & SA(s)cd       & H        & $-$       & $-$                 & 37.23        & $-$               & U &$-$      & $<$34.58 & $-$               \\
NGC~1023 & SB(rs)0-      &  ALG     & 204.5 & 7.62$^{*}$      & $-$      & $-$      & U &$-$      & $<$34.57 & $-$          \\
NGC~1058 & SA(rs)c        & L              & 31.0    & 4.88            & 35.95    & $-$3.48 & U  & $-$    & $<$34.41  &  $-$             \\
NGC~1156  & IB(s)m         & H              & 35.9  & 5.13            & 39.15    & $-$0.53  & U &$-$     & $<$34.15 &  $-$         \\
NGC~1161  & SA0             &  L$^{RL}$              & 258.4 & 8.58            & 38.13    & $-$5.00  &  I  & A      & 36.57    & 36.69         \\
NGC~1167   & SA0-           &  L$^{RL}$              & 216.9 & 8.27            & 40.17    & $-$2.65  & I  & A     & 39.65    & 39.76           \\
NGC~1169  & SAB(r)b        &  L        & 181.3 & 7.96           & 38.58    & $-$3.93  &  U &$-$     &  $<$35.66  &    $-$       \\
NGC~1186  & SB(r)bc:       & H              & 119.8 & 7.24            & 38.95    & $-$2.84   &  I  & E     &  35.91   & 36.78         \\
NGC~1275  & Pec             & L$^{RL}$         & 258.9 & 8.98$^{*}$      & 41.61    & $-$1.92      &  I  & A     & 40.97   & 41.62           \\
IC~342       & SAB(rs)cd    & H               & 74.0  & 6.41$^{*}$      &  35.99 & $-$4.97   & unI &$-$     & $<$33.86 &   $-$         \\
IC~356       & SA(s)ab pec & H         & 156.6 & 7.70            & 37.57    & $-$4.68     & I   & A      & 35.52   & 35.74     \\
NGC~1569 & IBm            & H               & 44.0    & 5.49           & 36.49 & $-$3.55   & unI &$-$     & $<$33.18 & $-$         \\
NGC~1560 & SA(s)d spin & H               & 33.9  & 5.03           & 37.38    & $-$2.20  & I   & A      & 33.78    & 34.12      \\
NGC~1961  & SAB(rs)c      &  L$^{RL}$         & 241.3 & 8.29$^{*}$      & 39.11    & $-$3.73 & I   & A      & 37.15    & 37.25     \\
NGC~2146 & SB(s)ab pec  &  H              & 126.8 & 7.33           & 38.36    & $-$3.52   & I   & E      & 36.34    & 36.86      \\
NGC~2273  & SBa            &  S         &   148.9     &  7.61         &  40.43   & $-$1.77    &   I   & E     &  36.64   & 37.98  \\
NGC~2342  & S pec         & H         &   147.3    &  7.60           &  39.71   &  $-$2.48  &   I  & E     &  36.55   &  37.64  \\
UGC~3714 & S? pec           & H               & 104.0   & 6.99           & 38.49    & $-$3.05   & U  & $-$     & $<$35.90 & $-$         \\
NGC~2268  & SABbc        &   H         &    143.3   &  7.55          &  39.27   & $-$2.87   &  U &  $-$   &  $<$35.76  & $-$   \\
UGC~3828  & SABb        &    j\,H          &   73.9     & 6.39           &  38.84    &  $-$2.14  &  I  & C       & 36.63   & 38.84  \\
NGC~2336 & SAB(r)bc       &  L              & 116.2 & 7.18           & 38.20    & $-$3.54   & U   & $-$     & $<$35.75 & $-$         \\
NGC~2276 & SABc            &  H          &    83.5   &  6.61          &  38.17   &  $-$3.03   & U & $-$     & $<$35.90 & $-$      \\
NGC~2366 & IB(s)m          & H        & $-$   & $-$            & $-$      & $-$     & U   &$-$      & $<$33.64 & $-$        \\
IC~467       & SAB(s)c:       &  H              & 64.2  & 6.15           & 37.15    & $-$3.55  & unI &$-$     & $<$35.56 & $-$         \\
NGC~2300  & SA0            & ALG      &   261.1   &  8.60           &  $-$    &  $-$        &    I  & A &  36.23   &   36.41  \\
NGC~2403 & SAB(s)cd       &  H        & 68.4  & 6.26           & $<$35.91    & $<-$4.90 & U   &$-$      & $<$33.96 & $-$        \\
UGC~4028  & SABc?         &    j\,H        &   80.5   &  6.54        &  38.88  &  $-$2.25   &   I   & C  &  36.22  &  36.70    \\
NGC~2500  & SBd          &    H          &  47.1    &  5.61        &    36.55  &  $-$3.65   &  U  & $-$     & $<$34.71 & $-$    \\
NGC~2543  & SBb         &    H          &  112.4   &  7.12        &   38.55   &  $-$3.16   &  U &   $-$     & $<$35.65 & $-$   \\
NGC~2537  & SBm pec   & H           &   63.0       &  6.11        &  38.72    &  $-$1.98    &  U & $-$    &   $<$34.60  &  $-$  \\
NGC~2541  & SAcd     &    H          &   53.0        & 5.81         &   36.80   &  $-$3.60    &  U & $-$    &   $<$34.72  &  $-$   \\
NGC~2549 & SA(r)0 spin   & ALG       & 142.6 & 7.16$^{*}$     & $-$      & $-$      & U   & $-$      & $<$35.26 & $-$        \\
NGC~2639  & SAa?         &    L$^{RL}$          &  179.3   & 7.94          &  39.60    &  $-$2.93   &  I   & C   &  37.61   &  38.50   \\
NGC~2634  & E1            &   ALG      &  181.1   &  7.96        &   $-$       &  $-$         &   I   & A    &  35.81   &  35.87  \\
NGC~2683 & SA(rs)b         & L               & 130.2 & 7.38           & 37.06    & $-$4.87   &  I  & A       & 34.49    &  34.65         \\
NGC~2681  & SAB0/a       & L           &    109.1   &  7.07       &   38.37   &  $-$3.29   &   I  & C  &   35.51   &   35.99   \\
IC~520 	  & SABab?       &    L        &    138.1    &   7.48      &  39.03    & $-$3.04    &  U &  $-$      & $<$36.08 & $-$   \\
NGC~2685 & (R)SB0+pec  & L               & 93.8  & 6.59$^{*}$      & 38.41    & $-$2.73  & U   &$-$      & $<$34.82  &  $-$          \\
NGC~2655  & SAB0/a       & L$^{RL}$         &    159.8    &  7.74          &   39.44   &  $-$2.89    &  I  & E    &  37.59  &  37.97  \\
NGC~2750 & SABc            & H               & 52.4  & 5.79            & 39.20    & $-$1.14   & unI & $-$      & $<$35.76  &  $-$          \\
NGC~2742 & SA(s)c:         & H               & 65.6  & 6.18            &  37.73    & $-$3.00  &    U   & $-$      & $<$35.37 & $-$        \\
NGC~2715  & SABc           &  H             &  84.6  &  6.63           &  37.79   &  $-$3.43  &   U &  $-$      & $<$35.32 & $-$  \\
NGC~2770 & SA(s)c:         & H               & 81.0    & 6.55            & 37.56    & $-$3.54  &    U   &$-$      & $<$35.46 &   $-$         \\
NGC~2768 & E6:               & L$^{RL}$              & 181.8 & 7.96            & 38.61    & $-$3.90  &      I   & A       & 37.10    & 37.19     \\
NGC~2776 & SAB(rs)c       & H               & 47.7  & 5.63            & 38.32    & $-$1.86   &     U   & $-$      & $<$35.61 &   $-$         \\
NGC~2748  & SAbc           & H            &    83.0     &  7.65$^{*}$   &  37.83  &  $-$4.41   &   U &    $-$      & $<$35.44 & $-$   \\
NGC~2782 & SAB(rs)a pec & j\,H               & 183.1 & 7.98            & 40.00    & $-$2.53   &      I   & A       &  36.79    &   37.16        \\
NGC~2787 & SB(r)0+        & L               & 202.0   & 7.61$^{*}$      & 38.37    & $-$3.79   &     I   & A        & 36.28    & 36.41    \\
NGC~2832 &  E+2:           & L            & 334.0   & 9.03             & $<$39.05  & $<-$4.53  &      I+unI   & A    &  36.83   &  37.05         \\
NGC~2841  & SAb            &  L            &   222.0  &  8.31           & 38.19    &  $-$4.63            &  I  & C   &  34.84  & 35.47 \\
NGC~2859 & (R)SB(r)0+    & L               & 188.2 & 8.02             & 38.57    & $-$4.00  &      U   &$-$       & $<$35.31 &  $-$          \\
NGC~2903 & SAB(rs)bc     & H               & 89.0    & 7.06$^{*}$       & 37.35    & $-$4.26   &     U   &$-$       & $<$34.32  &   $-$       \\
NGC~2950 & (R)SB(r)0       & ALG             & 163.0   & 7.77             & $-$      & $-$     &    U   &$-$       & $<$35.47 & $-$       \\
NGC~2964 & SAB(r)bc:      & H               & 109.4 & 6.73$^{*}$       & 38.71    & $-$2.57   &    I   & E        &  36.03   & 36.97          \\
NGC~2977 & SAb:             & H               & 104.5 & 7.00             & 38.20    & $-$3.35 &   U   &$-$       & $<$35.91 & $-$       \\
NGC~2976 & SAc pec        & H               & 36.0    & 5.14             & 36.58    & $-$3.11  &     U   &$-$       & $<$33.24 & $-$       \\
NGC~3003 & SAbc?           & H               & 44.1  & 5.49             & 38.90    & $-$1.14  &    U   &$-$       & $<$35.28 & $-$          \\
NGC~2985  & R')SA(rs)ab   & L         & 140.8 & 7.52            & 38.69    & $-$3.38  &          I   & A        & 35.81    & 36.03     \\
NGC~3031 & SA(s)ab        & L      & 161.6 & 7.81$^{*}$       & 37.72    & $-$4.64  &            I   & A        & 35.50    & 35.54   \\
NGC~3027 & SB(rs)d:        & H               & 25.6  & 4.54             & 37.69    & $-$1.40   &   U   &$-$       & $<$35.29 & $-$       \\
NGC~3034 & IAO spin      & H               & 129.5 & 7.37             & 38.33   & $-$3.59  &       unI &$-$       & $<$34.28 & $-$       \\
NGC~3043 & SAb: spin     & H               & 51.9  & 5.77             & 38.08    & $-$2.24   &     U   &$-$       & $<$35.91 & $-$       \\
NGC~3073 & SAB0-           & H         & 35.6  & 5.12             & 38.02    & $-$1.65  &        U   &$-$       & $<$35.26 & $-$       \\
NGC~3077  & AO pec        & H               & 32.4  & 4.95             & 37.15    & $-$2.35 &      I+unI & A      & 33.32    & 34.35     \\
NGC~3079 & SB(s)c spin   & L         & 182.3 & 6.40$^{*}$        & 37.67    & $-$3.28  &           I   & C        & 37.27    & 37.75       \\
NGC~3162 & SAB(rs)bc     & H               & 89.0    & 6.72             & 38.04    & $-$3.23  &     U   &$-$       & $<$35.19 & $-$       \\
NGC~3147 & SA(rs)bc       & L               & 219.8 & 8.29             & 39.54    & $-$3.30   &    I   & A        & 37.36    & 37.51    \\
NGC~3185 & (R)SB(r)0/a    & S               & 79.3  & 6.51             & 39.44    & $-$1.62   &    U   &$-$       & $<$35.15    & $-$       \\
NGC~3190 & SA(s)a pec spin  & L         & 188.1 & 8.02             & 38.71    & $-$3.86   &        U   &$-$       & $<$35.32 & $-$       \\
NGC~3184  & SABcd        & H              & 43.3     & 5.46            &  37.31   & $-$2.74    &    U &    $-$      & $<$34.64 & $-$  \\
NGC~3193 & E2               & L               & 194.3 & 8.08             & 38.42    & $-$4.21   &  U   &$-$       & $<$35.23 & $-$       \\
NGC~3198  & SBc            &  H            &  46.1    &  5.57            & 36.97    & $-$3.19  &   I     & A         &  35.02    &  35.19  \\
NGC~3245 & SA(r)0?         & H          & 209.9 & 8.38$^{*}$      & 38.70    & $-$4.23   &         I   & E        & 35.71    & 36.36  \\
IC~2574    & SAB(s)m       & H               & 33.9  & 5.03             & 35.55    & $-$4.03    & U   &$-$       & $<$33.66 & $-$       \\
NGC~3254 & SA(s)bc         & S               & 117.8 & 7.21             & 38.60    & $-$3.16   &     U   & $-$       & $<$35.27 & $-$       \\
NGC~3294  & SAc           &  H                &   56.4  &  5.92         &  38.33     &  $-$2.18   &   U  &   $-$      & $<$35.66 & $-$  \\
NGC~3301 & (R')SB(rs)0/a    & L         & 132.3 & 7.41             & $<$38.30 & $<-$3.66  &        I   & A        & 35.70    & 35.86    \\
NGC~3310 & SAB(r)bc pec   & H               & 84.0   & 6.70$^{*}$       & 38.43    & $-$2.82  &     I   & A        & 35.60    & 36.20       \\
NGC~3319  & SBcd           &   L             &  87.4    &  6.68        &   37.07   &  $-$4.20     &     U &   $-$      & $<$34.84 & $-$  \\
NGC~3344 & (R)SAB(r)bc     & H               & 73.6  & 6.38             & 38.24    & $-$2.69   &     U   &$-$       & $<$34.11 & $-$       \\
NGC~3359 & SAB(rs)cd: pec  & H               & 96.5  & 6.86             & 37.93   & $-$3.48    &    U   &$-$       &  $<$35.21 & $-$      \\
NGC~3348 & E0                 & ALG             & 236.4 & 8.42              & $-$      & $-$       &    I   & C        & 36.59    & 36.73    \\
NGC~3395 & SAB(rs)cd: pec  & H               & 96.5  & 6.86             & 37.93   & $-$3.48 &    U   &$-$       & $<$35.32  &  $-$     \\
NGC~3414  & SA0 pec         & L                &  236.8   &  8.40$^{*}$ &  39.06   &  $-$3.95  &  I   & C    &  36.20    &  36.33  \\
NGC~3430  & SABc              & H                &  50.4    &   5.72        &   37.74   & $-$2.57  &  I  & E      &   35.74   &  36.19  \\
NGC~3432  & SBm              &   H               &  37.0       &  5.18        &  38.00    &  $-$1.77   & I  & A    &  34.82   &  34.83   \\
NGC~3448 & IAO             & H               & 50.7  & 5.73              & 39.48    & $-$0.80  &    I   & A        & 35.55    & 36.35       \\
NGC~3486 & SAB(r)c        & S               & 65.0    & 6.17             & 37.93    & $-$2.79   &    U   &$-$       & $<$34.26 & $-$       \\
NGC~3504 & (R)SAB(s)ab  & j\,H               & 119.3 & 7.23             & 39.88    & $-$1.90   &    I   & A        & 37.30    & 37.65    \\
NGC~3516 & (R)SB(s)0:     & S               & 181.0   & 7.96             & 40.80    & $-$1.71  &    I   & C        & 36.84    & 36.95    \\
NGC~3556 & SB(s)cd spin  &  H              & 79.4  & 6.52             & 37.62    & $-$3.45   &    U   &$-$       & $<$34.92 & $-$       \\
NGC~3583  & SBb             &  H               &  131.7   & 7.40        &   38.26  &  $-$3.73    &   U &   $-$      &   $<$35.69  &  $-$ \\
NGC~3600  &  SAa?          &  H               &    49.8   &  5.70        &   38.21  &  $-$2.08    &  U  &   $-$      & $<$34.66 & $-$  \\
NGC~3610 & E5:              & ALG             & 161.2 & 7.75             & $-$      & $-$      &    U   &$-$       & $<$35.56 & $-$       \\
NGC~3613  & E6              & ALG             & 220.1 & 8.30             & $-$      & $-$      &    U   &$-$       & $<$35.69 & $-$       \\
NGC~3631 & SA(s)c          & H               & 43.9  & 5.48             & $<$38.01    & $<-$2.02  &    U   &$-$       & $<$35.23 & $-$       \\
NGC~3646 & Ring            & L               & 153.1 & 7.66             & 39.15    & $-$3.06   &    U   & $-$       & $<$35.98 & $-$       \\
NGC~3642 & SA(r)bc:        & L          & 85.0    & 7.42$^{*}$       & 38.96    & $-$3.01      &    U   & $-$        & $<$35.55 & $-$      \\
NGC~3652  & SAcd?         &  H         &   56.4   &  5.92          &  38.51    & $-$2.00      &    U  &   $-$      & $<$35.73 & $-$  \\ 
NGC~3665  & SA0            &  j\,H$^{RL}$         &   236.8  &  8.76$^{*}$   &  38.28  &  $-$4.73    &   I    & B  &    36.85   &  37.67 \\
NGC~3675  & SAb             &  L        &  108.0       &  7.26$^{*}$   &  37.79  &  $-$3.85  &  I    & A   &   34.96    &  35.17   \\
NGC~3690 & IBm pec       & H                & 47.6  & 5.62            & 37.04    & $-$3.13  &    unI &$-$       &  $<$36.21 & $-$          \\
UGC~6484 & SB(rs)c         & H               & 61.1  & 6.06             & 37.47    & $-$3.14   &   U   &$-$       & $<$35.67 & $-$       \\
NGC~3718 & SB(s)a pec    & L$^{RL}$         & 158.1 & 7.72             & 37.41    & $-$4.86 &    I   & A        & 36.78    &  36.78        \\
NGC~3726  & SBa pec       & H       &  41.5    &   5.38         &  37.80    & $-$2.17   &  U &   $-$      & $<$34.99 & $-$  \\
NGC~3729  & SB(r)a pec   & H                & 76.2  & 6.45             & 36.60    & $-$4.40 &      I   & A        & 35.85   &  36.34       \\
NGC~3738  & IAm            & H                & 49.1  & 5.68            & 37.77    & $-$2.46  &       U   &$-$       & $<$33.89 & $-$        \\
NGC~3735  & SAc: spin    & S                & 140.6 & 7.51            & 39.88    & $-$2.18  &      I   & A        & 36.24    & 36.50    \\
NGC~3756  & SAB(rs)bc   & H                & 47.6  & 5.62            & 37.04    & $-$3.13  &      U   &$-$       & $<$35.29  & $-$       \\
NGC~3780  & SA(s)c:     & L                & 89.8  & 6.73            & 37.43    & $-$3.85  &      U   &$-$       & $<$35.78  &  $-$         \\
NGC~3813   & SA(rs)b:    & H                & 72.1  & 6.35            &  37.59    & $-$3.31  &   U   &$-$       & $<$35.53 & $-$       \\
NGC~3838   & SA0/a?      & ALG         & 141.4 & 7.52            & $-$      & $-$           & unI &$-$       & $<$35.44 &   $-$       \\
NGC~3877  & SAc           &  H            &   86.1   & 6.66          &   37.86  & $-$3.39   &  U &   $-$      & $<$36.47 & $-$ \\
NGC~3884   & SA(r)0/a      & L                & 208.3 & 8.20            & 40.20    & $-$2.55       & I   & A        & 37.79    & 37.80    \\
NGC~3893  & SABc        &  H             &    85.3     &  6.64        &   37.44    & $-$3.79     &   U &  $-$      & $<$34.96 & $-$  \\
NGC~3900   & SA(r)0+    & ALG              & 139.2 & 7.50            & $<$38.35  & $<-$3.70   & U   &$-$       & $<$35.63 & $-$       \\
NGC~3898   & SA(s)ab   & L         & 206.5 & 8.19            & 38.52    & $-$4.03      & I   & A         & 35.84    &  36.11       \\
NGC~3917   & SA(s)ab  & L                & 38.0    & 5.23             & 36.90    & $-$2.88      & U   &$-$        & $<$35.04 & $-$      \\
NGC~3938  & SAc        &   H            &   29.1    &  4.76        &   37.61    & $-$1.74     &  I & A   &    35.16   &  35.49   \\
NGC~3941   & SB(s)0    & S           & 133.0   & 7.42            & 38.52    & $-$3.45  & I   & A         & 35.46    & 35.70   \\
NGC~3945   & (R)SB(rs)0+  & L                & 191.5 & 6.94$^{*}$       & 38.39    & $-$3.10       & I   & C         &  35.77  & 35.86    \\
NGC~3949  &  SAbc       &  H         & 82.0    &  6.57   &     37.44    & $-$3.72      &    U &  $-$      & $<$35.05 & $-$  \\
NGC~3953   & SB(r)bc        & L                & 116.0   & 7.33$^{*}$       & 37.90    & $-$3.98    & U   &$-$       & $<$35.07 & $-$      \\
NGC~3963   & SAB(rs)bc  & H                & 40.9  & 5.36             & 38.09    & $-$1.82     & I   & A         &  36.03  &  36.09       \\
NGC~3982   & SAB(r)b:   & S                & 73.0    & 6.95$^{*}$       & 39.83    & $-$1.67      & I   & A        &  36.15  &  36.36       \\
NGC~3992   & SB(rs)bc    &  L       & 148.4 & 7.51$^{*}$      & 37.10 & $-$4.96      & U   &$-$        & $<$35.07 & $-$     \\
NGC~3998    & SA(r)0?         & L$^{RL}$                 & 304.6 & 8.93$^{*}$       & 39.56 & $-$3.92        & I   & A        & 37.98     &  37.98     \\
NGC~4013  &  SAb            &   L            &   86.5   &  6.67  &    37.36   &  $-$3.90   & unI &  $-$      & $<$35.10 & $-$  \\
NGC~4026    & SA0 spin        &   L     & 177.2 & 8.26$^{*}$      & $-$      & $-$        & U   &$-$        & $<$35.11 & $-$     \\
NGC~4036   & SA0-            & L                & 215.1 & 7.89$^{*}$       & 39.16    & $-$3.28      & I   & C        & 35.99    & 36.94   \\
NGC~4041   & SA(rs)bc:    & H                & 95.0    & 6.83            & 38.24    & $-$3.14     & I   & E         & 35.50   & 36.51   \\
NGC~4051  &  SABbc       &   S              &     89.0     &  6.10$^{*}$  &  40.17  & $-$1.14    &     I & C    &   35.29   &   36.92   \\
NGC~4062   & SA(s)c     & H                & 93.2  & 6.80             & 36.51    & $-$4.84      & U   &$-$        & $<$34.63 & $-$      \\
NGC~4088   & SAB(rs)bc & H                & 77.0    & 6.79$^{*}$       & 37.48    & $-$3.86       & U   &$-$        & $<$35.10   &  $-$  \\
NGC~4096   & SAB(rs)c & H                & 79.5  & 6.52             & 36.15    & $-$4.92     & U   &$-$        & $<$34.53 & $-$      \\
NGC~4100   & (R')SA(rs)bc & H                & 75.5  & 6.43             &  37.95    & $-$3.03     & I   & A         & 35.36    & 35.76  \\
NGC~4102   & SAB(s)b?   & H                & 174.3 & 7.89             & 39.10    & $-$3.34      & I   & E         & 35.91    & 37.55   \\
NGC~4111   & SA(r)0+: spin & L                & 147.9 & 7.60             & 38.65    & $-$3.50       & I+unI   & A     & 35.41    & 35.83   \\
NGC~4125   & E6 pec & L           & 226.7 & 8.35            & 38.71    & $-$4.19      & U   &$-$        & $<$35.36 & $-$      \\
NGC~4136   & E6 pec & H                & 38.4  & 5.25            & 37.39    & $-$2.41       & U   &$-$       & $<$34.66 & $-$       \\
NGC~4138   & SA(r)0+       & L                & 120.9 & 7.25            & 38.75    & $-$3.05        & U   &$-$       & $<$35.29 & $-$       \\
NGC~4143  & SAB(s)0        & L                & 204.9 & 7.92$^{*}$       & 38.81    & $-$3.66     & I   & B      & 36.13   & 36.22   \\
NGC~4144   & SAB(s)cd? spin & H                & $<$64.3  & $<$6.15       & 36.26    & $>-$4.44     & U   &$-$      & $<$34.02 & $-$      \\
NGC~4145   & SAB(rs)d    & H          & $-$   & 5.33$^{*}$        & 36.80    & $-$3.08      & U   &$-$        & $<$35.18 & $-$      \\
NGC~4151   & (R')SAB(rs)ab:  & S                & 97.0    & 7.81$^{*}$       & 41.74    &  $-$0.62     & I   & C        & 37.75    & 38.35   \\
NGC~4150 & SA(r)0?         & L                & 87.0    & 5.94$^{*}$       & 35.83  & $-$4.66       & U   &$-$      & $<$34.62 & $-$      \\
NGC~4157 & SAB(s)b? spin   &  H               & 90.1  & 6.74             & 36.96    & $-$4.33        & U   &$-$        & $<$35.07 & $-$      \\
NGC~4162 & (R)SA(rs)bc     & H                & 76.1  & 6.44             & 38.44    & $-$2.55      & unI &$-$        & $<$35.98 & $-$      \\
NGC~4169 & SA0             & S                & 183   & 7.97              & 38.99    & $-$3.53    & U   &$-$        & $<$36.08 & $-$      \\
NGC~4183 & SA(s)cd? spin     & H                & 34.4  & 5.06              & 37.90    & $-$1.71   & U   &$-$        & $<$35.17 & $-$      \\
NGC~4203 & SAB0-:          &  L               & 167.0   & 7.82$^{*}$        & 38.28   & $-$4.09  & I   & A        & 36.09    & 36.11  \\
NGC~4214 & IAB(s)m         & H                & 51.6  & 5.76             & 38.71    & $-$1.60   & U   &$-$        & $<$33.72 & $-$      \\
NGC~4217 & SAb spin        & H                & 91.3  & 6.76             & 36.93    & $-$4.38     & I   & C        & 35.23    & 35.96    \\
NGC~4220 & SA(r)0+          & L                & 105.5 & 7.01              & 36.19    & $-$5.37  & I   & A        & 35.25    & 35.72    \\
NGC~4236 & SB(s)dm          & H                & $<$62.8  &  $<$6.11           & 36.24    & $>-$4.42     & unI &$-$   & $<$33.27  & $-$         \\
NGC~4244 & SA(s)cd: spin     & H                & 36.8  & 5.17             & 36.05    & $-$3.67   & I   & A        & 33.83 & 33.94        \\
NGC~4242 & SAB(s)dm       & H        & $-$   & $-$               & $<$36.30   &   $-$       & unI &$-$       & $<$34.42 & $-$            \\
NGC~4245 & SB(r)0/a:      & H                & 82.7  & 7.19$^{*}$        & 37.36    & $-$4.38    & U   &$-$     & $<$34.64 & $-$            \\
NGC~4251 & SB0? spin      & ALG              & 119.4 & 7.23               & $-$      & $-$     & U   &$-$       & $<$34.66 & $-$            \\
NGC~4258 & SAB(s)bc       & S                & 148.0   & 7.58$^{*}$        & 38.76    & $-$3.37    & I   & A      & 34.93    & 35.06         \\
NGC~4274 & (R)SB(r)ab      & L        & 96.6  & 6.86              & 38.49    & $-$2.92   & U   &$-$        & $<$34.62 & $-$           \\
NGC~4278 & E1+             & L$^{RL}$                  & 261.0   & 7.96$^{*}$        & 38.88    & $-$3.63   & I   & A     & 37.61    & 37.63         \\
NGC~4291 & E               &   ALG              & 285.3 & 8.99$^{*}$         & $-$      & $-$         & U   &$-$     & $<$35.50 & $-$            \\
NGC~4314 & SB(rs)a        & L         & 117.0   & 6.91$^{*}$        & 37.75    & $-$3.71    & U   &$-$    & $<$34.65 & $-$           \\
NGC~4346 & SA0 spin        &    L         & 146.5 & 7.59               & 37.88    & $-$4.26     & U   &$-$       & $<$35.28 & $-$            \\
NGC~4369 & (R)SA(rs)a      & H                & 71.6  & 6.34                & 38.83    & $-$2.06     & I   & A        &  35.18  &  35.40        \\
NGC~4395 & SA(s)m:         &    S                & 26.0    & 4.57                & 38.35    & $-$0.77    & I   & A        & 34.17    & 34.39         \\
NGC~4414 & SA(rs)c?        &  L         & 117.0   & 7.19                 & 36.46    & $-$5.28    & U   &$-$        & $<$34.68 & $-$           \\
NGC~4449 & IBm             & H                & 17.8  & 3.91                & 38.28    & $-$0.18  & U   &$-$       & $<$33.58 & $-$            \\
NGC~4448 & SB(r)ab         & H                & 119.8 & 7.24                & 37.34    & $-$4.45     & U   &$-$       & $<$34.47 & $-$            \\
NGC~4460 & SB(s)0+? spin   & H                & 39.8  & 5.31                  & 38.09    & $-$1.77      & U   &$-$       & $<$34.36 & $-$            \\
NGC~4485 & IB(s)m pec     & H           & 52.2  & 5.78                & 36.95    & $-$3.38    & U   &$-$        & $<$34.32 &  $-$          \\
NGC~4490 & SB(s)d pec      & H          & 45.1  & 5.53                 & 37.12    & $-$2.96   & U   &$-$       & $<$34.33 & $-$            \\
NGC~4494 & E1+             & L           & 145.0   & 7.57               & 37.35    & $-$4.77    & I   & A        & 34.69    & 34.92         \\
NGC~4559 & SAB(rs)cd        & H                 & 49.2  & 5.68               & 37.01    & $-$3.22      & U   &$-$      & $<$34.49 & $-$             \\
NGC~4565 & SA(s)b? spin     & S                 & 136.0   & 7.46               & 38.22    & $-$3.79    & I   & A       & 35.20    & 35.31          \\
NGC~4589 & E2               & L$^{RL}$                  & 224.3 & 8.33               & 38.78    & $-$4.10    & I   & C       & 37.45    & 37.60          \\
NGC~4605 & SB(s)c pec       & H                 & 26.1  & 4.57               & 36.63    & $-$2.49      & U   &$-$      & $<$33.77 & $-$             \\
NGC~4618 & SB(rs)m           & H                 & $<$54.6 & $<$5.86          & 37.97    & $>-$2.44     & U   &$-$   & $<$34.28 & $-$             \\
NGC~4648 & E3                & ALG               & 224.5 & 8.33                & $-$      & $-$       & U   &$-$      & $<$35.45 & $-$             \\
NGC~4631 & SB(s)d spin       & H                 & $<$71.9 & $<$6.34         & 37.12    & $>-$3.77        & unI &$-$        & $<$34.30 & $-$        \\
NGC~4656 & SB(s)m pec        & H                 & 70.4  & 6.31                & 37.93    & $-$2.93      & U   &$-$      & $<$34.32 & $-$             \\
NGC~4750 & (R)SA(rs)ab      & L                 & 136.0   & 7.46                 & 38.76    & $-$3.25      & I   & A       & 35.84    & 36.17          \\
NGC~4725 & SAB(r)ab pec      & S                 & 140.0   & 7.51                & 38.51    & $-$3.55       & U   &$-$      & $<$34.78 & $-$             \\
NGC~4736 & (R)SA(r)ab        &  L         & 112.0   & 7.12               & 37.33    & $-$4.08        & I   & A        & 34.82    & 35.29        \\
NGC~4800 & SA(rs)b           & H                 & 111.0   & 7.02$^{*}$          & 37.62    & $-$3.95        & U   &$-$    & $<$34.91 & $-$            \\
NGC~4793 & SAB(rs)c          & H                 & 26.6  & 4.61                 & $<$38.84    & $<-$0.32   & U   &$-$       & $<$35.78 & $-$            \\
NGC~4826 & (R)SA(rs)ab       & L          & 96.0    & 6.85                & 37.92 & $-$3.48          & I   & A         & 33.88    & 34.93        \\
NGC~4914 & E                     &   ALG      &     224.7   &  8.33     &    $-$       &    $-$     &  U &   $-$      & $<$36.17 & $-$  \\
NGC~5005  & SABbc          &   L           &   172.0       &   8.27$^{*}$        &  39.41  &  $-$3.05    & I  & D  &  36.29    &  37.54    \\
NGC~5012 & SAB(rs)c          & L                 & 141.4 & 7.52                & 38.58    & $-$3.49       & unI &$-$       & $<$35.84 & $-$            \\
NGC~5033 & SA(s)c          &  L                & 151.0   & 7.64                & 39.34    & $-$2.85      & I   & A        & 36.04   & 36.42         \\
NGC~5055  & SAbc          &   L               &   117.0   &  8.92$^{*}$  & 37.44   &  $-$6.07    &    U &  $-$      & $<$34.37 & $-$  \\
NGC~5112  & SBcd           & H               &  $<$60.8    &  $<$6.05    &   37.42   & $-$3.22   &   U &  $-$      & $<$35.40 & $-$  \\
NGC~5204 & SA(s)m            & H                 & 39.9  & 5.32                & 36.58    & $-$3.29     & U   &$-$       & $<$34.14 & $-$            \\
NGC~5194  & SAbc pec     & S                  &  96.0     &  6.85                &  38.91    & $-$2.53    &  I  & D  &  35.06   & 36.14    \\
NGC~5195  & IA0  pec      &  L               &    124.8   &  7.31             &  37.84    &  $-$4.06    &   I & C   &  34.90   &  35.59  \\
NGC~5273  & SA0            &   S            &    71.0          &  6.61$^{*}$    &   39.82    & $-$1.38     &   unI  & $-$      & $<$35.31 & $-$  \\
NGC~5297  & SABc          &   L           &     61.3       &   6.07            &   38.22    &   $-$2.44   &  U & $-$      & $<$35.89 & $-$  \\
NGC~5308 & SA0- spin      & ALG               & 249.0   & 8.51                & $-$      & $-$          & U   &$-$       & $<$35.82 & $-$            \\
NGC~5322 & E3+             & L$^{RL}$            & 232.2 & 8.39                & 38.20    & $-$4.74       & I   & B        & 36.96    & 37.51         \\
NGC~5353  & SA0             &   L$^{RL}$         &   286.4      &   8.76       &   38.73   & $-$4.62    &   I  & A   &  37.63     &  37.66   \\
NGC~5354 & SA0 spin        & L$^{RL}$                 & 217.4 & 8.28                 & 38.61    & $-$4.22       & I   & A        & 36.94    & 36.97              \\
NGC~5371  & SABbc           &   L             &  179.8   &  7.94              &  39.03   &  $-$3.50   &   U & $-$      & $<$35.79 & $-$  \\
NGC~5377  & SBa              &   L              &   169.7    &  7.84             &  38.81   &  $-$3.62   &  I & C  &  35.71  &  36.35    \\
NGC~5383  & SBb pec       &  H              &   96.5      &  6.86             &  38.07   & $-$3.38     &   U &  $-$      & $<$35.82 & $-$  \\
NGC~5395  & SAb pec       &   L             &   145.5     &  7.57            &  38.66   &  $-$3.50    &   U & $-$      & $<$35.96 & $-$  \\
NGC~5448  & SABa            &    L            &   124.5     &  7.30           &   38.55  &  $-$3.34    &  I   & C   &  35.73   & 36.46  \\
NGC~5457 & SAB(rs)cd         & H                 & 23.6  & 6.41$^{*}$          & 36.99    & $-$3.97       & U   &$-$       & $<$34.28 & $-$           \\
NGC~5473 & SAB(s)0-:       & ALG               & 220.5 & 8.30                & $-$      & $-$            & U   &$-$       &  $<$35.82 & $-$           \\
NGC~5474 & SA(s)cd pec       & H                 & 29.0    & 4.76                & 37.35    & $-$1.96       & U   &$-$       & $<$34.38 & $-$            \\
NGC~5485 & SA0 pec           & L$^{RL}$            & 207.5 & 8.19                 & 38.31    & $-$4.43      & I   & A        & 36.44    & 36.44         \\
NGC~5523  & SAcd             &  H         &   30.1     &   4.82            &   37.25    &  $-$2.16      & U   &  $-$      & $<$35.29 & $-$ \\
NGC~5548 & (R')SA(s)0/a    & S                 & 291   & 7.70$^{*}$          & 41.60    & $-$0.65     & I   & B        &  36.96   & 37.61          \\
NGC~5557  & E1                &   ALG           &    295.3  &  8.81             &   $-$     &   $-$        &    U   &  $-$      & $<$35.94 & $-$  \\
NGC~5585 & SAB(s)d           & H                 & 42.0    & 5.41                & 37.68    & $-$2.28     & U   &$-$       & $<$34.53 & $-$            \\
NGC~5631 & SA(s)0            & L          & 168.1 & 7.83                & 38.76    & $-$3.62     & U   &$-$        & $<$35.89 & $-$           \\
NGC~5660  & SABc            &   H         &   60.7      &  6.05         &   38.10    &  $-$2.54   &   U  & $-$      & $<$35.76 & $-$  \\
NGC~5656  & SAab            &    L         &   116.7    &  7.19         &   37.99    &  $-$3.79   &  U   &  $-$      & $<$35.92 & $-$ \\
NGC~5678 & SAB(rs)b          & H                 & 132.8 & 7.42                & 38.27    & $-$3.70       & U   &$-$       & $<$35.93 & $-$            \\
NGC~5676  & SAbc              &  H               &   116.7    &  7.19         &  37.96    &  $-$3.82     & unI  & $-$      & $<$35.74 & $-$  \\
NGC~5866  & SA0                &  L$^{RL}$               &   169.1    & 7.84           &   37.50    &  $-$4.93    &   I    & D  &  36.50   &  36.76    \\
NGC~5879  & SAbc              &   L            &   73.9       &   6.62$^{*}$  &  37.89   &  $-$3.09    & I  & A   &   35.27    &   35.39   \\
NGC~5905  & SBb                &   H          &   174.6      &  7.89           &   39.03    &  $-$3.45    &  U  &  $-$      & $<$35.90 & $-$  \\
NGC~5907  & SAc              &     H         &  120.2       &  7.24          &   36.88     &  $-$4.94     & unI  & $-$      & $<$35.03 & $-$  \\
NGC~5982  & E3                &  ALG        &    239.4     &  8.44          &  $<$38.55   & $<-$4.48 &   U & $-$      & $<$35.80 & $-$  \\
NGC~5985  & SABb            &   L           &  157.6        &  7.71          &   38.76      &   $-$3.54   &  I  & D   &  35.76   & 36.30    \\
NGC~6015  & SAcd           &  H            &  43.5         &   5.47        &   37.18      &  $-$2.88      & unI & $-$      & $<$35.04 & $-$  \\
NGC~6140  & SBcd pec    &    H           &   49.4       &   5.69        &   37.52      &   $-$2.76     &   U &  $-$      & $<$35.15 & $-$  \\
NGC~6217 & (R)SB(rs)bc       & H                 & 70.3  & 6.30                & 39.25    & $-$1.60        & I   & A        & 35.66    & 36.59         \\
NGC~6207 & SA(s)c            & H                 & 92.1  & 6.78                & 38.30    & $-$3.03        & U   &$-$       & $<$35.19 & $-$            \\
NGC~6236 & SAB(s)cd          & H                 & 46.1  & 5.57                & 37.93    & $-$2.19        & U   &$-$       & $<$35.40 & $-$            \\
NGC~6340 & SA(s)0/a          & L                 & 143.9 & 7.56                & 38.31    & $-$3.80        & I   & A        & 35.53    & 36.03         \\
NGC~6412 & SA(s)c            & H                 & 49.9  & 5.71                & 37.59    & $-$2.67        & U   &$-$       & $<$35.40 & $-$            \\
NGC~6503 & SA(s)cd          & L                & 46.0    & 6.30$^{*}$           & $-$      & $-$        & U   &$-$      & $<$34.30 & $-$           \\
NGC~6482 & E:                 & L           & 310.4 & 8.90                 & $-$      & $-$       & I   & A         & 36.37 & 36.74           \\
NGC~6643 & SA(rs)c         & H                 & 95.4  & 6.84                & 37.52    & $-$3.87          & U   &$-$        & $<$35.47 & $-$           \\
NGC~6654 & (R')SB(s)0/a   & ALG               & 172.2 & 7.87                & $-$      & $-$        & U   &$-$        & $<$35.61 & $-$           \\
NGC~6689 & SAd? spin       & L                 & 26.0    & 4.57                & 37.17    & $-$1.95    & U   &$-$        & $<$34.85 & $-$           \\
NGC~6702  &  E               &  ALG             &  173.6   &  7.88              &  $-$      &   $-$       &    I  & C    &   36.19   &  36.85   \\
NGC~6703  & SA0           &  L                   &   179.9  &   7.95          &   38.46   &  $-$4.08   &  I & A   &  35.78    &   35.91   \\
NGC~6946  & SABcd       &   H                 &   55.8    &   5.90          &  37.03    &  $-$3.46   &   I & E    &   34.43   & 35.73   \\ 
NGC~6951  & SABbc       &   L                 &   127.8   &  6.93$^{*}$  &  38.69   &  $-$3.25   &   I & C    &    35.42  & 36.02    \\
NGC~7080 & SB(r)b         & H                  & 95.3  & 6.84                 & $-$      & $-$             &  I & A         & 36.50     &  36.91            \\   
NGC~7217  & SAab       &  L                    &  141.4    &  7.52           &  38.31   &  $-$3.80    &   I & C    &   35.09   &  35.87  \\
NGC~7331  & SAb         &   L                  &   137.2    &  8.02$^{*}$   &    38.30  &  $-$3.76     &    U  &  $-$ & $<$34.95    & $-$    \\
NGC~7332  &  SA0 pec    &    ALG        &    124.1     &  7.08$^{*}$   &   $-$     &  $-$       &   U &   $-$      & $<$35.18 & $-$  \\
NGC~7457  & SA0?        &   ALG             &   69.4      &  6.95$^{*}$  &  $-$      &   $-$       &    U &  $-$      & $<$34.88 & $-$   \\
NGC~7640  & H            &   H                 &   48.1      &   5.64          &   36.84  &  $-$3.39  &   U &  $-$      & $<$34.51 & $-$  \\
NGC~7741  & SBcd       &  H                 &  29.4        &  4.78           &    37.91    & $-$1.46  & U &  $-$      & $<$34.70 & $-$ \\
NGC~7798  &  S            &  j\,H                &   75.1        &  6.42          &  38.64    &  $-$2.37   &  I  & C    &  35.64     &  36.54  \\
 \hline
\hline
\end{longtable}
\end{center}
\vspace{-1.2cm}
Column description: (1) source name; (2)   morphological galaxy type taken from RC3
\citep{devaucouleurs91}; (3) optical spectroscopic classification based on BPT diagrams
and from the literature. H=H{\sc ii}, S=Seyfert, L=LINER, and ALG=Absorption line galaxy. `j\,H' marks the jetted H{\sc ii} galaxies and `RL' identifies the RL AGN; (4) stellar velocity dispersion $\sigma$ (\kms) from
\citet{ho09}; (5) logarithm of BH mass ($M_{\odot}$) determined
from $\sigma$ \citep{tremaine02} or from direct BH mass measurements
(galaxies marked with $^{*}$, \citealt{vandenbosch16}) ; (6) logarithm of [\ion{O}{iii}] luminosities from \citet{ho97a} or from the literature (non corrected for extinction, see Paper~II for references); (7) logarithm of Eddington ratio ($L_{\rm Bol}$/$L_{\rm Edd}$); (8) radio detection
status: `I' = detected and core identified; `U' =
undetected; `unI' = detected but core unidentified; `I+unI' =  detected and core identified with additional unknown source(s) in the field; (9) radio
morphological class: A = core/core--jet; B = one-sided jet; C =
triple; D = doubled-lobed ; E = complex; (10)--(11) logarithm of radio core and total luminosities at 1.5 GHz (erg s$^{-1}$).  To convert the radio luminosities in erg s$^{-1}$ to W Hz$^{-1}$ at 1.5 GHz, an amount of +16.18 should be subtracted from log $L_{\rm core}$ and log $L_{\rm Total}$.
\twocolumn

\label{lastpage}
\end{document}